\documentclass[hyper,a4paper]{JHEP3} 

\usepackage{epsfig}
\usepackage{latexsym}
\usepackage{graphicx}
\usepackage{amsmath}
\usepackage{amsfonts}   
\usepackage{amssymb}    

\def\mt{{\ifmmode M^{eff}_T\else $M^{eff}_T$\fi}}

\def\e{\epsilon}

\def\ra{\rangle}

\def\e3{$\epsilon_3$}

\def\ch2{$\chi2$}

\def\co#1{{\ifmmode{\cal O}_{#1}\else${\cal O}_{#1}$\fi}}

\newdimen\unit
\def\point#1 #2 #3{\vbox to0pt{\kern-#2\unit
  \hbox{\kern#1\unit#3}\vss}
 \nointerlineskip}
\newcommand{\be}{\begin{equation}}
\newcommand{\ee}{\end{equation}}
\newcommand{\bea}{\begin{eqnarray}}
\newcommand{\eea}{\end{eqnarray}}

\newcommand{\mev}{\mbox{ MeV }}
\newcommand{\gev}{\mbox{ GeV}}
\newcommand{\tev}{\mbox{ TeV}}

\newcommand{\cl}{\text{CL}}

\newcommand{\dasusy}{\delta a_{\mu}^{\text{SUSY}}}
\newcommand{\alphaemmz}{\alpha_{\text{em}}(M_Z)^{\overline{MS}}}
\newcommand{\alphas}{\alpha_s(M_Z)^{\overline{MS}}}

\newcommand{\data}{d}

\newcount\hour
\newcount\minute
\newtoks\amorpm
\hour=\time\divide\hour by60 \minute=\time{\multiply\hour by60
\global\advance\minute by- \hour}
\edef\standardtime{{\ifnum\hour<12 \global\amorpm={am}%
    \else\global\amorpm={pm}\advance\hour by-12 \fi
    \ifnum\hour=0 \hour=12 \fi
    \number\hour:\ifnum\minute<100\fi\number\minute\the\amorpm}}
\edef\militarytime{\number\hour:\ifnum\minute<100\fi\number\minute}
\def\bold#1{\setbox0=\hbox{$#1$}%
     \kern-.025em\copy0\kern-\wd0
     \kern.05em\copy0\kern-\wd0
     \kern-.025em\raise.0433em\box0 }

\newcommand{\newc}{\newcommand}
\newc\eg{{\it {e.g.}}}  \newc\etal{{\it {et al.}}} \newc\ie{{\it i.e.}}
\newc\etc{{\it {etc}}}
\newcommand\lsim{\mathrel{\rlap{\lower4pt\hbox{\hskip1pt$\sim$}}
    \raise1pt\hbox{$<$}}}
\newcommand\gsim{\mathrel{\rlap{\lower4pt\hbox{\hskip1pt$\sim$}}
    \raise1pt\hbox{$>$}}}
\newc{\mhalf}{m_{1/2}}      \newc{\mzero}{m_0}
\newc{\tanb}{\tan\beta}
\newc{\azero}{A_0}
\newc{\at}{A_t} \newc{\ab}{A_b} \newc{\atau}{A_\tau}
\newc{\bmu}{B\mu}           \newc{\sgn}{{\rm sgn}}
\newc{\mone}{M_1}           \newc{\mtwo}{M_2}
\newc{\charone}{\chi_1^\pm} \newc{\mcharone}{m_{\chi_1^\pm}}
\newc{\hl}{h}               \newc{\mhl}{m_{\hl}}
\newc{\hh}{H}               \newc{\mhh}{m_{\hh}}
\newc{\ha}{A}               \newc{\mha}{m_{\ha}}
\newc{\hc}{H^{\pm}}         \newc{\mhc}{m_{\hc}}
\newc{\qzero}{Q_0}          \newc{\qstop}{Q_{\widetilde t}}
\newc{\amu}{a_{\mu}}        \newc{\amususy}{a_{\mu}^{\rm SUSY}}
\newc{\amuexpt}{a_{\mu}^{\rm expt}}        \newc{\amusm}{a_{\mu}^{\rm SM}}
\newc{\deltaamususy}{\Delta a_{\mu}^{\rm SUSY}}
\newc\gmtwo{(g-2)_{\mu}} \newc\deltaamu{\Delta a_{\mu}}
\newc{\msbar}{\overline {\rm MS}} \newc{\drbar}{\overline {\rm DR}}
\newc{\yt}{h_t} \newc{\yb}{h_b} \newc{\ytau}{h_{\tau}}
\newc{\mtpole}{M_t}
\newc{\mtaupole}{m_{\tau}^{\rm pole}}
\newc{\mtmtsmmsbar}{m_t(m_t)^{\msbar}_{{\rm SM}}}
\newc{\mtmtsmdrbar}{m_t(m_t)^{\drbar}_{{\rm SM}}}
\newc{\mtmtmssmdrbar}{m_t(m_t)^{\drbar}_{{\rm SUSY}}}
\newc{\mbmbsmmsbar}{m_b(m_b)^{\msbar}_{{\rm SM}}}
\newc{\mbmzsmmsbar}{m_b(\mz)^{\msbar}_{{\rm SM}}}
\newc{\mbmzsmdrbar}{m_b(\mz)^{\drbar}_{{\rm SM}}}
\newc{\mbmzmssmdrbar}{m_b(\mz)^{\drbar}_{{\rm SUSY}}}
\newc{\mtaumzsmmsbar}{m_{\tau}(\mz)^{\msbar}_{{\rm SM}}}
\newc{\mtaumzsmdrbar}{m_{\tau}(\mz)^{\drbar}_{{\rm SM}}}
\newc{\mtaumzmssmdrbar}{m_{\tau}(\mz)^{\drbar}_{{\rm SUSY}}}
\newc{\mgut}{M_{\rm GUT}}
\newc{\mplanck}{M_{\rm P}}      \newc{\mpl}{M_{\rm Pl}}
\newc{\msusy}{M_{\rm SUSY}}      \newc{\ms}{M_{\rm S}}
\newc{\jxf}{J({\xf})}
\newc{\jxfexact}{J_{\rm exact}({\xf})}  \newc{\jxfexp}{J_{\rm exp}({\xf})}
\newc{\VEV}[1]{\langle #1 \rangle}
\newc{\xf}{x_f}
\newc\vrel{v_{\rm rel}}
\newcommand\mchi{m_{\chi}}              
\newc\sell{{\widetilde e}_L}      \newc\msell{m_{\sell}}
\newc\selr{{\widetilde e}_R}      \newc\mselr{m_{\selr}}
\newc\snue{{\widetilde \nu}_e}      \newc\msnue{m_{\snue}}
\newc\snutau{{\widetilde \nu}_\tau}      \newc\msnutau{m_{\snutau}}
\newc\supl{{\widetilde u}_L}      \newc\msupl{m_{\supl}}
\newc\supr{{\widetilde u}_R}      \newc\msupr{m_{\supr}}
\newc\sdl{{\widetilde d}_L}      \newc\msdl{m_{\sdl}}
\newc\sdr{{\widetilde d}_R}      \newc\msdr{m_{\sdr}}
   
\newcommand\stopone{{\widetilde t}_1}

\newcommand\stauone{{\widetilde \tau}_1}   
   
\newcommand\gluino{\widetilde g}
\newcommand\mgluino{m_{\widetilde g}}

\newc\hpm{H^\pm} \newc\hp{H^+} \newc\hm{H^-}
\newc\sfermion{\tilde f}  \newc\msfermion{m_{\sfermion}}
\newc\kmeter{{\mbox km}}
\newc\second{{\rm sec}}
\newc{\gstar}{g_\ast}           \newc{\gsstar}{g_{s\ast}}
\newc{\geff}{g_{\rm eff}}
\newcommand\mz{m_{Z}}

\newc{\sthw}{\sin\theta_W}              \newc{\cthw}{\cos\theta_W}
\newc{\bino}{\widetilde B}              \newc{\wino}{\widetilde W_30}
\newc{\higgsinob}{{\widetilde H}^0_b}   \newc{\higgsinot}{{\widetilde H}^0_t}
\newc{\abund}{\Omega h^2}
\newc{\abundchi}{\Omega_\chi h^2}
\newc{\abundcdm}{\Omega_{\text{CDM}} h^2}
\newc{\omegam}{\Omega_{M}}       \newc{\abundm}{\Omega_{M} h2}
\newc{\omegab}{\Omega_{b}}       \newc{\abundb}{\Omega_{b} h2}
\newc{\omegacdm}{\Omega_{CDM}}
\newc{\omegatot}{\Omega_{TOT}}
\newc{\rhocrit}{\rho_{crit}}
\newc{\rhochi}{\rho_{\chi}}
\newcommand\pb{\,\mbox{pb}}
\newcommand\fb{\,\mbox{fb}}

\newcommand\mpc{\mbox{Mpc}}
\newc\br{\mbox{BR}}
\newcommand{\BRg}{BR({\bar B}\rightarrow X_s\gamma)}
\newc{\beq}{\begin{equation}}
\newc{\eeq}{\end{equation}}
\renewcommand\({\left(}
\renewcommand\){\right)}

\newc\stoponetwo{{\widetilde t}_{1,2}}
\newc\sbotonetwo{{\widetilde b}_{1,2}}
\newc\stauonetwo{{\widetilde \tau}_{1,2}}

\newc\bsgamma{b\ra s \gamma }
\newc\brbsgamma{\br( B\rightarrow X_s \gamma )}
\newc{\sigsip}{\sigma^{SI}_{p}} \newc{\sigsin}{\sigma^{SI}_{n}}
\newc{\sigsiN}{\sigma^{SI}_{N}}
\newc{\sigsdp}{\sigma^{SD}_{p}} \newc{\sigsdn}{\sigma^{SD}_{n}}
\newc{\sigsiA}{\sigma^{SI}_{A}}

\newc\xilim{\xi_{\rm lim}} 
\newc\tlim{t_{\rm lim}} 
\newc\zetalim{\zeta_{\rm lim}} 

\newc\zetah{\zeta_h}
\newc{\relprobone}[1]{p({#1} \vert d)}
\newc{\relprobtwo}[2]{p({#1},{#2} \vert d)}
\renewcommand\({\left(}
\renewcommand\){\right)}

\long\def\begincomment#1\endcomment{%
        \begingroup\sf\baselineskip12pt#1\endgroup}

\newc\AJ[3]
                {{\em Astrophys. J.} {\bf #1} (#2) #3}
\newc\AP[3]
                {{\em Ann. Phys.} {\bf #1} (#2) #3}
\newc\APJ[3]
                {{\em Astropart. J.} {\bf #1} (#2) #3}
\newc\APP[3]
                {{\em Astropart. Phys.} {\bf #1} (#2) #3}
\newc\APS[3]
                {{\em Astrophys. J. Suppl.} {\bf #1} (#2) #3}
\newc\ARNPS[3]
                {{\em Ann. Rev. Nucl. Part. Sci.} {\bf C#1} (#2) #3}
\newc\CPC[3]
                {{\em Comput. Phys. Commun.} {\bf C#1} (#2) #3}
\newc\EPJ[3]
                {{\em Euro. Phys. Journ.} {\bf C#1} (#2) #3}
\newc\JCAP[3]
                {{\em JCAP} {\bf #1} (#2) #3}
\newc\JHEP[3]
                {{\em JHEP} {\bf #1} (#2) #3}
\newc\IJMP[3]
                {{\em Int. J. Mod. Phys.} {\bf A#1} (#2) #3}
\newc\MNRAS[3]
                 {{\em Mon. Not. Roy. Astron. Soc.} {\bf #1} (#2) #3}
\newc\MPL[3]
                {{\em Mod. Phys. Lett.} {\bf A#1} (#2) #3}
\newc\NCA[3]
                {{\em Nuovo Cimento} {\bf #1} (#2) #3}
\newc\NIM[3]
                {{\em Nucl. Instrum. Methods} {\bf #1} (#2) #3}
\newc\NIMA[3]
                {{\em Nucl. Instrum. Methods} {\bf A#1} (#2) #3}
\newc\NAT[3]
                {{\em Nature} {\bf #1} (#2) #3}
\newc\NPB[3]
                {{\em Nucl. Phys.} {\bf B#1} (#2) #3}
\newc\PL[3]
                {{\em Phys. Lett.} {\bf B#1} (#2) #3}
\newc\PLB[3]
                {{\em Phys. Lett.} {\bf B#1} (#2) #3}
\newc\PR[3]
                {{\em Phys. Rep.} {\bf #1} (#2) #3}
\newc\PRL[3]
                {{\em Phys. Rev. Lett.} {\bf B#1} (#2) #3}
\newc\PRD[3]
                {{\em Phys. Rev.} {\bf D#1} (#2) #3}
\newc\PTP[3]
                {{\em Prog. Theor. Phys.} {\bf #1} (#2) #3}
\newc\RMP[3]
                {{\em Rev. Mod. Phys.} {\bf #1} (#2) #3 }
\newc\RPP[3]
                {{\em Rept. Prog. Phys.} {\bf #1} (#2) #3 }
\newc\SC[3]
                {{\em Science} {\bf #1} (#2) #3 }
\newc\ZPC[3]
                {{\em Z. Phys.} {\bf C#1} (#2) #3}
\newc\Err[3]
           {{\em Erratum-ibid.} {\bf #1} (#2) #3 }

\newcommand{\squishlist}{
   \begin{list}{$\bullet$}
    { \setlength{\itemsep}{0pt}      \setlength{\parsep}{3pt}
      \setlength{\topsep}{3pt}       \setlength{\partopsep}{0pt}
      \setlength{\leftmargin}{1.em} \setlength{\labelwidth}{1em}
      \setlength{\labelsep}{0.5em} } }
\newcommand{\squishend}{
    \end{list}  }
        
\newcommand{\nuis}{\psi} 
\newcommand{\params}{\theta} 
\newcommand{\basis}{\eta}
\newcommand{\derived}{\xi}
\newcommand{\trued}{\hat{\xi}}
\newcommand{\sineff}[1]{\protect{\sin^2\theta_{\text{eff}}^{#1}}}

\title{A Markov Chain Monte Carlo Analysis of the CMSSM}
\author{Roberto Ruiz de Austri\\
        Departamento de F\'{\i}sica Te\'{o}rica C-XI
        and Instituto de F\'{\i}sica Te\'{o}rica C-XVI,\\
        Universidad Aut\'{o}noma de Madrid, Cantoblanco,
        28049 Madrid, Spain\\
        E-mail: \email{rruiz@delta.ft.uam.es}}
\author{Roberto Trotta\\
        Astrophysics Department, Oxford University \\
        Denys Wilkinson Building,  Keble Road, Oxford OX1 3RH, United Kingdom\\
         E-mail: \email{rxt@astro.ox.ac.uk}}
\author{Leszek Roszkowski\\
        Department of Physics and Astronomy, University of Sheffield,\\
        Sheffield S3 7RH, England\\
        E-mail: \email{L.Roszkowski@sheffield.ac.uk}}

\abstract{\small We perform a comprehensive exploration of the
  Constrained MSSM parameter space employing a Markov Chain Monte
  Carlo technique and a Bayesian analysis. We compute superpartner
  masses and other collider observables, as well as a cold dark matter
  abundance, and compare them with experimental data. We include
  uncertainties arising from theoretical approximations as well as
  from residual experimental errors of relevant Standard Model
  parameters.  We delineate probability distributions of the CMSSM
  parameters, the collider and cosmological observables as well as a
  dark matter direct detection cross section. The 68\% probability
  intervals of the CMSSM parameters are: $0.52\tev<\mhalf< 1.26\tev$,
  $\mzero<2.10\tev$, $-0.34\tev<\azero<2.41\tev$ and
  $38.5<\tanb<54.6$.  Generally, large fractions of high probability
  ranges of the superpartner masses will be probed at the LHC.  For
  example, we find that the probability of $\mgluino< 2.7\tev$ is 78\%, of
  $m_{\tilde{q}_R}<2.5\tev$ is 85\% and of $\mcharone< 0.8\tev$ is
  65\%. As regards the other observables, for example at 68\%
  probability we find $3.5\times10^{-9}< BR(B_s \rightarrow \mu^+
  \mu^-)< 1.7 \times10^{-8}$, $1.9\times 10^{-10} < \dasusy <9.9\times
  10^{-10}$ and $ 1 \times 10^{-10}\pb <\sigsip< 1 \times 10^{-8}\pb$
  for direct WIMP detection. We highlight a complementarity between
  LHC and WIMP dark matter searches in exploring the CMSSM parameter
  space.  We further expose a number of correlations among the
  observables, in particular between $BR(B_s \rightarrow \mu^+ \mu^-)$
  and $\BRg$ or $\sigsip$. Once SUSY is discovered, this and other
  correlations may prove helpful in distinguishing the CMSSM from
  other supersymmetric models. We investigate the robustness of our
  results in terms of the assumed ranges of CMSSM parameters and the
  effect of the $\gmtwo$ anomaly which shows some tension with the
  other observables. We find that the results for $\mzero$, and the
  observables which strongly depend on it, are sensitive to our
  assumptions, while our conclusions for the other variables are
  robust.  }

\preprint{}
\begin{document}


\section{Introduction}\label{sec:intro}
Two of the most challenging questions facing particle physics
today are the instability of the Higgs mass against radiative
corrections (known as the ``fine--tuning problem'') and the nature
of dark matter (DM). Unlike in the Standard Model (SM), both find
plausible solutions in the framework of weak scale softly broken
supersymmetry (SUSY).
Firstly, the fine--tuning problem is addressed via the cancellation
of quadratic divergences in the radiative corrections to the Higgs
mass.  Secondly, assuming $R$--parity, the lightest supersymmetric
particle (LSP) is a leading weakly interactive massive particle
(WIMP) candidate for cold DM (CDM). On the other hand, despite these and other
attractive features, without a reference to grand unified
theories (GUTs), low energy SUSY models suffer from the lack of
predictivity due to a large number of free parameters (\eg, over 120 in the
Minimal Supersymmetric Standard Model (MSSM)), most of
which arise from the SUSY breaking sector. At present, experimental constraints on
superpartner masses from direct SUSY searches at LEP and the
Tevatron remain fairly mild, although a dramatic improvement is
expected once the LHC comes into operation in 2007.  Indirect
limits from bounds on CP--violation and flavor changing neutral
currents are generally much stronger (except for the
2nd--3rd generation mixings) but even these can be evaded, for example if SUSY
breaking is ``universal''.

The MSSM with one particularly popular choice of universal
boundary conditions at the grand unification scale is called the
Constrained Minimal Supersymmetric Standard Model
(CMSSM)~\cite{kkrw94}. The CMSSM is defined in terms of five free
parameters: common scalar ($\mzero$), gaugino ($\mhalf$) and
tri--linear ($\azero$) mass parameters (all specified at the GUT
scale) plus the ratio $\tanb$ of Higgs vacuum expectation values
and $\text{sign}(\mu)$, where $\mu$ is the Higgs/higgsino mass
parameter whose square is computed from the conditions of
radiative electroweak symmetry breaking (EWSB). The economy of
parameters in this scheme makes it a useful tool for exploring
SUSY phenomenology. In addition to experimental
limits on Higgs and superpartner masses, and strong bounds on SUSY
contributions to $\BRg$ and the anomalous magnetic moment of the muon
$\gmtwo$, a very strong constraint
limiting the mass parameters of the model from above is provided
by the relic abundance of the LSP. Within the CMSSM the neutral
LSP is the (bino--like) lightest
neutralino~\cite{an93,rr93,kkrw94}.\footnote{When the CMSSM is
extended to include an axionic sector, a natural candidate for the
LSP and CDM is an axino~\cite{axino-lr}, the fermionic partner of the
axion. Likewise, when the CMSSM is coupled to
supergravity, a gravitino, the fermionic partner of the graviton,
arises as a possible choice for the LSP and CDM. (For some recent
work see, \eg,~\cite{gravitino-cdm,gravitino-lr}.)  In contrast to the
neutralino, in both of these cases $R$--parity does not have to be
conserved to provide a solution to the DM problem because of 
their very strongly suppressed interactions to ordinary matter.} It is well
known that recent precise determinations of the relic density
$\abundcdm$ of non--baryonic CDM obtained by combining WMAP data with other
observations of cosmic microwave background (CMB) anisotropies and
large scale structure data, provide an important and often tight
constraint on the CMSSM parameter space.

It is worth remembering that the CMSSM is not the only viable and
economical framework providing well defined and strongly
constrained ranges of parameters. 
A virtue of the CMSSM lies in the particularly simple boundary
conditions at the unification scale and, as a result, in a minimal
number of parameters. A drawback is that the CMSSM, as a
framework, is not broad enough to accommodate a richer structure
of GUT--scale physics, in particular a non--minimal flavor
structure and many realistic fermion family patterns. One
attractive and well studied model is the MSSM with boundary
conditions at the unification scale imposed by consistency with a
minimal $SO(10)$ GUT model~\cite{raby-s010,drrr1+2}.

The most popular approach to exploring and delimiting viable
regions of the parameter space of the CMSSM and other SUSY models
has been a usual method of evaluating the goodness--of--fit of points scanned
using fixed grids~\cite{grid-cmssm}. Such scans have the advantage
of pre--determining the ranges and step size for each parameter
and thus of being able to control exactly which points in the parameter
space will be probed. On the other hand, the method has several
strong limitations. Firstly, the number of points required scales
as $k^N$, where $N$ is the number of the model's parameters and $k$ the
number of points for each of them. Therefore this approach
becomes highly inefficient for exploring with sufficient
resolution  parameter spaces of even modest dimensionality, say
$N>3$. Secondly, narrow ``wedges'' and similar features
of parameter space can easily be missed by not setting a fine
enough resolution (which, on the other hand, are likely to be completely
unnecessary outside such special regions). Thirdly, extra sources of
uncertainties (\eg, those due to the lack of  precise
knowledge of SM parameter values) and relevant external information
(\eg, about the parameter range) are difficult to accommodate.

In contrast, Bayesian statistics formalism linked to a Markov Chain Monte Carlo
(MCMC) method of exploring a multi--dimensional parameter space offer
several advantages.  In the Bayesian context, the required
computational power scales very favorably with the dimensionality of
parameter space, $N$. The details depend on the problem under
consideration, but roughly the number of points needed scales
approximately linearly with $N$. Secondly, it is straightforward to
include in the analysis all sources of uncertainty
that are present in the highly complex problem of comparing
theoretical predictions with current collider and cosmological
measurements. For instance, our imperfect knowledge of the
relevant SM parameters has an impact on our statistical
conclusions (\ie, inferences) about the values of the quantities of
interest, in our case the CMSSM parameters. This source of
uncertainty  can fully and easily be accounted for in Bayesian statistics.
Thirdly, 
probability distributions can be  computed for any function
of the CMSSM parameters, and in particular for all interesting
physical observables. 

The MCMC approach has been widely used in several branches of science
with much success and is gaining popularity in the astrophysics and
cosmology community.  (For some recent applications, see
\eg~\cite{Lewis:2002ah} for cosmological data
analysis,~\cite{Loredo:2001rx} in the astrophysical context, and 
\eg~\cite{StatBooks} for a general introduction to Bayesian methods.)
Within the context of softly broken low energy SUSY, limited
random (not MCMC) scans were first interpretted in the language of
statistical analysis in~\cite{py04} in the MSSM with diagonal flavor
mass entries, and in the CMSSM in~\cite{eoss-scan04,ehow04}. The MCMC
method was first applied to the CMSSM in~\cite{bg04} (with some
modifications) and more recently in~\cite{al05,allanach06}.

In the present work, we employ the  MCMC algorithm to
explore the parameter space of the CMSSM. The model is constrained
with data coming from present collider data and cosmological
observations of the CDM abundance. We compute the $W$ boson pole mass
$M_W$, $\sineff{}$, Higgs
and superpartner masses, $\BRg$, the anomalous magnetic moment of
the muon $\gmtwo$, $BR(B_s \rightarrow \mu^ + \mu^-)$ and
$\abundchi$, and compare them with current experimental data. We
also compute a spin--independent dark matter WIMP elastic
scattering cross section on a free proton, $\sigsip$, but we do not
enforce upper experimental limits because of the uncertainties in the
structure of the Galactic halo as well as in the
values of some hadronic matrix elements entering the computation
of $\sigsip$. We shall see that the current
experimental limit lies just above the high--probability regions
of parameter space.

Our analysis goes beyond other recent works in several aspects. We
include both experimental and theoretical uncertainties in treating
relevant SM quantities. Instead of applying a sharp cut at
experimental limits, we smear them out by incorporating theoretical
and experimental uncertainties. Furthermore, we improve the accuracy
of CMSSM predictions for the neutralino relic abundance by including
previously neglected dependence on the remaining uncertainty in the
fine structure constant. Our analysis covers larger values of $\mzero$
than in~\cite{al05}, thus allowing us to explore the focus point (FP)
region~\cite{fprefs}.  We present inferences on the high probability
regions for the CMSSM parameters and superpartner masses, and for
the other observables listed above. We emphasize that, in some
cases, inferences on the favored intervals of some parameters do
depend on the assumed prior ranges, while in the other the inferred
high--probability regions are fairly robust.  As we will see, this
often affects resulting conclusions about prospects for SUSY discovery
at the LHC or in DM searches. We point out the difference between
posterior probability (Bayesian statistics) and the quality--of--fit
statistics, and emphasize that a discrepancy between the two methods
can be only resolved with better data.

The paper is organized as follows. In section~\ref{sec:prob} we
briefly review some elements of Bayesian statistics and the MCMC
method, and introduce our 8--dimensional parameter space which we
explore in the following. In section~\ref{sec:data} we describe
the current collider and astrophysical data used in the
analysis to put constraints on the CMSSM. We present and discuss our results
in section~\ref{sec:results}. In section~\ref{sec:summary} we
summarize our main findings and conclusions.

\section{Bayesian statistics and the CMSSM}
\label{sec:prob}

In this section we introduce some basic concepts of the Bayesian
statistics and apply them to the CMSSM.

\subsection{Parameters and probabilities \label{sec:bayesian}}

We are interested in delineating high probability regions of the
CMSSM parameter space.  We fix $\text{sign}(\mu)= +1$ throughout
(see below) and denote the remaining four free CMSSM parameters by
the set
\be \label{indeppars:eq}
\params =  (\mzero, \mhalf, \azero, \tanb ).
\ee
In most of previous analyses, the values of the SM parameters,
such as the top quark mass, which strongly influences some of the
CMSSM predictions, have been fixed at their central values.
However, the statistical uncertainty associated with our imperfect
knowledge of the values of relevant SM parameters must be taken
into account in order to obtain correct statistical conclusions on
the regions of high probability for the CMSSM parameters. This can
easily be done in a Bayesian framework by introducing a set
$\nuis$ of so--called {\em ``nuisance parameters''}. For the
purpose of this analysis the most relevant ones are
\be \label{nuipars:eq} \nuis = ( \mtpole,
m_b(m_b)^{\overline{MS}}, \alphaemmz, \alphas ), \ee
where $\mtpole$ is the pole top quark mass, 
$m_b(m_b)^{\overline{MS}}$ is the bottom quark mass at $m_b$, while
$\alphaemmz$ and $\alphas$ are the electromagnetic and the strong
coupling constants at the $Z$ pole mass $M_Z$. The last three
parameters are evaluated in
the $\overline{MS}$ scheme.

The set of parameters $\params$ and $\nuis$ form an 8--dimensional
set $\basis$ of our {\em ``basis parameters''}
\be \label{eq:tuple} \basis = (\params, \nuis). \ee

In the following, we shall specify a set $\derived$ of several
collider and cosmological observables which we call {\em ``derived
variables''},
\be \label{eq:derived} \derived = (\derived_1, \derived_2, \ldots,
\derived_m). \ee
Their values depend on the CMSSM and SM parameters $\basis$
sampled in our MCMC analysis, $\derived(\basis)$. Some of the
observables will be used to compare CMSSM predictions with
experimental set of data $\data$ which is currently available either
in the form of positive measurements or as limits.

The central quantity which constitutes the basis of all
probabilistic inferences is the {\em posterior probability density
function} (pdf) $p(\basis | \data)$ for the basis parameters
$\basis$. The posterior pdf represents our state of knowledge
about the parameters $\basis$ after we have
taken the data into consideration (hence the name). Using
Bayes' theorem, the posterior pdf is given by
\be \label{eq:bayes}
 p(\basis | \data) = \frac{p(\data |
\derived) \pi(\basis)}{p(\data)}. \ee
On the rhs of Eq.~\eqref{eq:bayes}, the quantity $p(\data |
\derived)$, taken as a function of $\data$ for a given $\eta$, and
hence a given $\derived(\basis)$, is called  a ``sampling
distribution''. It represents the probability of reproducing the data
$\data$ for a fixed value of $\derived(\basis)$. Considered instead as
a function of $\derived$ for {\em fixed data} $\data$,
$p(\data|\derived)$ is called the {\em likelihood} (where the
dependence of $\derived$ on $\basis$ is understood).  The likelihood
supplies the information provided by the data.  In
section~\ref{sec:like} we explain in detail how it is constructed in
our analysis. The quantity $\pi(\basis)$ denotes a {\em prior
probability density function} (hereafter called simply {\em a prior})
which encodes our state of knowledge about the values of the
parameters in $\basis$ before we see the data. The state of knowledge
is then updated to the posterior via the likelihood.  Finally, the
quantity in the denominator is called {\em evidence} or {\em model
likelihood}. In the context of this analysis it is only a
normalization constant, independent of $\basis$, and therefore will be
dropped in the following.  For further details about the terminology
of Bayesian statistics, see \eg,~\cite{Trotta:2005ar,StatBooks}.

Since in this work we are not interested in the nuisance parameters
$\nuis$ themselves, at the end we simply marginalize over them by
integrating $p(\basis | \data)$ over their values. This procedure
gives a posterior pdf for the interesting CMSSM parameters $\params$
which takes full account of the uncertainties in $\nuis$
\be \label{eq:marginalization}
 p(\params | \data) = \int p(\params, \nuis \vert \data)\, d^4\nuis.
 \ee
Note that all pdf's should normally be  normalized so that the
total probability is unity. However, for the parameter estimation
procedure presented here only the {\em relative} posterior pdf's
are relevant, and in the following we shall plot pdf's normalized
in such a way that their maximum value is one. In practice, we
will present pdf's for only one or two variables at the time, with
the remaining ones integrated over. We will introduce and discuss
several of them in Section~\ref{sec:results} where we
present our results.

The purpose of the MCMC exploration of
parameter space is to obtain a series of points (called a {\em
``chain''}), whose density distribution is proportional to the
posterior pdf on the rhs of Eq.~\eqref{eq:bayes}. Further details
about the MCMC procedure are given in appendix~\ref{appx:mcmc}.
From the samples in the chain, it is straightforward to obtain all
pdf's of interest by plotting histograms of the number of samples
as a function of the parameter values that one wants to examine. In
particular, one does not need to carry out the marginalization
integral in Eq.~\eqref{eq:marginalization} explicitly. It is
sufficient to ignore the coordinates of the samples in parameter
space along the marginalized directions. This is one more major
advantage of the MCMC method.

Another useful feature is that from the
posterior pdf $p(\basis | \data)$ we can obtain the posterior pdf for {\em any}
function $f$ of basis parameters, using the fact that
 \be \label{eq:funcparams}
 p(f, \basis | \data) = p(f | \basis, \data) p(\basis | \data)
 = \delta(f(\basis) - f) p(\basis | \data),
 \ee
where $\delta(x)$ denotes the delta--function. Therefore for every
sample in the MC chain, one simply computes $f(\basis$) and the
resulting density of points in the $(f, \eta)$ space is
proportional to the posterior pdf $p(f, \basis | \data)$. From
this pdf one can then obtain by marginalization the pdf for any
subset of $(f,\basis)$. In particular, if we take $f(\basis) =
\derived(\basis)$, we can then investigate probability
distributions for any combination of the basis and the derived
parameters, as well as their correlations. This is investigated in
section~\ref{sec:results}.

Before we can proceed to delineating high probability regions in
the CMSSM parameter space, first we must choose our priors and
specify the likelihood. This is the subject of the next two
sections.

\subsection{The choice of prior probabilities \label{sec:prior}}

It is clear from the rhs of Eq.~\eqref{eq:bayes} that in the Bayesian
approach the first step is to specify the functional form of the
prior pdf. This is equivalent to assigning a probability measure
to parameter space.
The principle of indifference states that one should assign equal
probabilities to equal states of knowledge before seeing the data.
In our case, the basis parameters are {\em location parameters}
over which it is appropriate to set a {\em flat prior}
\be \label{eq:flatprior}
 \pi(\basis) =
  \left\{
 \begin{array}{l l}
 \begin{aligned}
  & \text{const} \quad &  \text{ for } \basis_\text{min} < \basis <
 \basis_\text{max} \\
  & 0 & \text{otherwise},
 \end{aligned}
 \end{array}
 \right.
 \ee
where the constant is determined by the requirement that the prior
integrates to probability one. Since we assume no correlation of
priors between the SM and the CMSSM parameters, the joint prior
can be written as
 \be
 \pi(\basis) = \pi(\params)\pi(\nuis).
 \ee
Flat priors are thus characterized by their ranges
$[\basis_\text{min},\basis_\text{max}]$. One alternative possibility
would be to employ a ``naturalness'' prior which gives more weight to
points exhibiting less fine--tuning~\cite{allanach06}.

\begin{table}
\centering
\begin{tabular}{|c|c|}
 \hline
\multicolumn{2}{|c|}{CMSSM parameters $\params$}       \\ \hline
 ``2\tev\  range''                                &      ``4\tev\  range'' \\\hline
 $50\gev < \mzero < 2 \tev$                  & $50\gev < \mzero < 4 \tev$\\
 $50\gev < \mhalf < 2 \tev$                  & $50\gev < \mhalf < 4 \tev$\\
 $|\azero| <  5\tev$                     & $|\azero| < 7\tev$  \\
 \multicolumn{2}{|c|}{$2 < \tanb < 62$}       \\ \hline
\multicolumn{2}{|c|}{SM (nuisance) parameters $\nuis$}       \\
\hline
 \multicolumn{2}{|c|}{$160\gev < \mtpole < 190 \gev$}\\
 \multicolumn{2}{|c|}{$4\gev < m_b(m_b)^{\overline{MS}} < 5 \gev$} \\
  \multicolumn{2}{|c|}{$127.5 < 1/\alphaemmz  < 128.5$} \\
 \multicolumn{2}{|c|}{$0.10 < \alphas  < 0.13$} \\ \hline
\end{tabular}
\caption{Initial ranges for our basis parameters
  $\basis=(\params,\nuis)$, with flat prior
probability distributions assumed. }
\label{table:prior}
\end{table}

Our SUSY and SM parameter priors are summarized in
Table~\ref{table:prior}. In this work we consider two initial
ranges of CMSSM parameters. In one, which we call a ``2\tev\
range'', we assume $50\gev<\mzero, \mhalf<2\tev$ and $|\azero| <
5\tev$. This choice is motivated by an expected LHC reach in
exploring superpartner mass and by a general ``naturalness''
argument of SUSY mass parameters to preferably lie within
$\mathcal{O}(1\tev)$. In the other case, called a ``4\tev\ range'', we
assume $50\gev<\mzero, \mhalf<4\tev$ and $|\azero| < 7\tev$, which
goes far beyond the LHC reach. (The larger range will include the
focus point region, along with various
uncertainties involved. We will discuss this point later.)  We
will compare our findings for both ranges in order to see to what
extent statistical conclusions depend on our preconceived
expectation that a SUSY signal might lie within reach of the LHC
(represented by the ``2\tev\ range'').  Such a
sensitivity test is essential to establish the extent to which
inferences depend on the initial range one chooses, \ie\ on
the prior. The lower bounds on $\mzero$ and $\mhalf$ come from the
negative results of sparticle searches. We allow a
rather generous range for $\azero$, in part to see to what extent
this choice would allow one to reduce~\cite{shbp05} the impact of the cosmological
constraint, which in the usually explored case of $\azero=0$ is
very tight. For both sets we further assume $2 < \tanb < 62$. The
lower bound comes from negative Higgs searches~\cite{lephiggs}.
Very large values of $\tanb \gsim 60$ are in conflict with
theoretical considerations, \eg\ they would make it extremely
difficult to achieve radiative electroweak symmetry
breaking~\cite{bbo94}. Furthermore, at such large $\tanb$ large
uncertainties arise in the computation of the SUSY spectrum,
leading to unreliable predictions. On the other hand, since the
SM nuisance parameters are well measured, it turns out that their
prior ranges are irrelevant for the outcome of the analysis.

Before closing this section, we comment that the necessity of choosing
priors is often (incorrectly) regarded as a limitation to the
``objectivity'' of the Bayesian approach. This can be easily dispelled
by noting that two scientists in the same state of knowledge before
seeing the data (\ie, who have assumed the same priors) will
necessarily reach the same conclusions.  When the choice of priors
makes a difference in drawing the final inference (given by the
posterior pdf), this is a ``health warning'' that the data is not
informative enough, \eg\ the likelihood is not sufficiently peaked to
override the assumed prior distribution. In this case, the inference
on the parameters must rely either on external relevant information in
the form of the prior (\eg, a theoretical ``naturalness criterion''
investigated in Ref.~\cite{allanach06}), or of better and more
constraining data. In the present study, this will be the case for the
parameter $\mzero$, as we discuss in section~\ref{sec:susymasses}.

\section{Collider and cosmological observables \label{sec:data}}

In this section, we first define the likelihood function for the
CMSSM.
Next, we introduce the data that we use for the nuisance
parameters, and then proceed to describe electroweak and dark
matter observables. We give details about their calculation, the
theoretical uncertainties involved and  the experimental errors in
their determination.

\subsection{Constructing the likelihood for the CMSSM\label{sec:like}}

The likelihood is a key element of our analysis. It encodes
the information from the observational data and therefore
particular care must be taken in constructing it. In the Bayesian
framework it would be easy to incorporate the full likelihood
functions from various experimental measurements if they were
available. However, even though the actual measurements
contain much more useful information, most measurements in particle physics
experiments are presented only by the mean and the standard
deviation, while upper or lower exclusion bounds are usually given
in terms of the $95\%$ exclusion CL. 

Uncertainties in the observable quantities can be split into two
categories. The first is an experimental uncertainty, the second
is a theoretical one and is a consequence of making some
approximations (\eg, neglecting higher order loop corrections), a
limited numerical precision in the code, \etc. In practice, the
theoretical uncertainty can be modelled in a Bayesian context by
considering that our mapping from the basis parameters $\basis$ to
the derived quantities $\derived$ is imperfect: instead of an
``exact'' mapping $\trued(\basis)$ we actually have only an
imperfect version $\derived(\basis)$ which suffers from the sort
of uncertainties outlined above.
The likelihood $p(\data | \derived)$ introduced in
Eq.~\eqref{eq:bayes} can then be written as
\be \label{eq:uncert}
 p(\data | \derived) = \int p(\data | \trued) p(\trued | \derived)
  d^m\trued,
\ee
where we have integrated out the true (and unknown) mapping. The
pdf $p(\trued | \derived)$ encodes the estimated uncertainty of
our mapping. Usually we only have (at best) an
estimate of the theoretical errors, which means that we only have
information on the scale of the associated uncertainty. This  is
described by a multi--normal distribution of a general form
 \be \label{eq:mdimg}
 p(\trued | \derived) = \frac{1}{(2\pi)^{m/2}|C|^{1/2}}
 \exp\left(-\frac{1}{2}(\derived-\trued)C^{-1}(\derived-\trued)^T\right),
 \ee
where $C$ is an $m\times m$ covariance matrix describing the error
of the mapping ($m$ being the number of elements in $\derived$)
and $|C|$ denotes its determinant. If one assumes that the theoretical
errors $\tau_i$ ($i=1,\ldots, m$) for the different quantities are uncorrelated then 
$C$ is diagonal, $C = \text{diag}\left(\tau_1, \ldots,
\tau_m\right)$. Furthermore, if the likelihood $p(\data | \trued)$
is also a multi--dimensional Gaussian function with diagonal
covariance matrix, then we have
 \be \label{eq:mdimgtwo}
 p(\data | \trued) = \frac{1}{(2\pi)^{m/2}|D|^{1/2}}
 \exp\left(-\frac{1}{2}(\data-\trued)D^{-1}(\data-\trued)^T\right),
 \ee
where $D = \text{diag}\left(\sigma_1, \ldots, \sigma_m\right)$ and 
$\sigma_i$ denotes the  experimental standard error. In
this simple case Eq.~\eqref{eq:uncert} reduces to the usual rule
of adding theoretical and experimental errors in quadrature for
each derived observable, \ie, the total error in each direction is
$s_i=\sqrt{\sigma_i^2 + \tau_i^2}$. This familiar result is a
special case of the more general treatment given above, which
shows that in the Bayesian framework all sources of uncertainty
specified in the model can be fully taken into account in a
systematic way.

In this work we model the likelihood of all observables for which
there exists a positive measurement by an uncorrelated,
multi--dimensional Gaussian function. The experimental means and
standard deviations for SM quantities are summarized in
Table~\ref{tab:meas} and will be discussed in more
detail in section~\ref{sec:dataSM}. Since the SM quantities are at the same time
input quantities of our analysis, there are no theoretical
uncertainties associated with them. In Table~\ref{tab:measderived}
we display the experimental and theoretical errors for derived
cosmological and collider quantities, see section~\ref{sec:dataLE}
for a further discussion.

For the quantities for which only lower or upper bounds are
available (\eg, superpartner masses or $BR(B_s \rightarrow \mu^+
  \mu^-)$), a usual procedure is to
simply discard points in the parameter space for which such limits
are violated at some confidence level (\eg, $1\sigma$ or 95\%),
essentially using a step function as a likelihood. This procedure
is not totally rigorous since it does not take into account the
amount of uncertainty associated with the theoretical error
(denoted by $\tau$). In our analysis instead we use
Eq.~\eqref{eq:uncert} to incorporate our estimate of the
theoretical uncertainty of the mapping in the limits that we use,
as explained below.

As an illustration, let us consider a 1--dimensional case involving only one
observable $\xi$. A lower bound on the (exact) mapping $\hat{\xi}$
can be described by the following likelihood function (replacing
$\data \rightarrow (\sigma, \xilim)$):
\be\label{eq:likefunction}
 p(\sigma, \xilim | \hat{\xi}) =
\left\{
 \begin{array}{l l}
 \begin{aligned}
  & \frac{1}{\sqrt{2\pi}\sigma} \quad &  \text{ for } \hat{\xi} \geq \xilim, \\
  & \frac{1}{\sqrt{2\pi}\sigma}\exp -\frac{(\hat{\xi} -
  \xilim)^2}{2\sigma^2} & \text{ for } \hat{\xi} < \xilim, \\
 \end{aligned}
 \end{array}
 \right.
 \ee
where a Gaussian function has been used to model the drop of the
likelihood function below the experimental bound $\xilim$. As
explained above, the theoretical uncertainty of the mapping is
described by a 1--dimensional Gaussian of standard deviation
$\tau$ (see Eq.~\eqref{eq:mdimgtwo} for $m=1$). Then from the
integral in Eq.~\eqref{eq:uncert} we obtain
\be \label{eq:limerr}
 p(\sigma, \xilim | \params, \tau) = \frac{1}{\sqrt{2\pi(\sigma^2 + \tau^2)}}
\exp \left[ - \frac{(\xilim - \xi)^2}{2(\sigma^2+\tau^2)} \right]
\left[1 - Z(\tlim) \right] + Z\left(\frac{\xilim - \xi}{\tau}
\right),
 \ee
where $\xi = \xi(\params)$ and we have defined
\be \label{tstar:eq} \tlim \equiv \frac{\sigma}{\tau}\frac{\xilim
- \xi}{\sqrt{\sigma^2+\tau^2}}, \hspace*{1cm} Z(\tlim)  \equiv
\frac{1}{\sqrt{2\pi}} \int_{\tlim}^\infty
 d\,x \exp(-x^2/2).
\ee
%

\begin{figure}[t!]
\centering
\epsfig{file=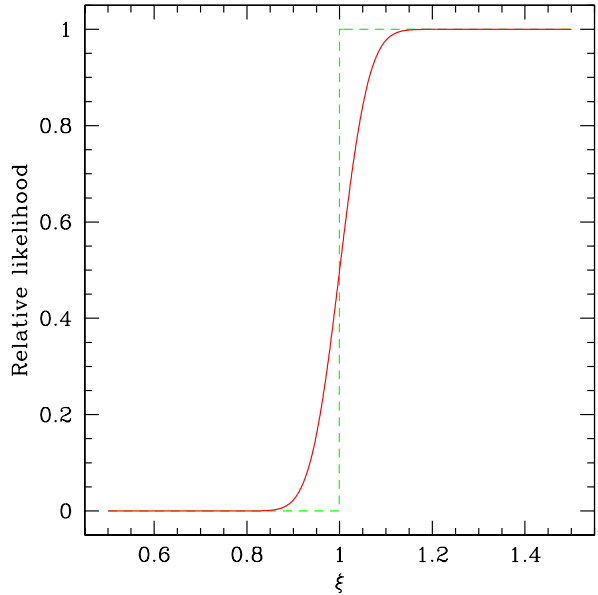,height=3.5in} 
\caption{An illustration of the likelihood function $p(\sigma,
\xilim | \params, \tau)$ used for a quantity for which only a lower
bound is available, including the theoretical
uncertainty $\tau$ and setting the experimental error $\sigma = 0$. The dashed green line
is the sharp 95\%~\cl\ bound ($\xilim = 1$), the solid red curve
includes the theoretical uncertainty of $\tau=0.05$ which  smears out
the limit.}
 \label{fig:limlike}
\end{figure}
This form of the likelihood encodes the uncertainty associated
with our imperfect mapping, as described by $\tau$. The effect of
including the theoretical uncertainty is to smear out the drop of
the likelihood function: the scale of the drop goes from $\sigma$
to $\sqrt{\sigma^2 + \tau^2}$. Unfortunately, experimental bounds
are usually given in the form of the lower or upper $95\%$
confidence limit only, without the possibility of deriving a form
of the likelihood analogous to \eqref{eq:limerr}. In the absence
of fuller information about the experimental value of $\sigma$, we can
take in~\eqref{eq:limerr} the limit $\sigma \ll \tau$ and simply
use the 95\%~\cl\  bound as $\xilim$. This procedure leads to the
likelihood function plotted in Fig.~\ref{fig:limlike}, showing
that the inclusion of the theoretical uncertainty smears the 95\%
bound over the scale $\tau$. This procedure is more
conservative than the usual method of simply rejecting all points
below $\xilim$.

We use \eqref{eq:limerr} as the likelihood function for all of the
upper/lower bounds listed in Table~\ref{table:lims} along with our
estimates of the corresponding theoretical uncertainty. We also
discard, \ie, assign zero likelihood to all unphysical points in
parameter space, \ie, those for 
which any of the masses becomes tachyonic or the conditions of
EWSB are not satisfied. We do the same for the cases where the
lightest neutralino is not the LSP.

\subsection{Inputs for SM quantities \label{sec:dataSM}}

As explained in section~\ref{sec:prob}, the uncertainties coming
from experimental errors in the SM parameters are incorporated through
four nuisance parameters: $\mtpole$, $m_b(m_b)^{\overline{MS}}$,
$\alphaemmz$ and $\alphas$.  These quantities and their
uncertainties are summarized in Table~\ref{tab:meas}.

\begin{table}
\centering
\begin{tabular}{|l | l l | l|}
\hline
Nuisance parameter  &   Mean value  & \multicolumn{1}{c|}{Uncertainty} & Ref. \\
 &   $\mu$      & ${\sigma}$ (exper.)  &  \\ \hline
$\mtpole$           &  172.7 GeV    & 2.9 GeV&  \cite{mtop} \\
$m_b (m_b)^{\overline{MS}}$ &4.24 GeV  & 0.11 GeV &  \cite{mbmb} \\
$\alphas$       &   0.1186   & 0.002 &  \cite{PDG}\\
$1/\alphaemmz$  & 127.958 & 0.048 &  \cite{lepwwg}\\ \hline
\end{tabular}
\caption{Experimental mean $\mu$ and standard deviation $\sigma$
for the nuisance parameters used in the analysis. For all these
quantities we use a Gaussian likelihood function with
the mean $\mu$ and the standard deviation $\sigma$. \label{tab:meas}}
\end{table}

Note that for the running bottom quark mass
$m_b(m_b)^{\overline{MS}}$ the range adopted in
Table~\ref{tab:meas} is rather conservative.  It arises from
combining measurements from $\bar{b} b$ systems, b--flavored
hadrons and high--energy processes. More recently, a much more
precise determination of $m_b (m_b)^{\overline{MS}} = 4.19 \pm
0.06\gev$ has been obtained from a renormalization group improved
sum rule analysis~\cite{ps06}. As for the fine structure
constant, the dominant error comes from the hadronic contribution
$\Delta \alpha_{had} = 0.02758 \pm 0.00035$~\cite{lepwwg}. Note
that, because of this uncertainty $\alphaemmz$ is not as precisely
known at the scale $M_Z$ as at zero momentum transfer. Not
included in the list of nuisance parameters are the pole mass of
the $Z$ boson $M_Z=91.1876(21)\gev$, as well as the Fermi constant
$G_F=1.16637(1)\times10^{-5}\gev^{-2}$~\cite{PDG}. As one can see,
they are now known with very high precision and we fix them at
their central values. Including them as nuisance parameters would
not have any appreciable effect.

\subsection{Derived observables} \label{sec:dataLE}

We now present the derived observables which in
section~\ref{sec:bayesian} were denoted by a general symbol
$\derived$. In our analysis they are computed in terms of the
CMSSM parameters: $\mzero$, $\mhalf$, $\azero$, $\tanb$, as well as the
nuisance parameters discussed above.

In Table~\ref{tab:measderived} we summarize the derived variables for
which positive measurements have been made while in
Table~\ref{table:lims} we list the ones for which currently only
experimental bounds exist. Lower bounds on sfermion masses have been
obtained in the context of the MSSM at LEP-II and
Tevatron Run~II. Below we comment on some of the entries and on our
procedure for their computation. Generally, to calculate Higgs and SUSY mass spectra
we use the package SOFTSUSY~v1.9~\cite{softsusy} which employs 2--loop
RGEs for couplings (both gauge and Yukawa) as well as for gaugino and
sfermion masses. 

\begin{table}
\centering
\begin{tabular}{|l | l l l | l|}
\hline
Derived observable &   Mean value & \multicolumn{2}{c|}{Uncertainties} & Ref. \\
 &   $\mu$      & ${\sigma}$ (exper.)  & $\tau$ (theor.) & \\\hline
 $M_W$          &  $80.425 \gev$     & $34\mev$ & $13\mev$ & \cite{lepwwg} \\
 $\sineff{}$    &  $0.23150$      & $16\times10^{-5}$
                &$25\times10^{-5}$ &  \cite{lepwwg}  \\
$\dasusy \times 10^{10}$       &  $25.2 $ & 9.2 &  $1$ & \cite{bnl,gm2} \\
 $\text{BR}(\bar{B} \rightarrow X_s \gamma) \times 10^{4}$ &
 $3.39$ & $0.30$ & $0.30$ & \cite{bsgexp} \\
 $\abundchi$ &  0.119 & 0.009 & $0.1\,\abundchi$& This work \\\hline
\end{tabular}
\caption{Summary of derived observables used in the analysis for
which positive measurements have been made. As explained in the
text, for each quantity we use a likelihood function with mean
$\mu$ and standard deviation $s = \sqrt{\sigma^2+ \tau^2}$, where
$\sigma$ is the experimental uncertainty and $\tau$ our estimate
of the theoretical uncertainty. \label{tab:measderived}}
\end{table}

\begin{table}
\centering
\begin{tabular}{| l | c | l r | l |}
\hline
Derived observable &  & \multicolumn{2}{c|}{Constraints} &  Ref. \\
                &  &  $\xilim$ & $\tau$ (theor.) & \\ \hline
 $\sigsip$       & UL & WIMP mass dependent     & $\sim100\%$ &
\cite{cdms-sep05} \\
 $\text{BR}({B}_s \rightarrow \mu^+\mu^-)$ & UL & $1.5 \times 10^{-7}$
                   & 14\% & \cite{bsgexp}\\
$m_h$          & LL   & $114.4\gev$\ ($91.0\gev$) & $3 \gev$ & \cite{lephiggs} \\
 $\zetah^2$      & UL   & $f(m_h)$       &  3\% & \cite{lephiggs} \\
 $m_{\chi   }$   & LL   & $50 \gev$ & 5\% & \cite{lsp} (\cite{aleph,l3}) \\
 $m_{\chi^\pm_1}$& LL  & $103.5\gev$\ ($92.4\gev$) & 5\% &\cite{lepsusy} (\cite{aleph,l3}) \\
 $m_{\tilde{e}_R}$ & LL & $100 \gev$\ ($73 \gev$)& 5\% &\cite{lepsusy} (\cite{aleph,l3})  \\
 $m_{\tilde{\mu}_R}$ & LL & $95 \gev$\ ($73 \gev$)& 5\% &\cite{lepsusy} (\cite{aleph,l3})  \\
 $m_{\tilde{\tau}_1}$ & LL & $87 \gev$\ ($73 \gev$)& 5\% & \cite{lepsusy} (\cite{aleph,l3}) \\
 $m_{\tilde{\nu}}$ & LL & $94 \gev$\ ($43 \gev$) & 5\% & \cite{delphi} (\cite{PDG}) \\
 $m_{\tilde{t}_1}$ & LL & $95 \gev$\ ($65 \gev$) & 5\% &\cite{lepsusy} (\cite{aleph})\\
 $m_{\tilde{b}_1}$ & LL & $95 \gev$\ ($59 \gev$) & 5\% &\cite{lepsusy} (\cite{aleph})\\
 $m_{\tilde{q}}$ & LL & $318 \gev$     & 5\% & \cite{D0} \\
 $m_{\tilde{g}}$ & LL & $233 \gev$     & 5\% & \cite{D0} \\ \hline
\end{tabular}
\caption{ \label{table:lims} Summary of derived observables for
which only limits exist, with UL = upper limit, LL = lower limit
(at $\xilim$~{\protect\eqref{eq:likefunction}}, 95\%~\cl, unless
otherwise stated). The experimental limit on the spin--independent
cross section for a WIMP elastic scattering on a free proton $\sigsip$
is {\em not} included in the likelihood, as explained in the text.
Since the precise form of the likelihood is not available, we use
the conservative procedure of including at least an estimated
theoretical uncertainty $\tau$. The likelihood is then given by
Eq.~{\protect\eqref{eq:limerr}}, in the limit where $\sigma \ll
\tau$. The value in parenthesis indicates the more conservative
bound, see the text for details. Note that theoretical errors for
scalar masses are probably much larger in the focus point region, as
discussed in the text.
}
\end{table}

\paragraph{\boldmath $W$ gauge boson mass}
Intrinsic theoretical uncertainties coming from higher loop
effects to the $W$ boson pole mass are estimated to be $\tau(M_W)
= 13 \mev$~\cite{bmpz,mw_prosp}. The parametric uncertainties are
dominated by the experimental error of the top quark mass and by
the hadronic contribution to the shift of the fine structure
constant. Both uncertainties are fully taken into account (via
$\Delta \alpha_{had}$) by marginalizing over $\mtpole$  as a
nuisance parameter. We compute $M_W$ in the $\drbar$ scheme (from
gauge couplings relations), including full 1--loop
contributions~\cite{bmpz}.

\paragraph{Effective leptonic weak mixing angle}
The effective leptonic weak mixing angle $\sineff{}$ receives SM and
SUSY contributions. In our computation we include a full 1--loop SM
and full 1--loop universal $Z$--vertex supersymmetric
corrections~\cite{bmpz}. The net contribution of the non--universal
corrections is negligible~\cite{dlt}. Intrinsic theoretical
uncertainties come from higher--loop effects, which induce an
uncertainty\footnote{In Ref.~\cite{ehow04} a smaller uncertainty is
used (\ie, $\tau(\sineff{}) = 12\times 10^{-5}$) as a result of
including leading two--loop supersymmetric corrections.}  taken to
be~\cite{bmpz} $\tau(\sineff{}) = 25\times 10^{-5}$.

Note that we treat $\alphaemmz$ as a nuisance parameter but take
$M_W$ and $\sineff{}$ as derived quantities. This is because we
use a parametrization in which the former (along with $M_Z$ and
$G_F$) are taken as inputs from which one can compute the latter
and compare with experiment.

\paragraph{The anomalous magnetic moment of the muon}
The final measurement of the anomalous magnetic moment of the
muon, $a_\mu \equiv (g-2)_\mu $, by the Brookhaven E821
experiment~\cite{bnl} $a_\mu^{\text{exp}}=(11659208.0 \pm 5.8)
\times 10^{-10}$ 
remains in an apparent disagreement with  SM predictions. The
current SM theoretical value, based on $e^+ e^-$ low energy data,
is~\cite{gm2} $a_\mu^{\text{SM}}= 11659182.8 \pm 6.3_{\text{had}}
\pm 3.5_{\text{LBL}} \pm 0.3_{\text{QED+EW}}) \times 10^{-10}$. 
The discrepancy between the two, if confirmed, could be attributed
to an additional contribution from loops involving superpartners,
\begin{equation}
\label{dasusy:eq}
\dasusy = a_\mu^{\text{exp}}-a_\mu^{\text{SM}}=
 (25.2 \pm 9.2) \times 10^{-10}, 
\end{equation}
where the independent errors have been added in quadrature.

In calculating $\dasusy$ we have taken into account the full
one--loop SM+SUSY contributions~\cite{moroi95} and several
two--loop corrections. The first class of two--loop corrections
comprises the leading  one--loop diagrams with a photon in the
second loop~\cite{dg98}. The second class comprises diagrams with
a closed loop of SM fermions or scalar fermions~\cite{hsw03}. The
last class comes from diagrams containing a closed
chargino--neutralino loop~\cite{hsw04}. As a consequence, in the
CMSSM parameter space the intrinsic uncertainties are estimated to
be $\tau({\dasusy})=  1 \times 10^{-10}$~\cite{hhw}. The sign of
the SUSY contribution to $a_\mu$ is the same as the sign of $\mu$.
Since the former is positive, throughout this analysis we have
assumed $\text{sign}(\mu)= +1$.

Finally, we note that, the SM value evaluated using $\tau$--data is
$a_\mu^{\text{SM}}=(11659201.8 \pm 6.3$, leading to only a $0.7\sigma$
deviation~\cite{gm2-tau}. In light of this we do not feel it is
justified to apply a SM prediction based on combining the two sets of
data. Secondly, there appears to be potentially some discrepancy among
different sets of $e^+ e^-$ which again may put the claimed
deviation~\eqref{dasusy:eq} from the SM into question. Because of
these outstanding problems, in subsection~\ref{sec:nogm2} we will
perform a separate analysis where we exclude $a_\mu$ from the
likelihood.

\paragraph{\boldmath $BR(\bar{B} \rightarrow X_s \gamma)$}
The current experimental world average value  by the Heavy Flavour
Averaging Group (HFAG) is~\cite{bsgexp}
\begin{equation} \label{eq:bsgexp}
\BRg_{\text{exp}}= (3.39^{ +0.30}_{-0.27}) \times 10^{-4},
\end{equation}
which agrees rather well with the full NLO prediction of the
SM~\cite{bsgsm}
\begin{equation}
\BRg_\text{SM} = (3.70 \pm 0.30) \times 10^{-4}.
\end{equation}

This clearly imposes an important constraint on any additional
contributions.  In the CMSSM, the assumed universality of soft
mass terms leads to a particularly simple flavor structure in
which no additional sources of flavor mixings beyond those due to
the CKM matrix are present, the framework known as minimal flavor
violation (MFV). In this case SUSY contributions arise  from a
loop involving the top quark and the charged Higgs boson and one
of the stop--chargino exchange. In a more general flavor mixing
scenario additional one--loop contributions arise due to gluino
(or neutralino) and down--type squark exchange.

We compute SUSY contribution to $\BRg$ following the procedure
outlined in Refs.~\cite{dgg00, gm01} where, in addition to full
leading log corrections, large $\tanb$--enhanced terms arising
from corrections coming from beyond the leading order (BLO) have
been included.\footnote{An analogous and updated BLO--level
analysis in the case of general flavor mixing has been performed
in Refs.~\cite{or1+2,for1+2,for3} and shows in general a much
larger difference between LO and BLO results than in MFV.}
Furthermore, combined experimental constraints from $\BRg$ and
$BR(B \rightarrow X_s l^+ l^-)$ imply that, baring highly
non--standard scenarios~\cite{for3},  the sign of the total
amplitude for the decay $\bar{B}\to X_s\gamma$ has to be the same
as that of the SM~\cite{ghm04}.

We assume no significant additional theoretical uncertainty beyond
that coming from the SM calculation. This can in part be justified
by the fact that the overall SUSY contribution to $\BRg$ has to be
small, as noted above. More importantly, we estimate that most of
the uncertainties due to SUSY contributions are due to remaining
uncertainties in the values of SM parameters, especially the top
and bottom masses which are accounted for in
their treatment as nuisance parameters. Therefore we take
$\tau(\BRg) = 0.30 \times 10^{-4}$.

\paragraph{Cosmological constraints on the dark matter density}
A combination of the recent WMAP data~\cite{wmap} on the CMB
anisotropies with other cosmological observations, such as
measurements of the matter power spectrum, leads to tight constraints
on the cold dark matter (CDM) relic abundance. The exact numerical
value depends on a number of assumptions about the underlying
cosmology (\eg, the geometry of space, the adiabaticity of initial
conditions and a power--law, feature--free primordial power
spectrum). It is important to keep in mind that relaxing some or all
of these assumptions can considerably weaken the
constraints. Furthermore, different combinations of data sets also
lead to somewhat different values for the relic abundance. For
definiteness, we have performed a re--analysis of CMB data from the
first year of WMAP observations~\cite{wmap}, as well as from
CBI~\cite{Readhead:2004gy}, VSA~\cite{Dickinson:2004yr} and
ACBAR~\cite{Goldstein:2002gf}, and combined them with the real--space
power spectrum of galaxies from the SLOAN galaxy redshift survey
(SDSS)~\cite{Tegmark:2003uf}, restricted to scales over which the
fluctuations are assumed to be in the linear regime, \ie\ for $k < 0.1
h^{-1}\, \mpc$, with the Hubble Space Telescope
measurement~\cite{Freedman:2000cf} of the Hubble parameter $H_0$
($H_0=100\, h\, \kmeter/{\second}/\mpc$),
and with the latest supernovae observations data~\cite{Riess:2004nr}.
Assuming a flat Universe $\Lambda$CDM cosmology, the resulting
constraint on the dark matter relic abundance -- after marginalizing
over all other relevant cosmological parameters -- 
is well approximated by a Gaussian function with a mean and
standard deviation given by\footnote{The central value is slightly
different from $\abundcdm =0.113^{+0.008}_{-0.009}$ obtained by
the WMAP team in~\cite{wmap} because of the different
combination of data employed.}
\begin{equation} \label{eq:cdmrange}
\abundcdm = 0.119 \pm 0.009.
\end{equation}
We make use of this constraint and we  assume that all the
CDM is made up of stable neutralinos, but we also enlarge the
error to include a theoretical uncertainty in the computation of
$\abundcdm$ (see below).

A precise determination of the neutralino relic abundance
$\abundchi$ requires an accurate treatment of the neutralino
pair annihilation and coannihilation cross sections into all SM
particle final states. We employ exact expressions for neutralino
pair annihilation processes into all allowed final--state
channels which have been computed in~\cite{nrr1+2} and which are
valid both near and further away from 
resonances and thresholds. We further include the
neutralino coannihilation with the lightest chargino and
next--to--lightest neutralino~\cite{eg97} and with the lighter
stau~\cite{nrr3} with similar precision.  We include all
coannihilation channels, including coannihilation with light
stops~\cite{ellis-stop}. We compute $\abundchi$  by solving the
Boltzmann equation numerically as in~\cite{darksusy}.

As is well known from fixed--grid scans, in the CMSSM the values of
$\abundchi$ typically exceeds the range given in~\eqref{eq:cdmrange},
except in some rather special regions of the parameter space.
Firstly, at fairly small $\mhalf$ and $\mzero$, the neutralino
annihilation channel into SM fermion/boson pairs via $t/u$--channel
exchange of a superpartner opens up only a narrow band consistent with
$\abundcdm$ (the ``bulk'' region) which however is largely excluded by
a lower bound on the Higgs mass and/or by a strong upper bound on
allowed SUSY contribution to $\BRg$.  Secondly, if the LSP and the
next--to--LSP (NLSP) are closely degenerate in mass, then the LSP
coannihilations with NLSPs near freeze--out may reduce the LSP relic
density considerably. In the CMSSM, efficient coannihilation takes
place along the boundary dividing neutralino and lightest stau
$\stauone$ LSP regions. In addition, at large $\azero$ there are
limited cases where the lightest stop is the NLSP and is almost
degenerate in mass with the LSP~\cite{ellis-stop}. Thirdly, in some
cases neutralino annihilation can be enhanced via the process
$\chi\chi\rightarrow f\bar f$ involving an $s$--channel Higgs or $Z$
boson exchange. At large $\tanb\sim50$ the most significant effect
comes from the CP-odd Higgs $A$ resonance (around $m_A\simeq2 \mchi $,
where $\mchi$ is the lightest neutralino mass), since the $A$
couplings to down--type fermions are enhanced at large $\tanb$, and
because, in contrast to heavy scalar Higgs exchange, the process is
not $p$--wave suppressed.  Finally, at very large $\mzero$ of a
few$\tev$ the rapidly decreasing $\mu^2$ (from large positive to
negative values) causes the higgsino component of the LSP to increase
which, in a narrow focus point (FP) region (of still positive $\mu^2$)
allows $\abundchi$ to pass through the favored
range~\eqref{eq:cdmrange}, before becoming too small.

The theoretical uncertainty involved in computing $\abundchi$
varies greatly depending on the case. Errors come from computing
the Higgs and superpartner mass spectrum at finite (two--loop)
order in RGEs, a scale dependence, finite numerical accuracy in
solving the Boltzmann equation, some residual uncertainties in
computing the gauge couplings, as well as potentially much larger errors
in computing top and bottom Yukawa couplings and Higgs widths. The
choice of the scale at which one minimizes the effective potential has a
minor effect~\cite{al05}. A numerical 
algorithm of solving the Boltzmann equation is very accurate and in
the bulk region has an estimated error of a few per
cent~\cite{gondologelmini,nrr1+2}, which is comparable with the
observational error in~\eqref{eq:cdmrange}.

In the bulk region where $\abundchi$ primarily depends on the
$t/u$--channel exchange of sleptons, uncertainties in the calculation
of the SUSY spectrum are $\mathcal{O}(1\%)$ (for moderate values of
the CMSSM mass parameters, even at large $\tanb$) and therefore under
control~\cite{akp03}. On the other hand, the accuracy is much poorer
in the regions of special cosmological interest that have been
mentioned above. In the coannihilation regions $\abundchi$ is
sensitive to $e^{-\Delta m/T}$, where $\Delta m = m_{\rm NLSP} -
\mchi$ is the difference between the mass of the NLSP $m_{\rm NLSP}$
and $\mchi$.  Even a small variation of $\mathcal{O}(1\%)$ in $\Delta
m$ can lead to $\sim30 \%$ variations in
$\abundchi$~\cite{oh2susy}. This sort of error is inherent in current
determinations of the mass of $\stauone$ which is the NLSP in large
parts of the $(\mhalf,\mzero)$ plane. Larger uncertainties arise if
the NLSP is the $\stopone$ since the shift in its mass $10\%$ can lead
to order--of--magnitude discrepancies in the prediction of the relic
density~\cite{akp03}. In the CP--odd Higgs resonance region one finds
a strong suppression of $\abundchi$ for broad ranges of $\mhalf$ and
$\mzero$. In the CMSSM this occurs for large $\tanb$ where effects of
the bottom Yukawa coupling $h_b$ on the RGE running are important and
reduce $m_A$ compared to the low $\tanb$ regime.  The relic density is
further suppressed by the enhancement of the coupling $A b \bar{b}$
which is proportional to $h_b$. Special attention must also be paid to
the computation of the Higgs boson width which receives sizable
radiative corrections. We include one--loop QCD
corrections~\cite{djouadi95} as well as those due to (QCD corrected)
Yukawa vertices. Still, unknown two--loop corrections to $h_b$ may
cause an uncertainty of up to $30 \%$~\cite{oh2susy}. In the FP
region, a determination of $\mu$ is strongly sensitive to two--loop
corrections proportional to $h_t$, whose computation requires special
care, and can lead to $\sim100\%$ errors in the computation of
$\abundchi$~\cite{oh2susy}.

In addition to the theoretical errors discussed above, the value
of $\abundchi$ depends on the top and bottom masses, as well as
$\alphaemmz$. In particular, we have found that the seemingly
small experimental error in $\alphaemmz$ leads, indirectly, via
its effect on the SM gauge couplings $g_{1,2}$, to a variation in
$\abundchi$ of order $10\%$. All of these effects are accounted
for by our use of the nuisance parameters.

The error in the quantity $\abundchi$ can thus be very sensitive
to which (co)annihilation process is most efficient. This makes it
difficult to evaluate the theoretical uncertainty in $\abundchi$.
Given the above discussion we estimate $\tau(\abundchi)=10\%$ in
the bulk of the parameter space although we are aware that in the
FP region the error is almost certainly much larger.  We add this
error in quadrature to the observational error
$\sigma(\abundcdm)=0.009$ in~\eqref{eq:cdmrange} and obtain the
Gaussian with the following mean and standard deviation,
\bea
\abundchi&=& 0.119\pm \sqrt{(0.009)^2+ (0.1\,\abundchi)^2}\\
&=&0.119\pm 0.009\,\sqrt{1+ 1.75\,(\abundchi/0.119)^2}.
\label{eq:abundchitot} \eea
Note that the theoretical uncertainty is of the same order as the
uncertainty from current cosmological determinations of
$\abundcdm$.

\vspace*{0.75truecm}\noindent We now comment on some of the
derived variables for which the $95 \%$~\cl\ experimental
upper/lower limits have been presented in Table~\ref{table:lims}.
We start, however, with the only observable which is not
included in the likelihood.

\paragraph{Direct detection of dark matter}

Assuming the Galactic DM halo is mostly made up of neutralinos, it may
be possible to directly detect them via their elastic scatterings off
nuclei, or indirectly via their annihilation products. Direct DM
detection in SUSY frameworks has been investigated by many
authors~\cite{dddm}. The CDMS Collaboration (CDMS--II) has recently
improved their previous upper limit on the spin--independent dark
matter WIMP elastic scattering cross section on a free proton,
$\sigsip$, down to some $2 \times 10^{-7}\pb$ (at low WIMP
mass)~\cite{cdms04limit}. Comparable upper limits have also been set
up by the Edelweiss--I~\cite{edelweiss-one-final} and
ZEPLIN--I~\cite{zeplin-one-final} experiments.

However, several questions regarding the properties of the DM halo
(\eg, the existence of clumps of DM and the value of the local halo
mass density) remain unsettled.  Recent numerical N--body simulations
of large structure formation have revealed a large number of overdense
regions surviving until today~\cite{morehalo99}. It is possible that
an improved sensitivity of the simulations will reveal further, and
finer, clumpiness. The clumps may contain a sizable fraction of the
total dark matter halo, of the order of 10\%.  It is therefore not
unlikely that locally (at the Earth's location) the DM density may be
significantly different from the usually assumed average value of
$\rho_\chi=0.3\gev/{\mbox{cm}}^3$.

It is worth remembering that, in translating null experimental results
for an elastic WIMP--target cross section into upper limits on
$\sigsip$, not only the local DM density enters but also a WIMP
velocity distribution. This is usually taken to be Maxwellian with a
peak at $220~\kmeter/\second$ with an estimated error of between 20
and $50~\kmeter/\second$. While this leads to an additional
uncertainty in $\sigsip$, it is actually tiny compared with the
uncertainty of the actual WIMP density at the position of the Earth
caused by cuspiness.  In view of the above discussion, in our opinion
upper experimental limits on $\sigsip$ should not be used to constrain
supersymmetric parameters with the same degree of reliability as from
collider searches.

In computing $\sigsip$ in the CMSSM, we
include full supersymmetric contributions which have
been derived by several
groups~\cite{dn93scatt:ref,jkg96,bb98,efo00,knrr1}. $\sigsip$  can
be expressed as
\beq \sigsip= \frac{4}{\pi}\mu^2_p f_p^2 \label{sigsipdef} \eeq
where $\mu_p=m_p \mchi/(m_p+\mchi)$ is the reduced mass of the
WIMP--proton system. (For spin--independent interactions of neutralinos, and more
generally Majorana WIMPs, there is no need to consider an
analogous quantity on a free neutron since $\sigsin=\sigsip$.)

The coefficients $f_{p,n}$ can be expressed
as~\cite{dn93scatt:ref} 
\bea 
{f_p \over m_p} = \sum_{q=u,d,s}
{f_{Tq}^{(p)} \over m_q} f_q ~+~ {2\over 27} f^{(p)}_{TG}
\sum_{q=c,b,t} {f_q \over m_q} + ... \nonumber 
\eea 
where
$f^{(p)}_{TG} = 1 - \sum_{q=u,d,s} f^{(p)}_{Tq} $, and nuclear
form factors $f_{Tq}^{(p)}$ are defined via $\langle p|m_q \bar{q} q|p\rangle =
m_p f_{Tq}^{(p)}$ ($q=u,d,s$), and analogously for the neutron.
The masses and ratios $B_q=\langle p|{\bar q}q|p\rangle$ of light
constituent quarks in a nucleon come with some uncertainties. For
definiteness, we adopt the set of input parameter given
in~\cite{efo00} and assume $m_u/m_d= 0.553\pm0.043$, $m_s/m_d=
18.9\pm0.8$, and $B_d/B_u= 0.73\pm0.02$, as well as
\begin{eqnarray}
\label{fpvalues:eq} f^{(p)}_{Tu}=0.020\pm0.004,\quad
f^{(p)}_{Td}=0.026\pm0.005,\quad f^{(p)}_{Ts}=0.118\pm0.062,
\\
\label{fnvalues:eq} f^{(n)}_{Tu}=0.014\pm0.003,\quad
f^{(n)}_{Td}=0.036\pm0.008,\quad f^{(n)}_{Ts}=0.118\pm0.062,
\end{eqnarray}
and for the parton density functions we employ the CTEQ6L
set~\cite{cteq} evaluated at the QCD scale defined by the averaged
squark mass and neutralino mass $Q \approx
\sqrt{M^2_{\tilde{q}}-\mchi^2}$. The nuclear form factors
$f_{Ts}^{(p,n)}$ come with a large error and these are the ones
that provide the dominant contribution to $\sigsip$. This has a
rather significant impact on the size of $\sigsip$~\cite{sip_had},
unless $\tanb$ is very small.

\paragraph{\boldmath $BR(B_s \rightarrow \mu^+ \mu^-)$}
The latest $95 \%$~\cl\ experimental upper limits from the D\O\
Run II and CDF Run II experiments at Fermilab are,
respectively,
\begin{eqnarray}
BR(B_s \rightarrow \mu^+ \mu^-) &<& 2.0 \times 10^{-7}\ \ \ {\rm
(CDF)~\cite{cdfbsmm05}},\\
BR(B_s \rightarrow \mu^+ \mu^-) &<& 3.7 \times 10^{-7}\ \ \ {\rm
(D\O)~\cite{dzerobsmm05prelim}}.
\end{eqnarray}
A combined $95 \%$~\cl\ limit is $BR(B_s \rightarrow \mu^+ \mu^-)
< 1.5 \times 10^{-7}$~\cite{Tev:bsm}.
Ultimately, assuming the integrated luminosity of $8\fb^{-1}$, a combined CDF and
D\O\ limit is expected to reach some
$2\times 10^{-8}$~\cite{tevatronbsmmfuture} which is significantly
above the SM prediction~\cite{buras03-bsmmsm}
\begin{eqnarray}
BR(B_s \rightarrow \mu^+ \mu^-)_{\mathrm{SM}}=\left(3.42\pm
0.54\right)\times 10^{-9}. \label{eq:bsm-SM}
\end{eqnarray}

We compute the SUSY contribution to $BR(B_s \rightarrow \mu^+
\mu^-)$ by following a full one--loop calculation of
Ref.~\cite{beku01} which assumes MFV. Furthermore we include
$\tanb$ enhanced corrections to the bottom quark
mass~\cite{cgnw2000}. The parametric uncertainties are associated
with an error in the decay constant $f_{B_s}$, which arises from
lattice calculations, and an error in the bottom quark
mass~\cite{eos05}. The latter is accounted for by the MC
procedure, while the former is of order $10 \%$. Unknown higher
order corrections are of order $10 \%$~\cite{bbku01}. Therefore we
use a total theoretical error of 14\%, obtained by adding the
above uncertainties in quadrature.

\paragraph{The lightest MSSM Higgs boson mass}
A final LEP--II lower bound on the SM Higgs mass is $m_{H_{\text{SM}}}
> 114.4 \gev$ (95 \%~\cl)~\cite{lephiggs}.  The bound applies to the
lightest Higgs boson $h$ in the MSSM if its coupling to the $Z$ boson
is SM--like, \ie\ if $\zetah^2 \equiv g^2_{ZZh}/g^2_{ZZH_{\text{SM}}}
= \sin^2(\beta -\alpha) \simeq 1$.  This occurs in the decoupling
regime where $m_A \gg m_Z$. For arbitrary values of $m_A$, the LEP--II
Collaboration has set 95\%~\cl\ bounds on $m_h$ and $m_A$ as a function
of $\zetah^2$~\cite{lephiggs}, with the lower bound of $m_h > 91 \gev$
for $m_h \sim m_A$ and $\zetah^2\ll 1$~\cite{lephiggs}. In this
low--mass region we use a cubic spline to interpolate between some
selected points in $m_h$ and derive the corresponding 95\%~\cl\ bound,
which is then smeared with a theoretical uncertainty $\tau$ of 3\%.
The intrinsic theoretical error in computing $m_h$, after taking into
account effects of the renormalization scale dependence, in the CMSSM
has been estimated to be $\tau(m_h)= 3 \gev$~\cite{adkps,hhrw}.  The
parametric uncertainty coming from the errors in top quark mass and
the strong coupling constant are accounted for by taking them as
nuisance parameters.

As mentioned above, we compute the lightest Higgs mass following the SOFTSUSY~v1.9
package~\cite{softsusy}, where full one--loop and leading
two--loop corrections and two--loop effects on the EWSB conditions
are included.

\paragraph{Superpartner masses} Below we comment on some limits on
superpartner masses. Since currently there is no information
available on the likelihood function for sparticle masses, we make
use of the experimental error of 95\%~\cl\ and of the likelihood
given in Eq.~\eqref{eq:limerr}.

The parametric uncertainties in the sparticle masses coming from
the SM variables are accounted for via the nuisance parameters.
The authors of Ref.~\cite{akp03} have argued that the theoretical
uncertainties in the computation of SUSY masses are of order
$\mathcal{O}(1\%)$ except in some special regions of the parameter
space (such as the FP region) which require a separate
treatment. Conservatively we take $\tau=5\%$ for each computed
superpartner mass and use it to smear out the 95\%~\cl\
experimental lower limit on its mass, as explained in
section~\ref{sec:like}.

\paragraph{Neutralino LSP mass}
In the context of the CMSSM, LEP--II provides an absolute lower bound
on the mass of the lightest neutralino LSP $\chi_1$ (in this work denoted
simply by $\chi$)~\cite{lsp}
\begin{equation}
\mchi > 50 \gev.
\end{equation}

\paragraph{Chargino mass}
Chargino mass has been excluded up to $\mcharone > 103.5
\gev$~\cite{lepsusy}, provided that $m_{\tilde \nu} > 300 \gev$,
where $m_{\tilde \nu}$ stands for the lightest sneutrino mass
which in the CMSSM is ${\tilde
  \nu}_\tau$.
However, when the mass difference $\mcharone - m_{\chi} \lesssim
3 \gev$, as in the FP region, then the bound is relaxed to
$m_{\chi^\pm_1} > 92.4 \gev$~\cite{lepsusy}. The latter bound is
also applied when $m_{\tilde \nu} < 300 \gev$.

\paragraph{Slepton masses}
Combined slepton mass limits have been obtained by the LEP SUSY
working group~\cite{lepsusy}. The overall limits are
\begin{equation}
m_{\tilde{e}_R} > 100 \gev,\qquad m_{\tilde{\mu}_R} > 95
\gev,\qquad m_{\tilde{\tau}_1} > 87 \gev.
\end{equation}
These limits are valid provided that $m_{\tilde l} - \mchi >
10 \gev$. Otherwise we apply the more conservative bound
$m_{\tilde{l}_R} > 73 \gev$~\cite{aleph,l3}.

For the sneutrino, in the context of the CMSSM, the DELPHI
collaboration has obtained the following limit~\cite{delphi}
\begin{equation}
m_{\tilde{\nu}} > 94 \gev,
\end{equation}
provided  $m_{\tilde{\nu}} - \mchi > 10 \gev$. Otherwise we
apply $m_{\tilde{\nu}} > 43 \gev$~\cite{PDG}.

\paragraph{Squark masses}
Combined limits on the lightest stop and sbottom masses from LEP--II
are~\cite{lepsusy}
\begin{equation}
m_{\tilde t_1, \tilde b_1} > 95 \gev,
\end{equation}
provided  $m_{\tilde t_1, \tilde b_1} - \mchi > 10 \gev$.
Otherwise we apply the more conservative limits obtained by the
ALEPH Collaboration~\cite{aleph}
\begin{equation}
m_{\tilde{t}_1} > 65 \gev \quad \text{and} \quad m_{\tilde{b}_1} >
59 \gev.
\end{equation}
Finally, for the two first generations, the D\O\ Run II
Collaboration obtained~\cite{D0}
\begin{equation}
m_{\tilde{q}} >  318 \gev.
\end{equation}

\section{Results \label{sec:results}}

We now present the results of our study in terms of high relative
posterior probability regions for CMSSM parameters and superpartner
masses (section~\ref{sec:susymasses}) and the implications for other
observables (section~\ref{sec:otherobservables}). In
section~\ref{sec:fitquality} we compare those results with the mean
quality of fit statistics, while in section~\ref{sec:dd} the prospects
for direct detection of DM are presented. Correlations among
observables are depicted in section~\ref{sec:correlationsobs}, while
section~\ref{sec:nogm2} is concerned with the influence that the
measurement of the anomalous magnetic moment has on our conclusions.

Our statistical inferences are drawn from multiple MC chains,
which contain a total of about $3\times 10^5$ to $4\times10^5$
samples. For more details about our numerical implementation of
the MCMC algorithm, see appendix~\ref{appx:mcmc}.

\subsection{High probability regions for parameters and superpartners
 masses} \label{sec:susymasses}
\begin{figure}[t!]
\begin{center}
\begin{minipage}{6in} \begin{center}
\epsfig{file=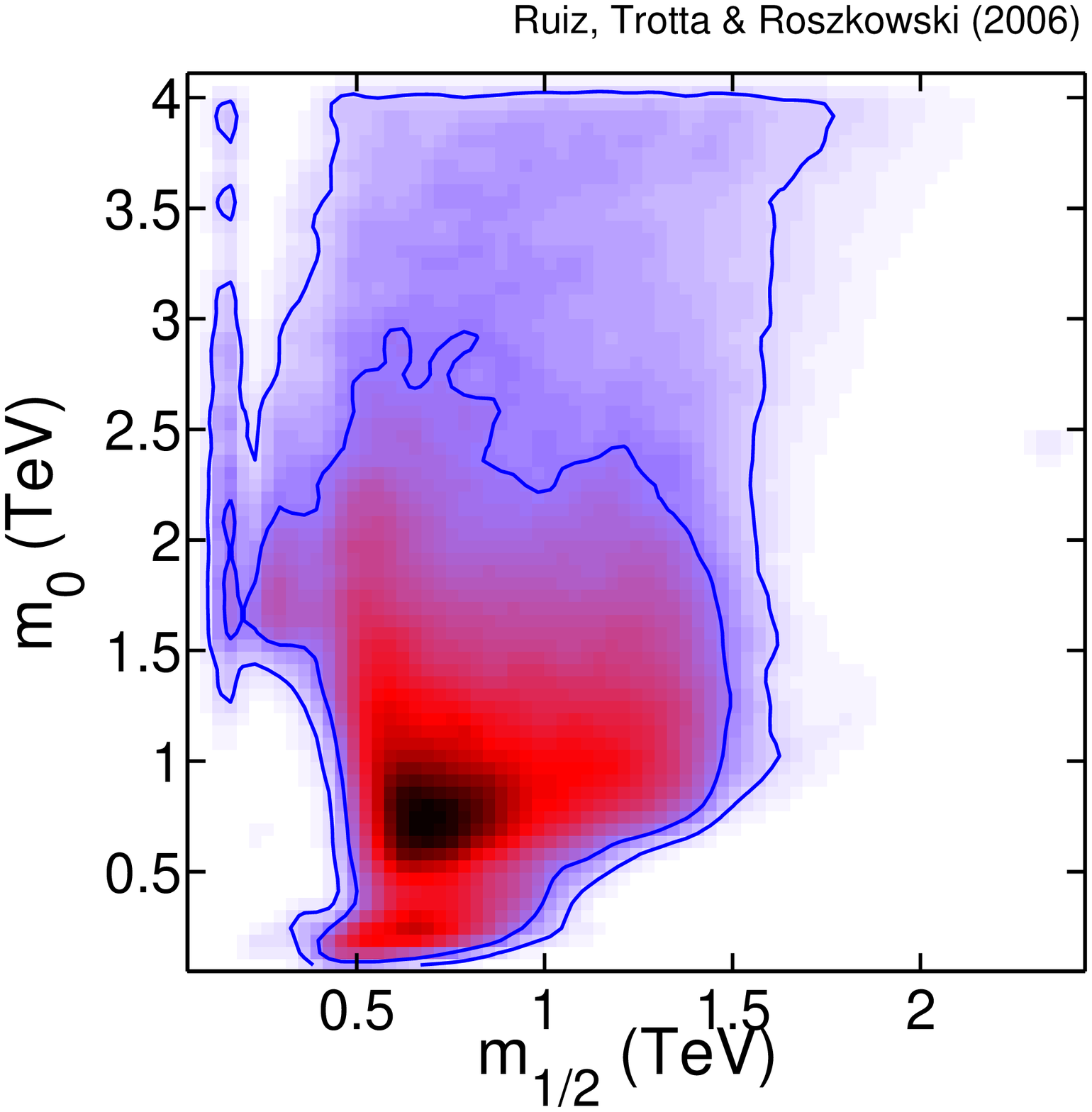,width=2.2in} 
\epsfig{file=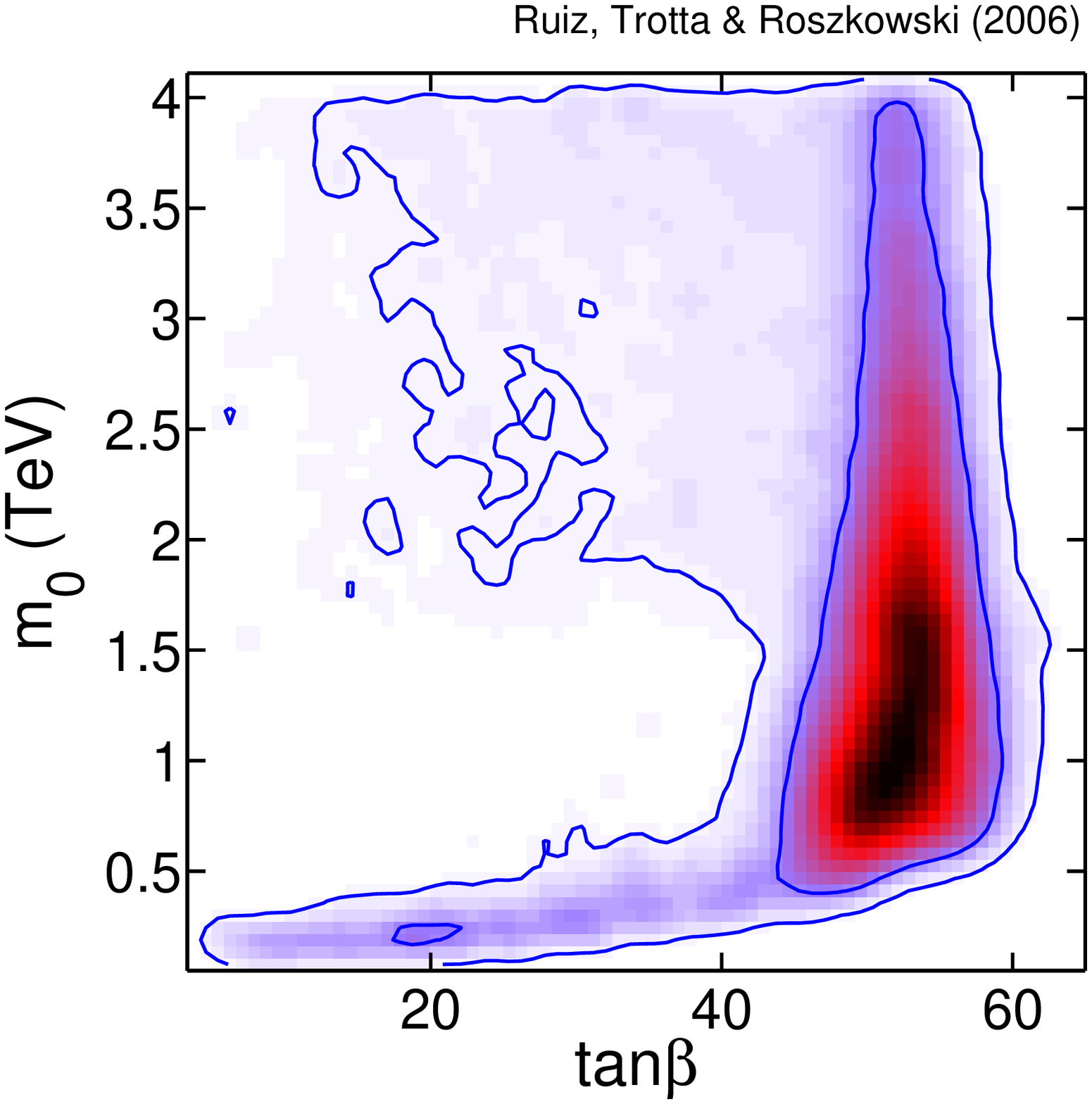,width=2.2in}
\end{center} \end{minipage}
\begin{minipage}{6in} \begin{center}
\epsfig{file=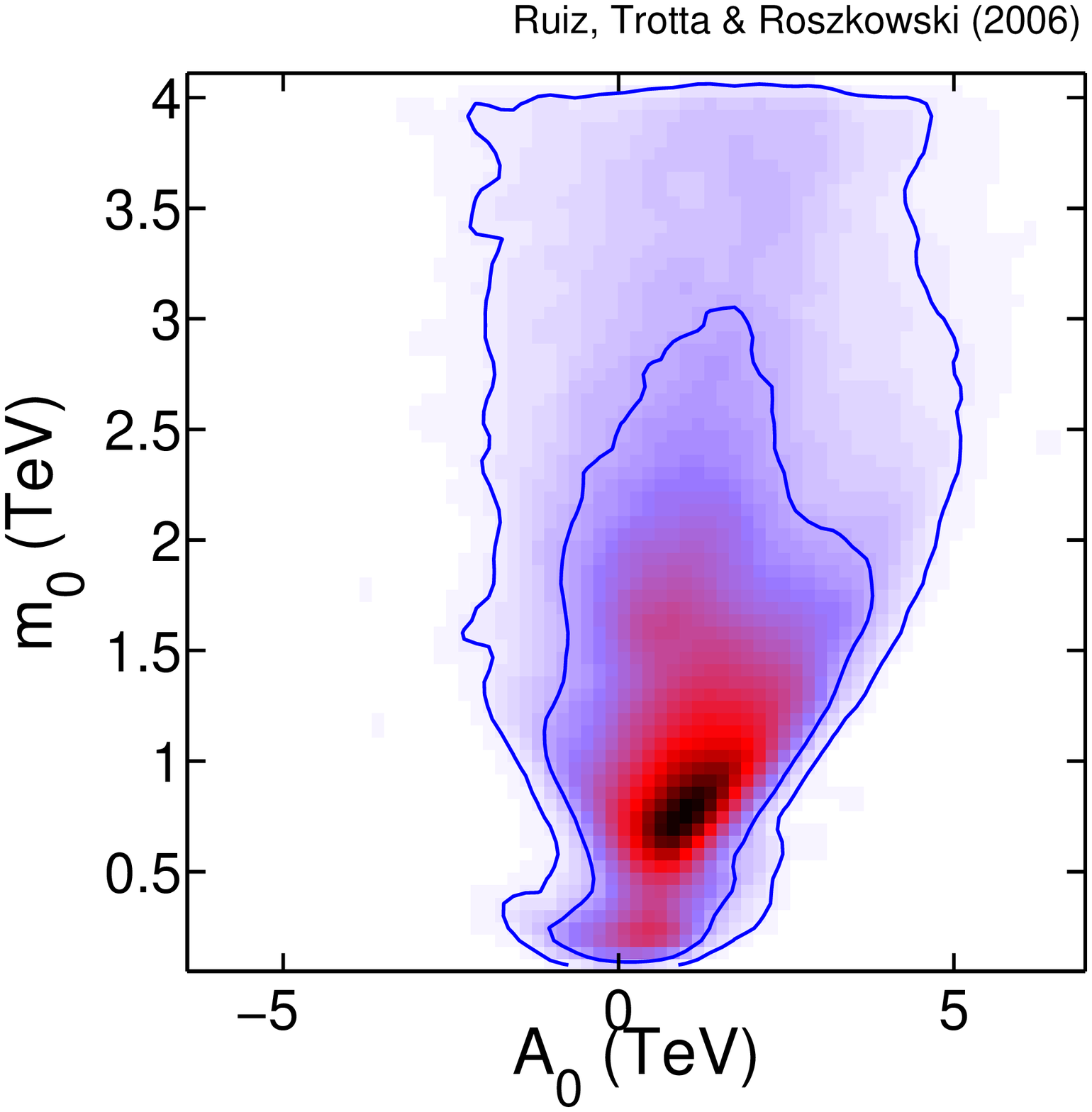,width=2.2in} 
\epsfig{file=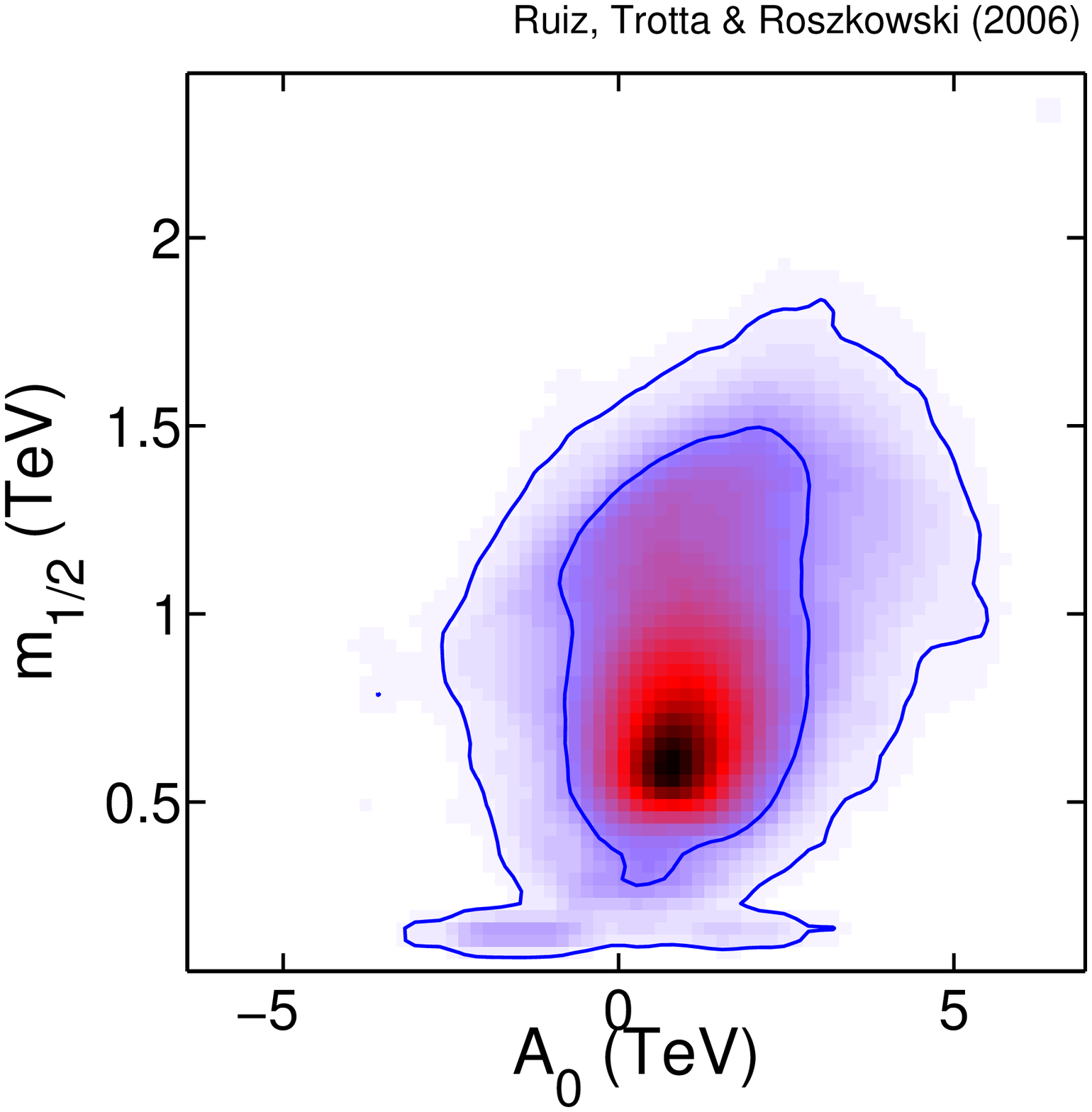,width=2.2in}
\end{center} \end{minipage}
\begin{minipage}{6in} \begin{center}
\epsfig{file=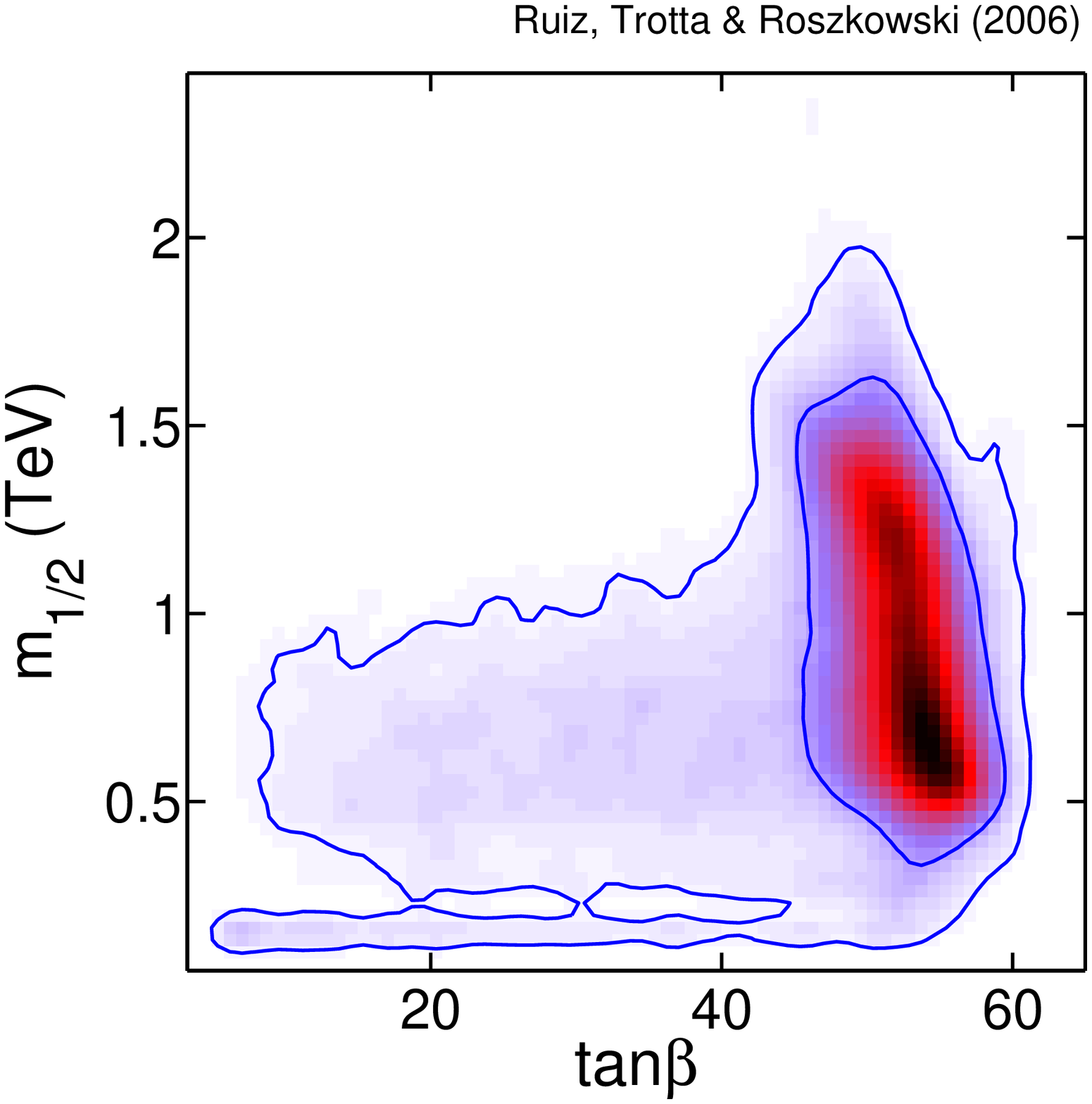,width=2.2in} 
\epsfig{file=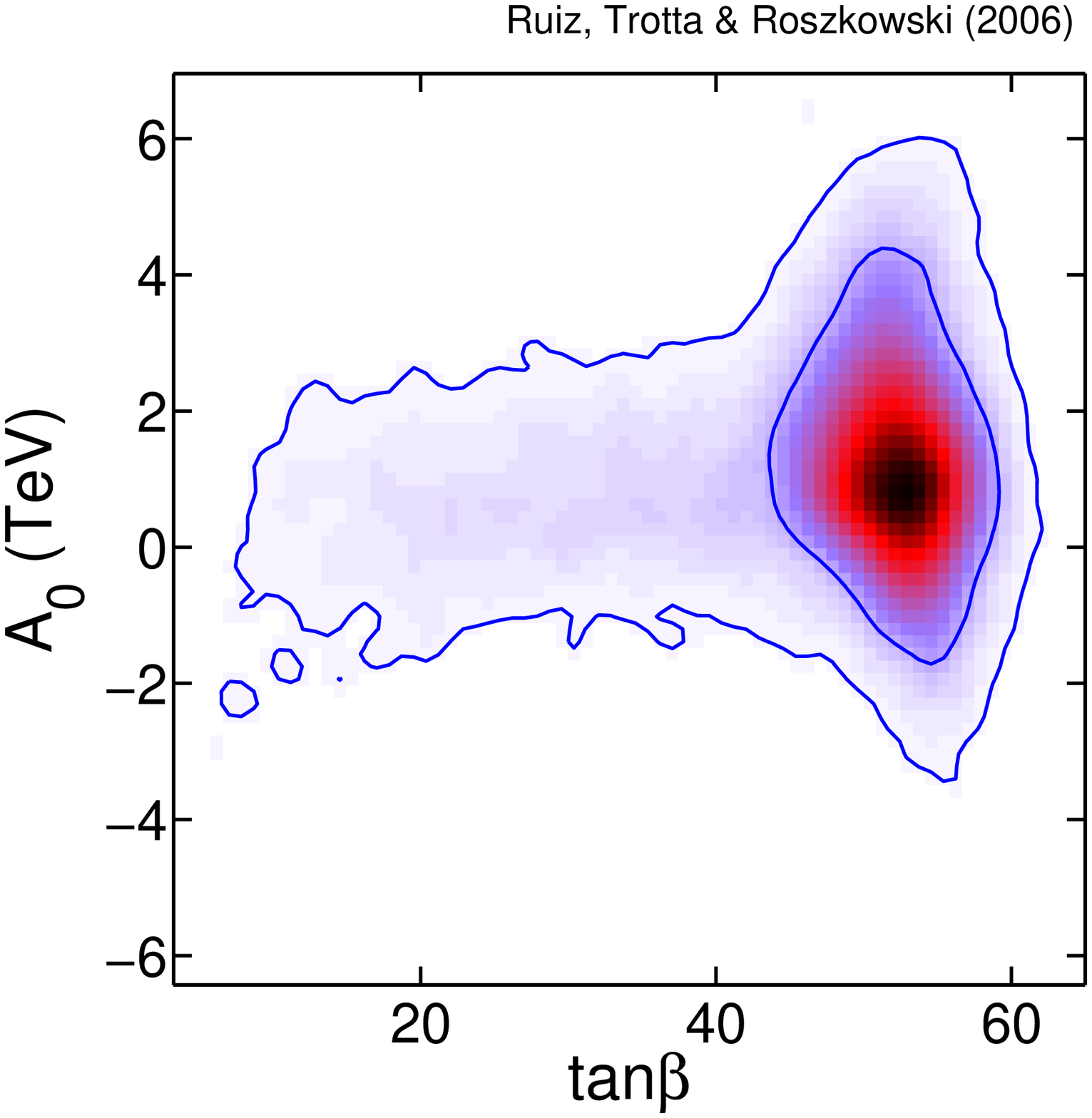,width=2.2in}
\end{center} \end{minipage}
\begin{minipage}{6in} \begin{center}
\epsfig{file=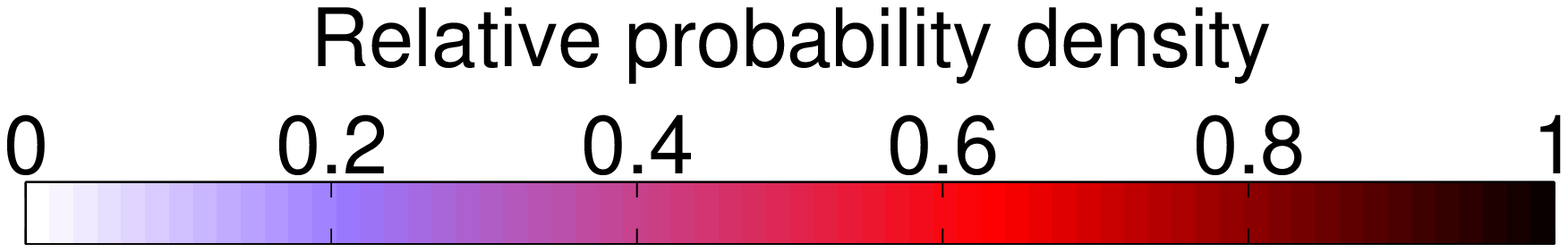,width=2.2in}
\end{center} \end{minipage}
\caption{\label{fig:contours4tev} {\small The 2--dimensional probability
densities in the planes spanned by the CMSSM parameters: $\mhalf$,
$\mzero$, $\azero$ and $\tanb$ for the ``4\tev\ range'' analysis
(see Table~\protect\ref{table:prior}). The pdf's are normalized to
unity at their peak. The inner (outer) blue solid contours delimit
regions encompassing 68\% and 95\% of the total probability,
respectively. All other parameters in each plane have been
marginalized over. } }
\end{center}
\end{figure}

\EPSFIGURE[t]{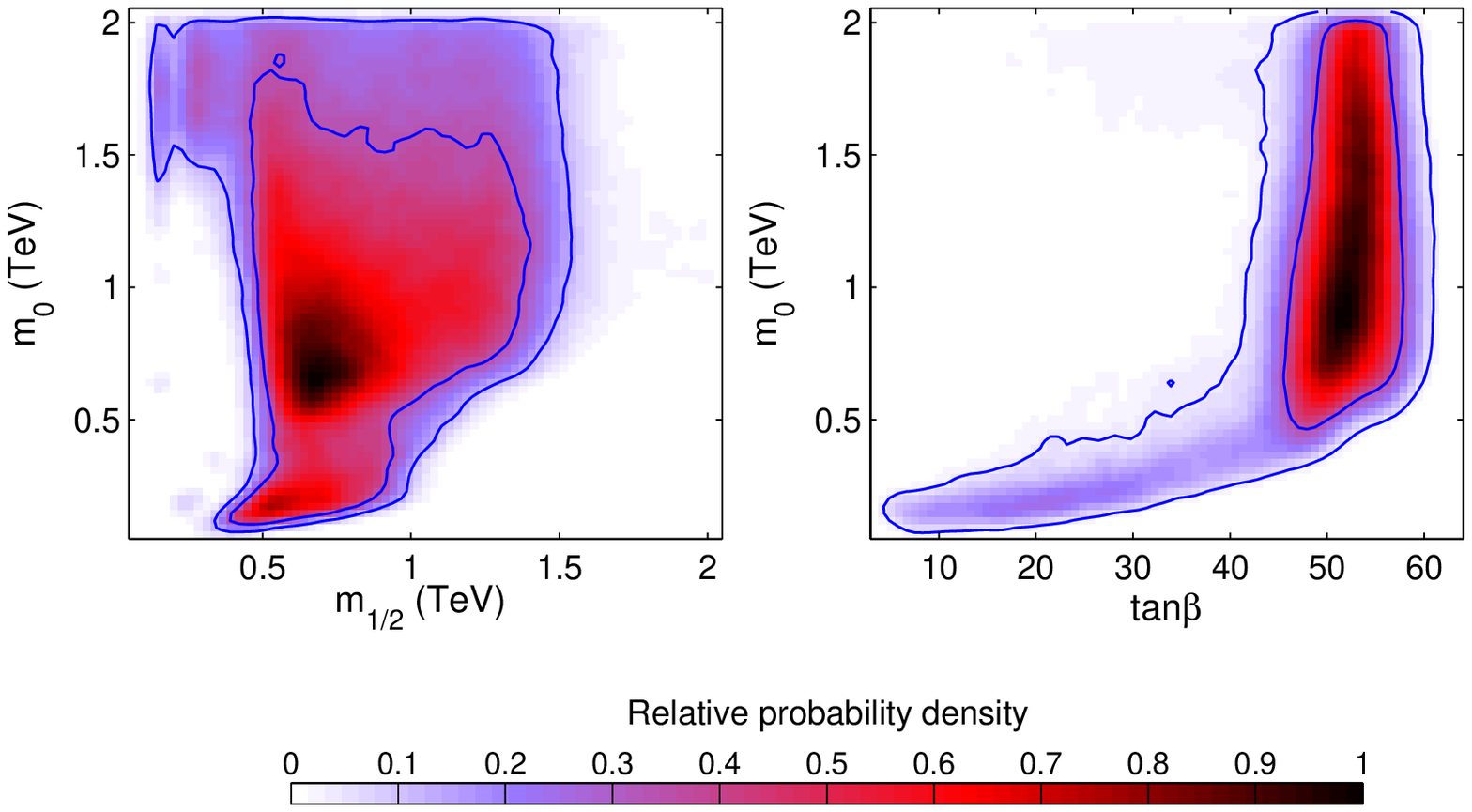,width=5in} {The 2--dimensional relative probability
densities as in Fig.~\protect\ref{fig:contours4tev} but for the
``2\tev\
  range'' analysis and for the planes $(\mhalf,\mzero)$ and
  $(\tanb,\mzero)$ only. Imposing a prior range $m_0 < 2 \tev$
  modifies substantially the inferences about the high probability
  regions for $m_0$.
\label{fig:contours2tev} }

In the six panels of Fig.~\ref{fig:contours4tev}, we show
the 2--dimensional posterior relative probability density functions
$p(\params_i,\params_j | \data)$, where
$(\params_i,\params_j)=(\mhalf,\mzero)$, $(\tanb,\mzero)$,
$(\azero,\mzero)$, $(\azero,\mhalf)$, $(\tanb,\mhalf)$ and
$(\tanb,\azero)$, for the ``4\tev\ range'' case which we will
treat as our default choice.  In each panel all other basis parameters have been
marginalized over. Redder (darker) regions
correspond to higher probability density. Inner and outer blue
(dark) solid contours delimit regions of 68\% and 95\% of the
total probability, respectively. In all the 2--dimensional plots, the
MC samples have been divided into $70\times 70$ bins. Jagged
contours are a result of a finite resolution of the MC chains.

It is clear that the structure of the parameter space is rather
complex. In particular, in the left top and bottom panels at
$\mhalf\simeq 0.2\tev$ we can see a narrow high--probability funnel
induced by the light Higgs boson resonance, which was also
observed in the analysis of Ref.~\cite{al05}. The presence of such
narrow wedges makes the exploration challenging for the MCMC
procedure, and much harder for a fixed--grid scan.

Values of $\mzero\lsim2\tev$ are favored, but larger values are
definitely not excluded. In particular, the 95\% probability
region extends up to the upper prior range for $m_0$, a clear sign
that the data is not powerful enough to constrain this parameter
sufficiently (we comment further on this issue below). The most
probable region in the $(\mhalf,\mzero)$ plane is centered around
the point
\begin{equation}\label{eq:2dimprob}
\mhalf\simeq0.7\tev,\qquad \mzero\simeq0.8\tev.
\end{equation}
The region encompassing $68\%$ of joint probability is roughly bounded
by $0.5\tev\lsim\mhalf\lsim 1.5\tev$, because of the efficiency of
both the stau coannihilation and/or the pseudoscalar resonance. A
sharp probability drop above $\mhalf\gsim1.5\tev$, almost independent
of $\mzero$, is caused by a combination of the relic abundance and the
$\dasusy$ constraints. At smaller $\mzero$ the boundary bends because
below it the neutralino is not the LSP. Large values of $\tanb$
between about 45 and 55 are definitely more favored but smaller values
are also allowed, in particular for small $\mzero\lsim0.5\tev$ (upper
right panel), as a consequence of the light Higgs boson resonance.

The large $\mzero$ region, starting from the upper part of the $68\%$
probability contour, corresponds to a wide range of possible positions
of the FP region. For each fixed choice of parameters, the FP region
consistent with the CDM abundance is actually very narrow, but its
position varies widely along $\mzero$ when we marginalize over all
the other parameters.

We observe that $\azero$ is largely uncorrelated with other
variables, and its pdf presents a strong peak around
$\azero\simeq0.8\tev$. The high probability contours for $\azero$
are well within the prior range ($|\azero| < 7 \tev$), which
indicates that this constraint is robust with respect to changes
to the prior.

In order to examine the sensitivity of these results to the assumed
ranges of CMSSM parameters, \ie, the prior used, in
Fig.~\ref{fig:contours2tev} we plot the 2--dimensional pdf's
$\relprobtwo{\mhalf}{\mzero}$ and $\relprobtwo{\tanb}{\mzero}$ for the
``2\tev\ range''. This is similar to the prior used by the authors of
Ref.~\cite{al05}, and for this case we find a fairly good general
agreement with their results. However, several differences in the
treatment of uncertainties, in the data employed and in the accuracy
of the theoretical predictions add up to appreciable differences in
the details of the results.\footnote{For instance, in the analysis of
Ref.~\cite{al05} $\mzero$ as small as 1\tev\ is allowed in the
vicinity of the resonance (compared to our $\mzero\gsim1.4\tev$) as a
result of employing a less restrictive chargino mass bound of only
$67.7\gev$.}

It is important to stress that by imposing an upper prior range
$\mzero < 2 \tev$ one cuts away a large region of parameter
space which is not excluded by the data (the FP region at large
$\mzero$, compare with the corresponding panels in
Fig.~\ref{fig:contours4tev}). Therefore, inferences on high probability regions
for large $\mzero$ are different for the ``2\tev\ range'' and the
``4\tev\ range'' cases. This means that current data is not
informative enough to strongly constrain the value of $\mzero$
independently of prior information, \ie\ the prior range one chooses
to use.  The main reason behind this is the presence of
the FP region which so far has not been investigated by MCMC
techniques.  On the other hand, results for the other CMSSM parameters
$\mhalf, \tanb$ and $\azero$ do {\em not} vary appreciably if one
changes the prior ranges, as we now discuss.

\EPSFIGURE[t]{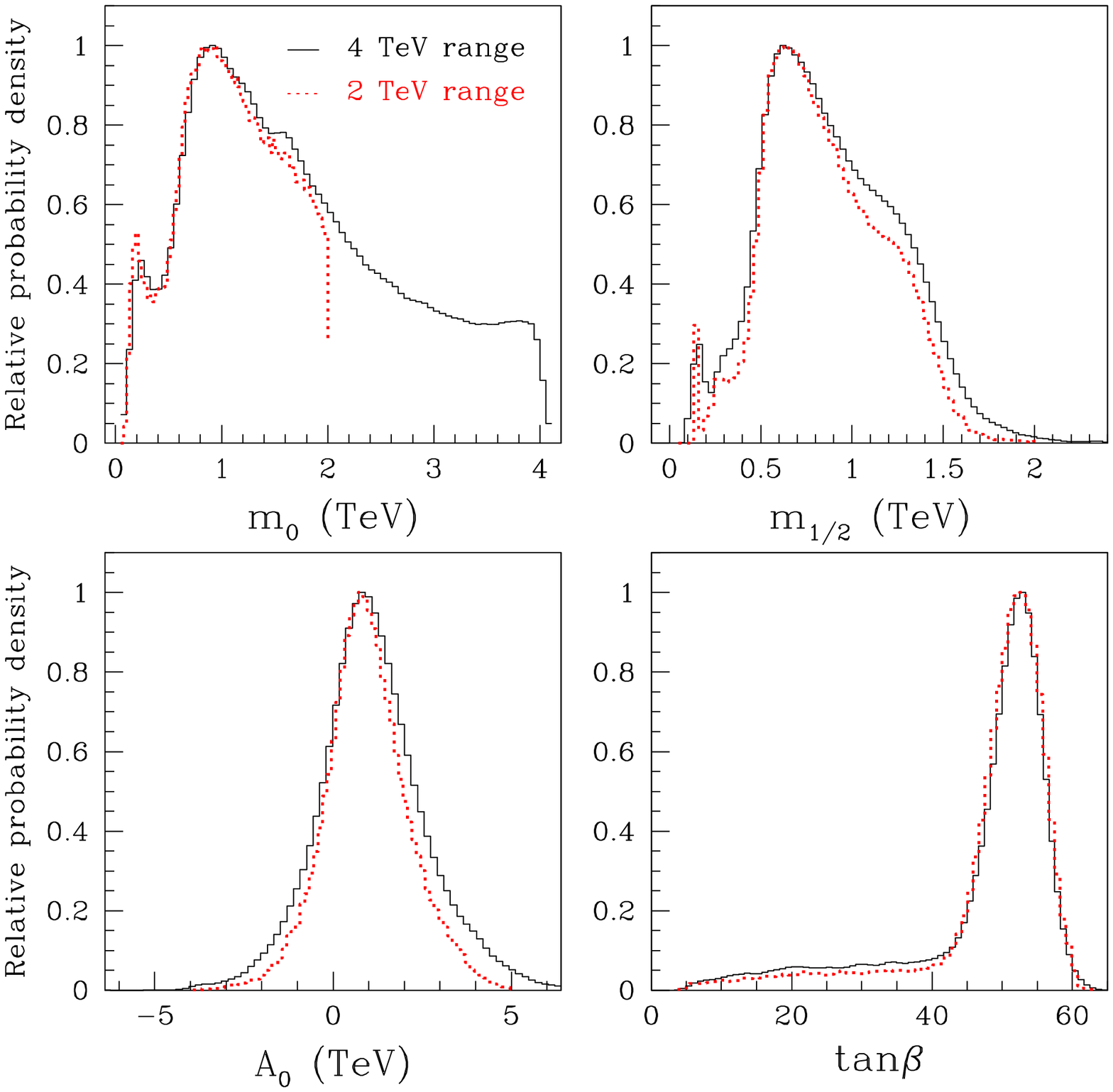,width=6in} {The 1--dimensional relative probability
densities $\relprobone{\params_i}$ for each of the CMSSM parameter,
$\params_i=\mzero$, $\mhalf$, $\azero$ and $\tanb$. All other
parameters have been marginalized over. The two curves compare the
results for two different prior ranges (see
Table~\ref{table:prior}).\label{fig:susyinputs2vs4tev} }
In Fig.~\ref{fig:susyinputs2vs4tev} we show the posterior
1--dimensional pdf $\relprobone{\params_i}$ for each of the CMSSM
parameters, with all the other basis parameters marginalized over.
The 1--dimensional pdf's contain the complete statistical
information about each of the CMSSM variables, fully accounting
for all sources of uncertainty included in the analysis. We plot
results for both the ``2\tev\ range'' and the ``4\tev\ range'' to
facilitate the comparison between the two. In the upper left
panel, we note that the pdf's for $\mzero \leq 2\tev$ are in
excellent agreement for both ranges, but above that value the
posterior pdf for the ``2\tev\ range'' is sharply cut by the
prior. On the contrary, the pdf for the ``4\tev\ range'' extends
all the way to $4\tev$. Since the posterior does not drop to zero
before reaching the prior range of $4\tev$, it is likely that by
allowing even larger values of $\mzero$ one would find
non--negligible probability densities even there. This is caused
by a high sensitivity of the position of the FP region along the
$\mzero$ axis to the Yukawa couplings $h_t$ and $h_b$. The effect
is partially accommodated by treating the top and bottom masses in
the nuisance parameters. However, as explained in the discussion
following Eq.~\eqref{eq:cdmrange}, the computation of $\mu$ and as
a result of $\abundchi$ is highly uncertain in the FP region. At
present this makes it difficult to make a more definitive
statement about the region $\mzero\gsim2\tev$ other than that
present data does not strongly constrain it. In fact, the upper
$95\%$ probability region for the ``4\tev\ range'' extends to
$\mzero < 3.66 \tev$. For smaller $\mzero$, the bulk of the pdf
lies around $0.5\tev\lsim\mzero \lsim 1.5 \tev$.

Constraints on the other CMSSM parameters are largely independent
of the adopted prior ranges. The bulk of the pdf for $\mhalf$ lies
around $\mhalf \approx 0.75 \tev$, with the 68\% region within
$0.52\tev <\mhalf<1.26\tev$, with again a narrow peak due to the
light Higgs resonance at smaller values. For the ``4\tev\ range''
this narrow peak is more pronounced, because in this case one
integrates over a larger range for $\mzero$, compare with the
upper left panel of Fig.~\ref{fig:contours4tev}. In the lower left
panel of Fig.~\ref{fig:susyinputs2vs4tev}, the peak around
$\azero\simeq0.8\tev$ is again clearly visible and almost
independent of the prior used. We notice that the 68\% region is
bounded by $-0.34\tev < \azero < 2.41\tev$ and thus $\azero = 0$
lies close to its boundary.  This is an interesting result
(which can also be seen in Ref.~\cite{al05})  in light
of the fact that most of fixed--grid scans (with a few
exceptions~\cite{shbp05,nonzeroazero}) have assumed
$\azero=0$.  Finally, in the last panel, the 1--dimensional pdf
for $\tanb$ shows a preference for large values, with the $68\%$
region given by $38.5 < \tanb < 54.6$.
Regions containing 68\% and 95\% of posterior probability for the
CMSSM parameters are summarized in Table~\ref{table:CMSSMtable}.

\begin{table}
 \centering
 \begin{tabular}{|l |c c| c c|}
  \hline
            & \multicolumn{2}{|c|}{``2\tev\  range''} &
 \multicolumn{2}{|c|}{``4\tev\  range'' }\\
  Parameter & 68\% region    & 95\% region            & 68\% region    & 95\%
            region\\\hline
  $\mzero$~(TeV)  &$<1.36$     &$<1.88$             &$<2.10$ &$<3.66\tev$        \\
  $\mhalf$~(TeV)  &$(0.54, 1.20)$  &$(0.26, 1.48)$  & $(0.52, 1.26)$ & $(0.20, 1.61)$\\
  $\azero$~(TeV)  &$(-0.18, 2.01)$ &$(-1.44, 3.40)$ &$(-0.34, 2.41)$ &$(-1.95, 4.31)$\\
  $\tanb $  &$(44.1, 54.9)$ &$(16.0, 58.0)$        &$(38.5, 54.6)$   &$(13.6, 57.8)$ \\

\hline
  \end{tabular} 
 \caption{CMSSM parameter ranges corresponding to 68\% and 95\% of
 posterior probability (all other parameters marginalized over) for the
 two different prior choices, the ``2\tev\  range'' and the
 ``4\tev\ range''. } \label{table:CMSSMtable}
 \end{table}

\begin{table}
\centering
\begin{tabular}{|c |c c| c c|}
 \hline
Super--  & \multicolumn{2}{|c|}{``2\tev\  range''} & \multicolumn{2}{|c|}{``4\tev\  range'' }\\
partner & 68\% & 95\% & 68\% & 95\% \\
           \hline
 $\chi_1^0$     & $(0.22, 0.52)$  & $(0.10, 0.64)$ & $(0.22, 0.55)$ & $(0.07, 0.70)$\\
 $\charone$     & $(0.42, 0.98)$  & $(0.16, 1.20)$ & $(0.36, 1.00)$ & $(0.11, 1.25)$\\
 $\gluino$      & $(1.27, 2.64)$  & $(0.70, 3.19)$ & $(1.25, 2.80)$ & $(0.54, 3.51)$\\
 $\tilde{e}_R$  & $(0.66, 1.69)$  & $(0.30, 1.97)$ & $(0.80, 2.93)$ & $(0.33, 3.85)$\\
 $\tilde{\nu}$  & $(0.68, 1.49)$  & $(0.42, 1.76)$ & $(0.79, 2.63)$ & $(0.46, 3.57)$\\
$\tilde{\tau}_1$& $(0.36, 1.03)$  & $(0.23, 1.41)$ & $(0.42, 2.12)$ & $(0.25, 3.31)$\\
 $\tilde{q}_R$  & $(1.47, 2.61)$  & $(1.10, 3.11)$ & $(1.60, 3.50)$ & $(1.18, 4.49)$\\
 $\tilde{t}_1$  & $(1.09, 2.04)$  & $(0.82, 2.48)$ & $(1.17, 2.44)$ & $(0.87, 3.22)$\\
 $\tilde{b}_1$  & $(1.23, 2.26)$  & $(0.98, 2.74)$ & $(1.33, 2.79)$ & $(1.03, 3.66)$\\
\hline
 \end{tabular}
\caption{Selected superpartner mass ranges (in TeV) containing
68\% and 95\% of posterior probability (all other parameters
marginalized) for the two different prior choices, the ``2\tev\
range'' and the ``4\tev\ range''.} \label{table:massesTable}
\end{table}
%

\EPSFIGURE[t]{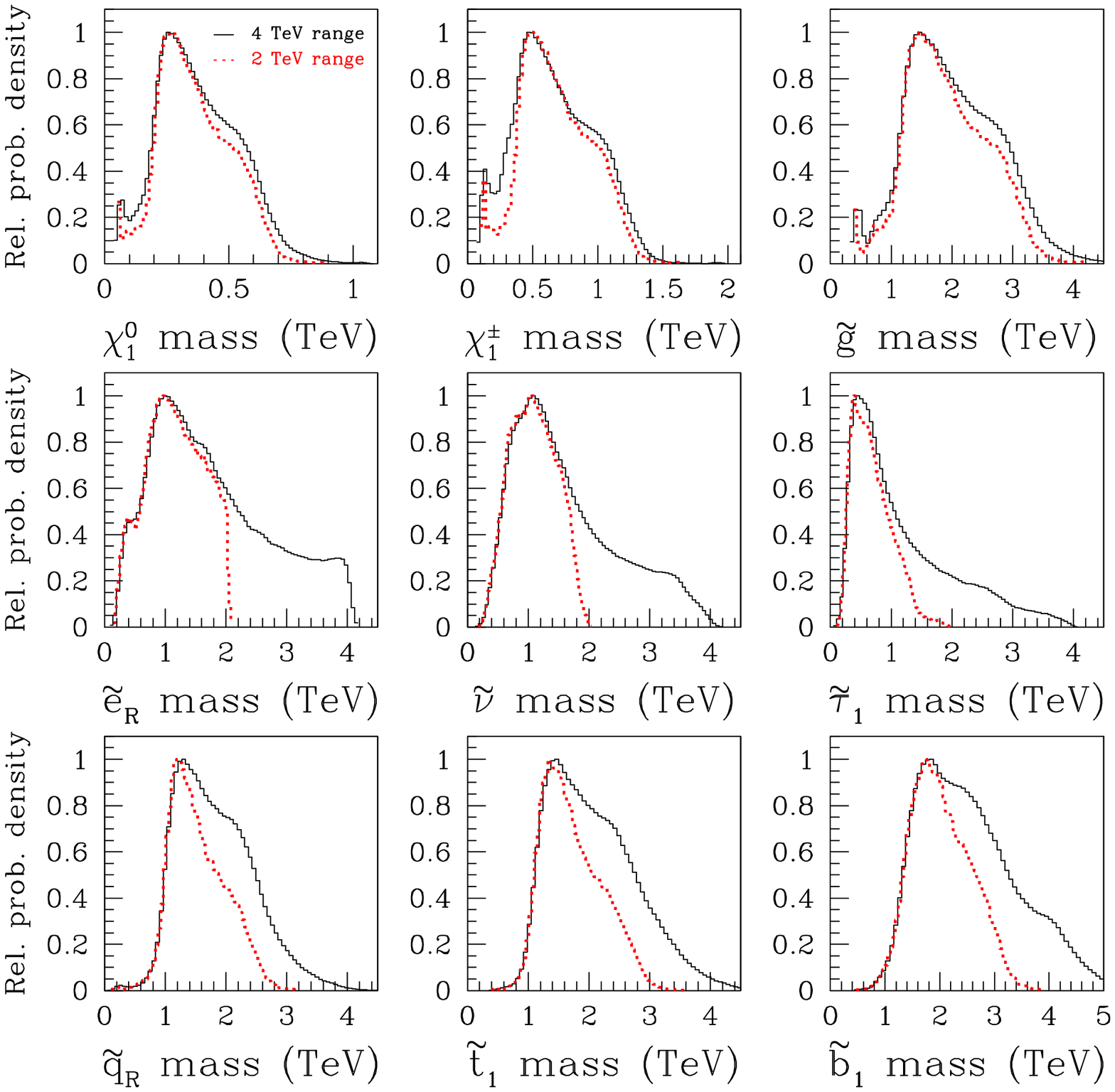,width=6in} {As in
Fig.~\protect\ref{fig:susyinputs2vs4tev}, but for the masses of
several representative superpartners.
\label{fig:susymasses2vs4tev} }

In Fig.~\ref{fig:susymasses2vs4tev} we show 1--dimensional pdf's for
several superpartners masses, and the corresponding 68\% and 95\%
probability regions are given in Table~\ref{table:massesTable}. Note
that the 95\%~\cl\ experimental bounds on the superpartner masses have
been included in the likelihood (and smeared out by corresponding
theoretical uncertainties), as explained in
section~\ref{sec:like}. The masses of the gluino $\gluino$, the
lightest chargino $\charone$ and the LSP neutralino $\chi$, which are
proportional mainly to $\mhalf$, are basically the same for both prior
ranges, in agreement with the result for $\mhalf$ displayed in
Fig.~\ref{fig:susyinputs2vs4tev}.  In contrast, the pdf's for the
masses of the sfermions exhibit a sharp cutoff in the ``2\tev\ range''
case, as a consequence of a basic mass relation $m^2_{{\tilde
f}_{L,R}}\simeq \mzero^2+ c_{{\tilde f}_{L,R}}\mhalf^2$. Therefore the
prior cut on $\mzero$ for ``2\tev\ range'' impacts on the posterior
probability distribution for the sfermions as well. Nevertheless, for
the squarks the relative probability peaks below some $2\tev$ which is
generally well within the LHC reach. As a result, the integrated
probability for $m_{\tilde{q}_R}<2.5\tev$ is 85\%. For comparison, we
find $\mgluino<2.7\tev$ with 78\% probability and $\mcharone<0.8\tev$
with 65\% probability. We will come come back to the issue of the
posterior probability distribution for the superpartners in
subsection~\ref{sec:nogm2}.

%
\subsection{High probability regions for other observables}
\label{sec:otherobservables}
%

\EPSFIGURE[t!]{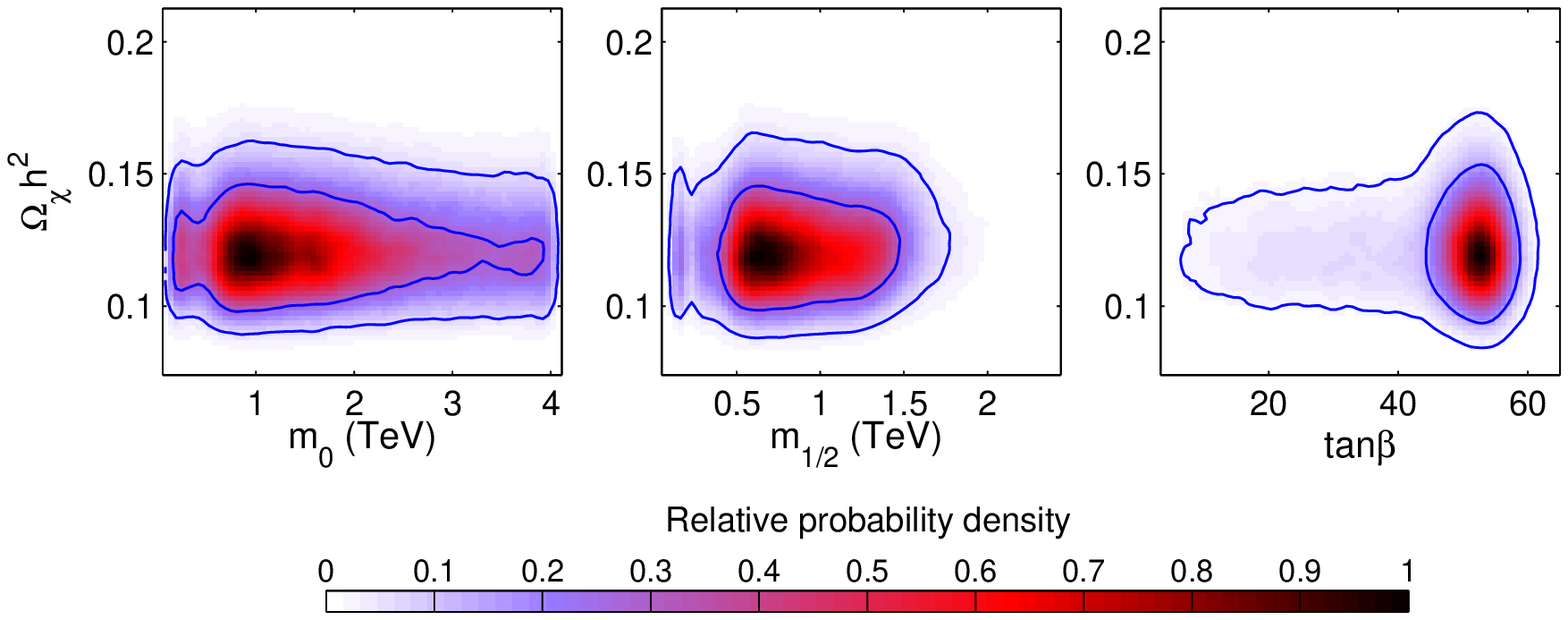,width=6in} {The 2--dim relative probability density
$\relprobtwo{\abundchi}{\theta_i}$, where $\theta_i=\mzero$,
$\mhalf$ and $\tanb$.  Note that the measured value of
$\abundchi$, Eq.~\protect\eqref{eq:cdmrange}, has been included in
computing the relative probability density. \label{fig:5b} }

\EPSFIGURE{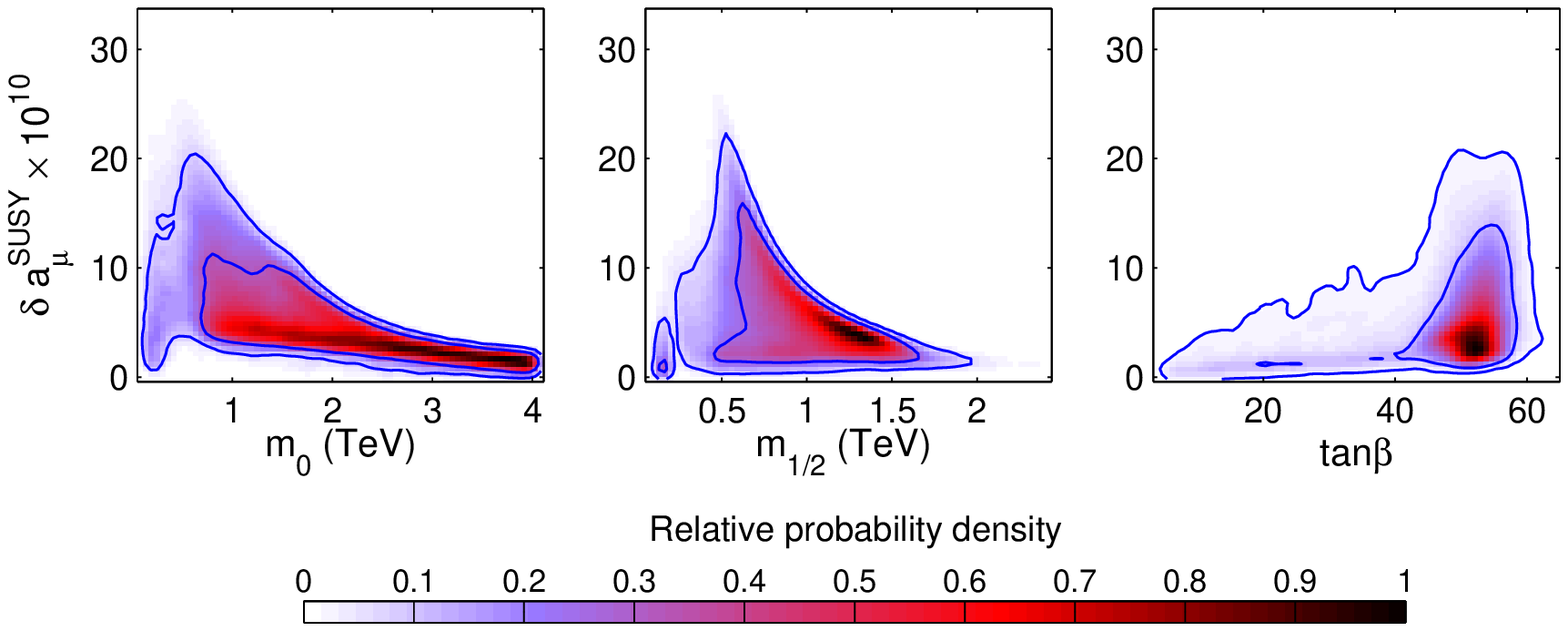,width=6in} {As in
Fig.~\protect\ref{fig:5b}, but for $\dasusy$. The measured value
of $\dasusy$, Eq.~\eqref{dasusy:eq}, has been included. \label{fig:5c} }

\EPSFIGURE{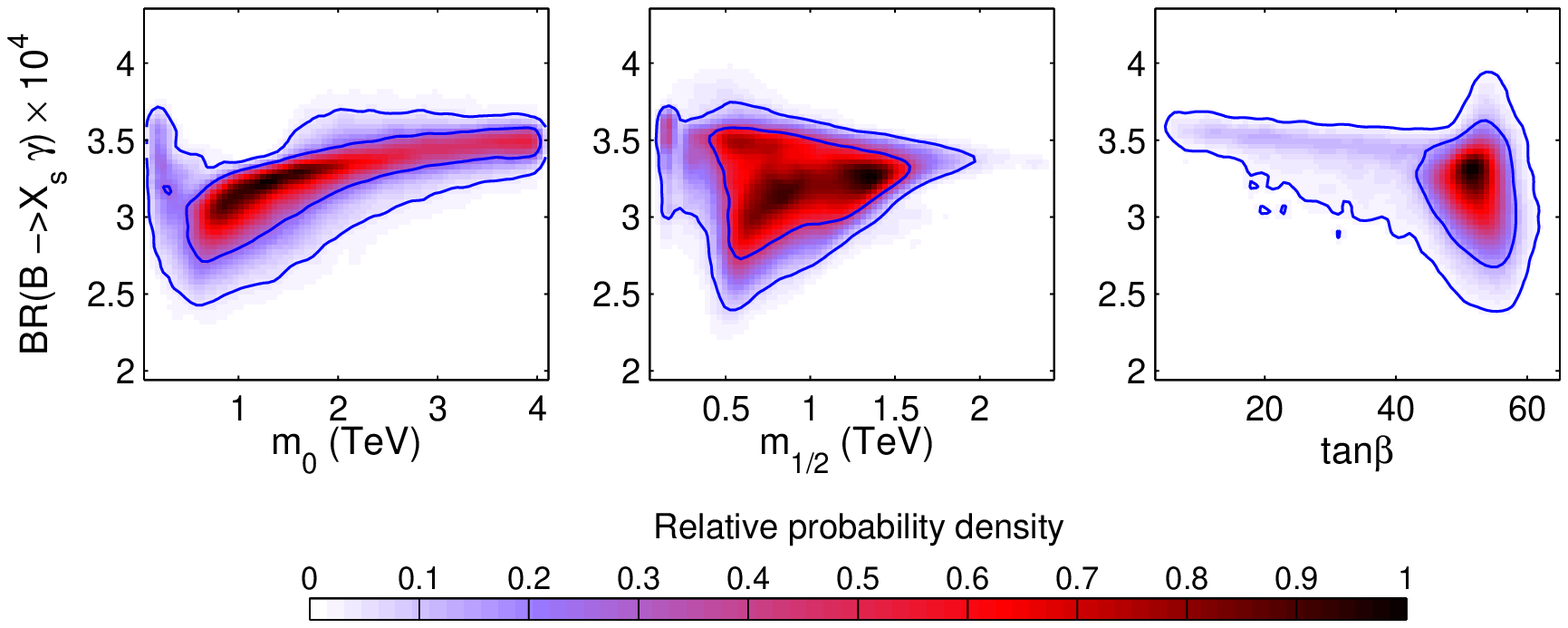,width=6in} {As in
Fig.~\protect\ref{fig:5b}, but for $\BRg$. The measured value
of $\BRg$, Eq.~\eqref{eq:bsgexp}, has been included. \label{fig:5d} }

Our MCMC approach allows us to investigate the joint posterior
probability distribution between CMSSM parameters and the various
observables, as explained in section~\ref{sec:bayesian}. In
Figs.~\ref{fig:5b} --~\ref{fig:5e} we plot the joint pdf for
$\abundchi$, $\dasusy$, $\BRg$, $BR(B_s \rightarrow \mu^+ \mu^-)$,
respectively, and $\mzero$, $\mhalf$ and $\tanb$ (all other parameters
in each panel are marginalized over). All plots correspond to the ``4
\tev\ range''. (See also column one in
Table~\ref{table:ResultsObservables}.)  We stress that the joint pdf
is obtained by taking into account observational constraints from all
the derived variables, including the one plotted along the vertical
axis.

In Fig.~\ref{fig:5b}, the pdf peaks around $\mzero \approx 1 \tev$ and
$\abundchi \approx 0.12$, with all values of $\mzero$ up to $4\tev$
compatible with the observed cosmological DM abundance. This is
another demonstration that the narrow ``WMAP strips'' -- which appear
when including only two parameters of the CMSSM at a time (see,
\eg,~\cite{rrn01,ehow04}) -- actually widen considerably to cover a
large region of parameter space when all the variables are taken
simultaneously into account. Not surprisingly therefore, we do not
find a strong correlation between $\abundchi$ and the CMSSM
parameters.

In the planes spanned by $\dasusy$ and the CMSSM parameters
(Fig.~\ref{fig:5c}), we observe a strong anti--correlation between
$\dasusy$ and both $\mzero$ and $\mhalf$. This is a simply a
reflection of the decreasing CMSSM contributions to $\dasusy$ with
increasing superpartner masses. In general, the posterior distribution
for $\dasusy$ is quite skewed with respect to the likelihood, which
represents the experimental measurement. The posterior pdf tends to
prefer values of $\dasusy$ close to zero. We comment further on this
below.

For $BR(\bar{B} \rightarrow X_s \gamma)$, Fig.~\ref{fig:5d} shows
a positive correlation with the masses, with the SM value
being reached in the asymptotic regime of large $\mhalf$ and $\mzero$.
We also see that the peak in the pdf is centered slightly
below the experimental central value given in
Eq.~\eqref{eq:bsgexp}.

\EPSFIGURE[t!]{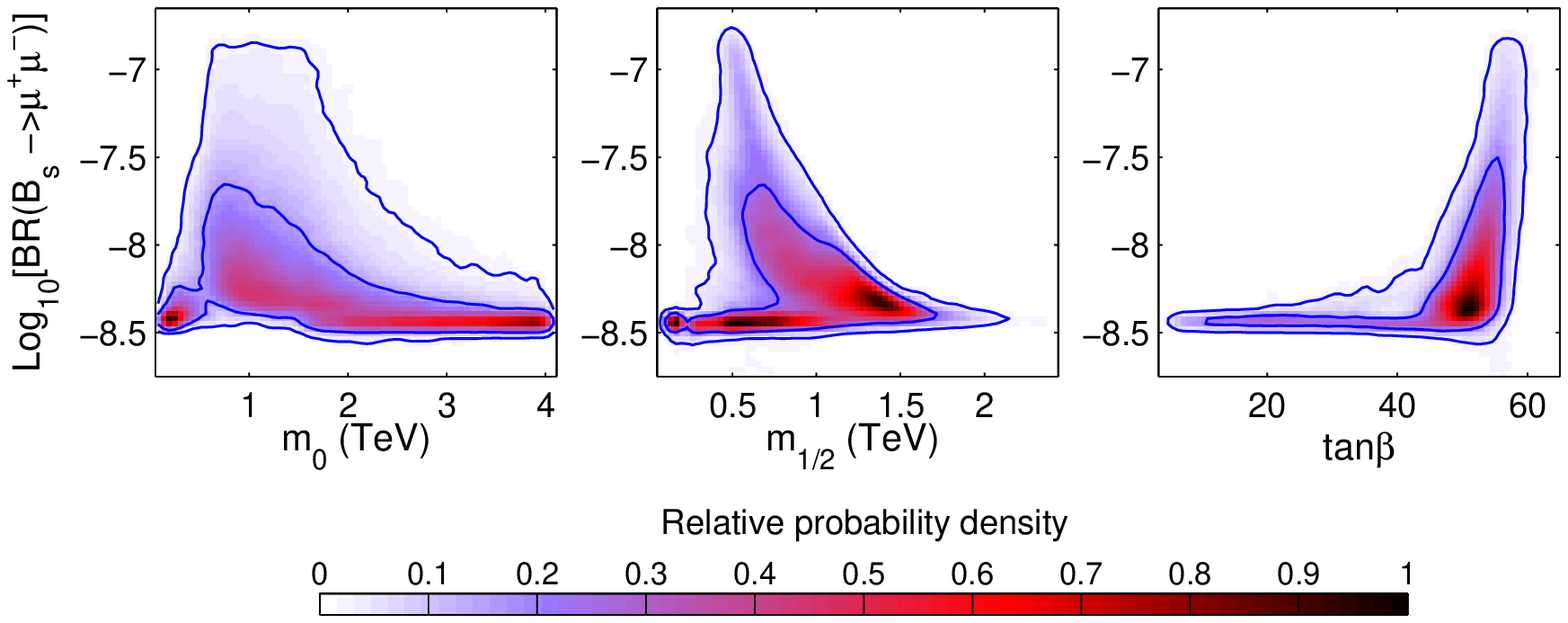,width=6in} {As in
Fig.~\protect\ref{fig:5b}, but for $BR(B_s \rightarrow \mu^+
\mu^-)$. The upper limit of $BR(B_s \rightarrow \mu^+ \mu^-)< 1.5
\times 10^{-7}$ has been included. \label{fig:5e} }

In Fig.~\ref{fig:5e} we show the 2--dimensional pdf for $BR(B_s
\rightarrow \mu^+ \mu^-)$ and $\mhalf$, $\mzero$ and $\tanb$. As
expected, the SUSY contribution decreases with increasing
superpartner masses and rapidly increases proportionally to
$\tan^6\beta$. Most of the high relative probability density lies close to the SM
prediction given in of
Eq.~\eqref{eq:bsm-SM}~\cite{buras03-bsmmsm}. The 1--dimensional (2
tails) regions encompassing 68\% and 95\% of probability are
\be
\begin{aligned}
 3.5 \times 10^{-9} < BR(B_s \rightarrow \mu^+ \mu^-) < 1.68
 \times 10^{-8} \quad & (68\% \text{ region}), \\
 3.3 \times 10^{-9} < BR(B_s \rightarrow \mu^+ \mu^-) < 7.50
\times 10^{-8} \quad & (95\% \text{ region}).
\end{aligned}
\ee
The current CDF and D\O\ limits are thus only approaching the 95\%
probability region. The 68\% (95\%) region extends just below (well
above) the Tevatron reach of about $2\times10^{-8}$.  Note also that a
positive measurement of the $BR$ at the Tevatron would imply
$\mzero\lsim 1.5\tev$ and would thus strongly disfavor the FP region.

\EPSFIGURE[t]{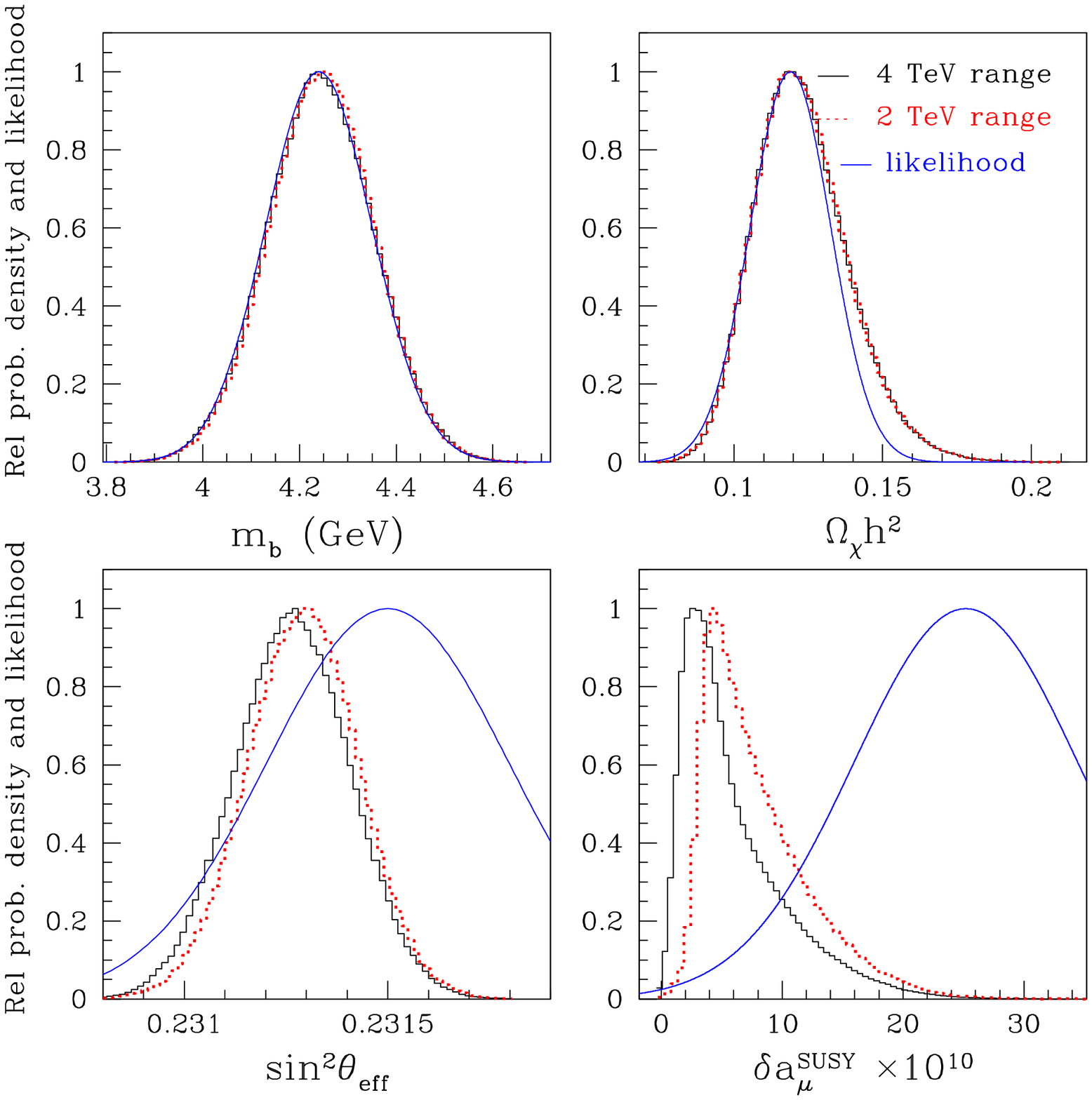,width=6in} {An illustration of the
tensions between different observables. While for
$m_b(m_b)^{\overline{MS}}$ and $\abundchi$ the likelihood and the
posterior pdf agree very well (top panels), for $\sineff{}$ and
$\dasusy$ (bottom panel) they exhibit a substantial discrepancy. The
latter case is a clear sign that the other constraints are strongly pulling
away the posterior pdf from the value preferred by the anomalous
magnetic moment measurement. \label{fig:datatension} }
When combining many different constraints, it is possible that some
combinations of them might be in conflict with each other. This has
been mentioned above for $\dasusy$, where we remarked that the
posterior pdf tends to prefer a value close to zero. This is further
highlighted in Fig.~\ref{fig:datatension} for a few representative
cases. For each observable, we plot the 1--dimensional marginalized
posterior pdf for the two prior ranges, along with the Gaussian
likelihood function used in the analysis. If all the observations
agreed with each other and in the absence of strong correlation among
variables, we would expect the posterior pdf and the likelihood to
overlap. This is the case for example for the bottom mass
$m_b(m_b)^{\overline{MS}}$ and for $\abundchi$ (upper panels). In the
latter case, the slightly skewed shape of the posterior is due to our
treatment of the theoretical uncertainty, which is larger for larger
values of $\abundchi$. However, the posterior pdf for $\sineff{}$ and
$\dasusy$ show a tension with the likelihood representing the
experimental result. We see a ``pull'' in the posterior towards values
of $\sineff{}$ lower than the measured mean (about one standard
deviation discrepancy). We notice a similar situation for $M_W$. This
tension is even more pronounced for the SUSY contribution to the
anomalous magnetic moment of the muon $\dasusy$, whose posterior pdf
peaks at values close to zero, in contrast to the experimental
measurement. This is a sign of a tension between the various
constraints used, with the other measurements pulling the posterior
pdf for $\dasusy$ towards the SM value. This motivates us to
investigate the dependence of our results on the inclusion of the
$\dasusy$ measurement, which will be presented in
section~\ref{sec:nogm2}.

\subsection{Mean quality of fit} \label{sec:fitquality}

In Bayesian statistics, the posterior pdf represents our state of
knowledge about the parameters in the problem after we have seen
the data and given our choice of priors, as we have explained in
some detail in section~\ref{sec:bayesian}. It is important to
remark that regions of high posterior probability do {\em not}
necessarily correspond to the best fitting points. The quality of
fit is defined in terms of an effective $\chi^2$, obtained from
the likelihood as
 \be \label{eq:effchisquare}
 \chi^2(\params) = -2 \text{ ln} \, p\left(\data | \derived(\params)
 \right).
 \ee
We can easily imagine a situation where a tiny multi--dimensional
region in parameter space -- let us call it a ``spike'' -- shows an
excellent quality of fit.  At the same time, there might be another
region where the quality of fit is slightly worse, but whose volume in
parameter space is much larger.  The much larger volume of the latter
region gives it a higher weight, since it is more generic and hence
less fine--tuned. In such a case, the posterior pdf would show a
strong peak in the large region, while for the spike it would be
suppressed due to its smallness. At the same time, the quality of fit
statistics would show a stronger peak in the spike.  We emphasize that
this is {\em not} a feature of the MCMC exploration of parameter
space, but rather a characteristics built into the Bayesian
formalism. As a consequence, our inferences automatically give more
weight to regions of parameter space showing less fine--tuning.

In the above scenario, an analysis performed using a fixed--grid
scan and a quality of fit statistics would reach potentially very
different conclusions.\footnote{A direct comparison between a
Bayesian and a fix--grid (frequentist) analysis would be difficult, since the
latter cannot easily handle nuisance parameters.} It can be
shown, however, that both methods lead to the same conclusions if
the data is informative enough, \ie, if their constraining power
is sufficient. Conversely, a discrepancy between the posterior pdf
and the quality of fit statistics is a useful indicator that the
above mechanism is at work.

In order to carry out such a comparison, we compute the {\em mean
quality of fit} in two dimensions. This is obtained from the posterior
pdf by adopting the effective $\chi^2$ defined in
Eq.~\eqref{eq:effchisquare} as a function of the parameters
$f(\params)$, as explained below Eq.~\eqref{eq:funcparams}, and
plotting it in the dimensions of interest in parameter space. In
Fig.~\ref{fig:fitquality} we plot the mean quality of fit in the same
six panels for which we presented the posterior pdf in
Fig.~\ref{fig:contours4tev}. The blue (solid) contours are the same as
in Fig.~\ref{fig:contours4tev} and are displayed to facilitate the
comparison between the two quantities.

\EPSFIGURE[t!]{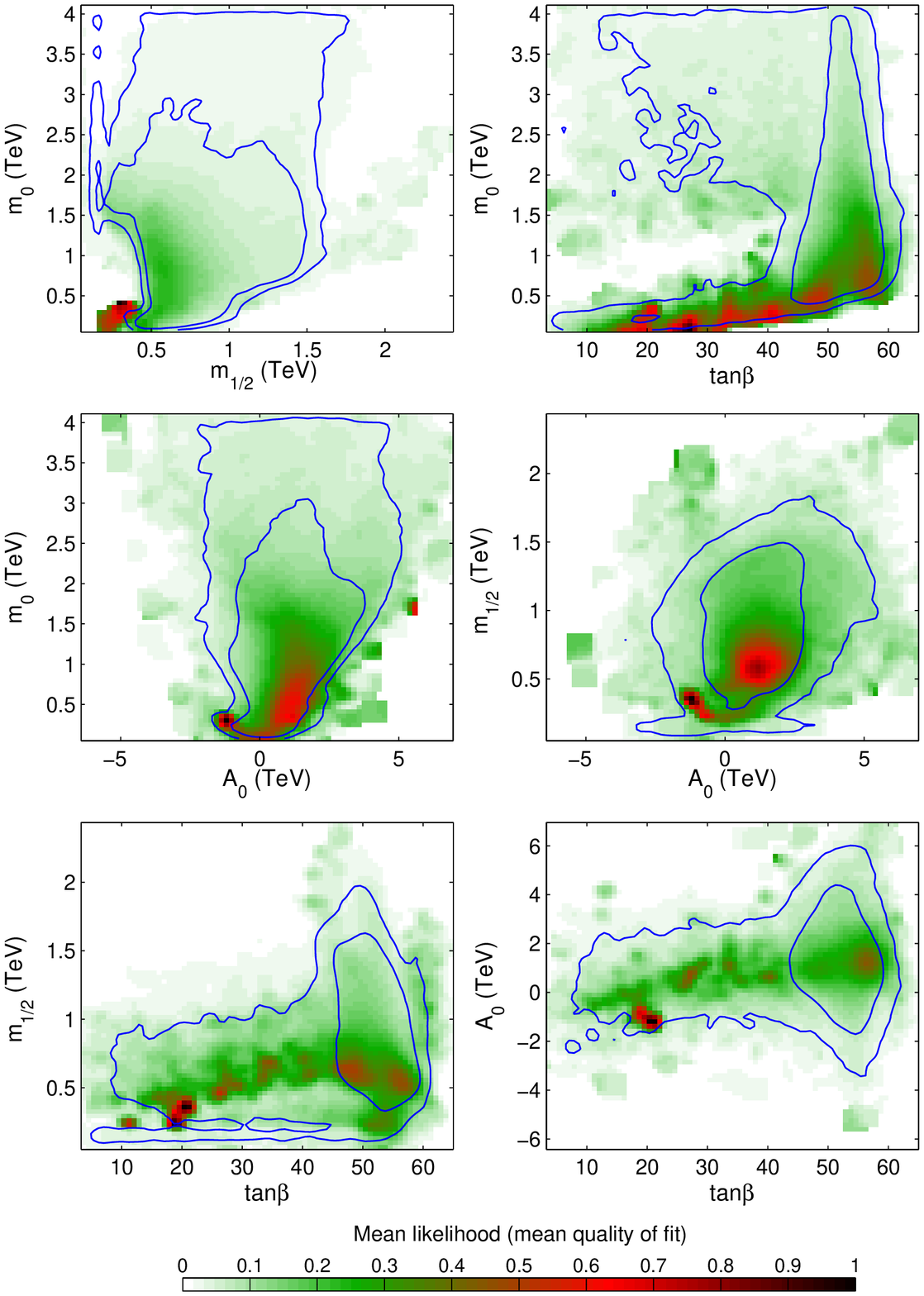,width=5in} {The quality of fit in the
planes spanned by the CMSSM parameters: $\mhalf$, $\mzero$,
$\azero$ and $\tanb$ for the ``4\tev\ range'' scan. This figure
should be compared with Fig.~\protect\ref{fig:contours4tev} where
the relative 2--dim joint relative probabilities are plotted for the same
pairs of variables. \label{fig:fitquality} }

%
In some panels, the best fitting points (represented by dark red
regions) are much more localized then the high--probability pdf.
For example in the $(\mhalf,\mzero)$ plane we find the best
quality of fit in the region $\mzero, \mhalf \lsim 0.5\tev$. In the
$(\tanb, \mzero)$ plane, on the other hand, we 
observe that good fitting point exist for almost all values of
$\tanb$, down to $\tanb \approx 15$. Comparing with the
corresponding panel in Fig.~\ref{fig:contours4tev}, we conclude
that the strong preference for a large $\tanb$ shown by the
posterior pdf does not imply that all the best fitting points lie
in that region of parameter space. This issue can only be resolved
once  better data becomes available.

%
\EPSFIGURE[t!]{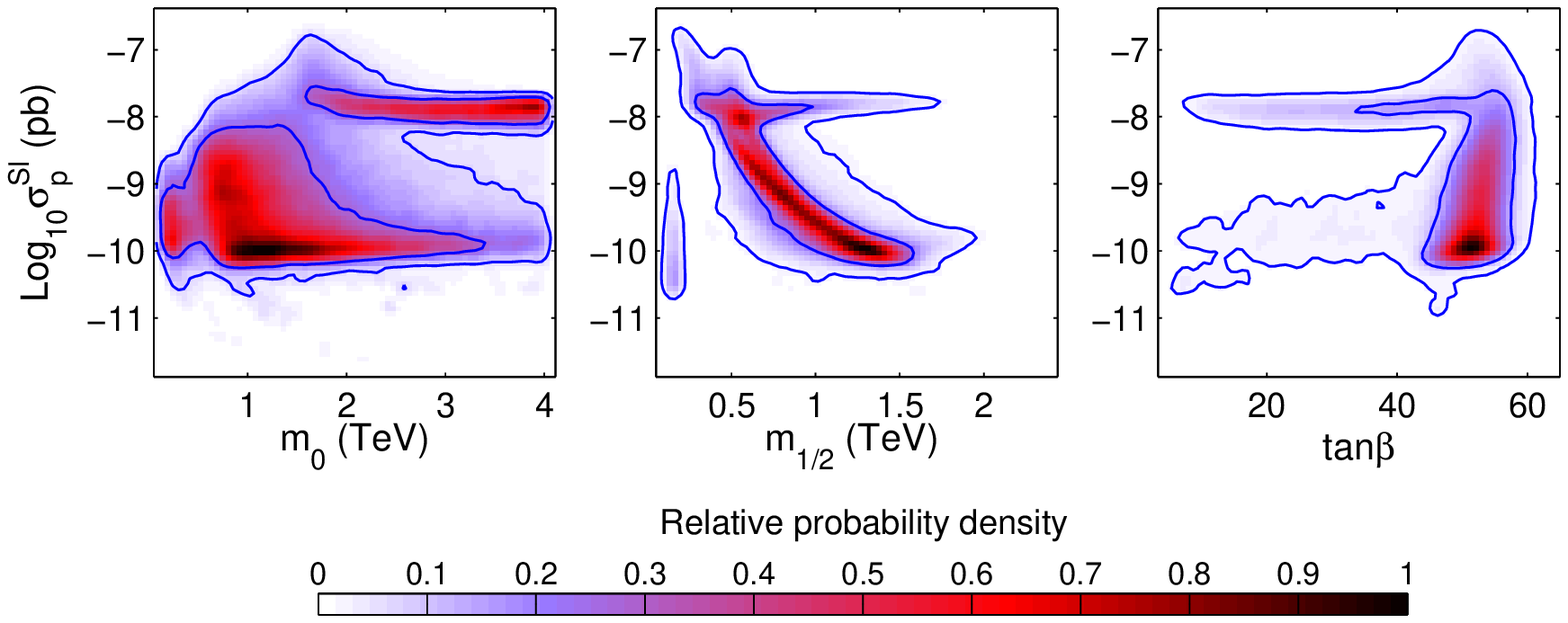,width=6in} {As in
 Fig.~\protect\ref{fig:5b}, but for $\sigsip$.
\label{fig:5g} }

\begin{figure}
\begin{center}
\begin{minipage}{3.5in}
\epsfig{file=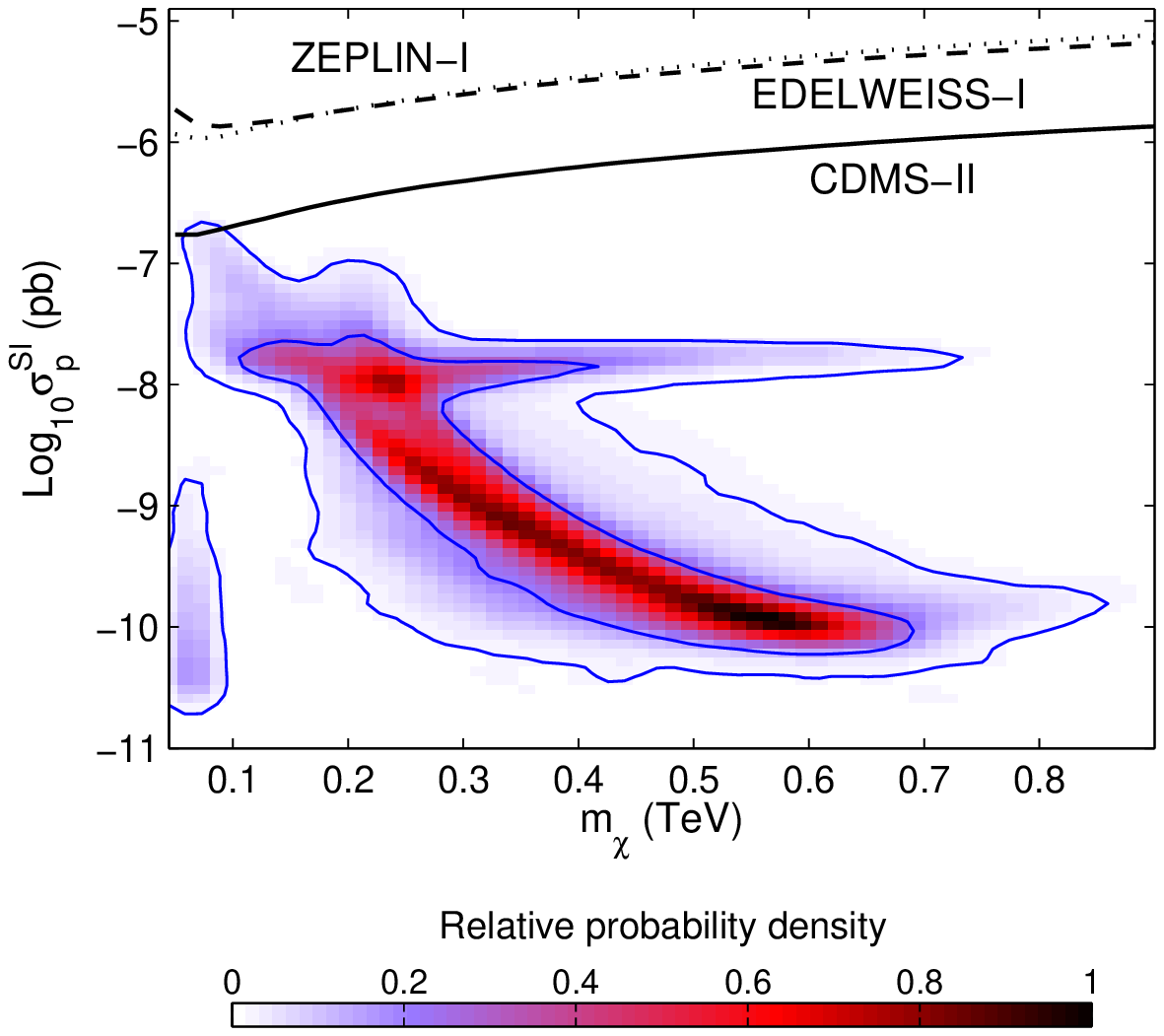,width=3.5in} \vspace*{0.2in}
\end{minipage}
\end{center}
\begin{center}
\begin{minipage}{3.5in}
\epsfig{file=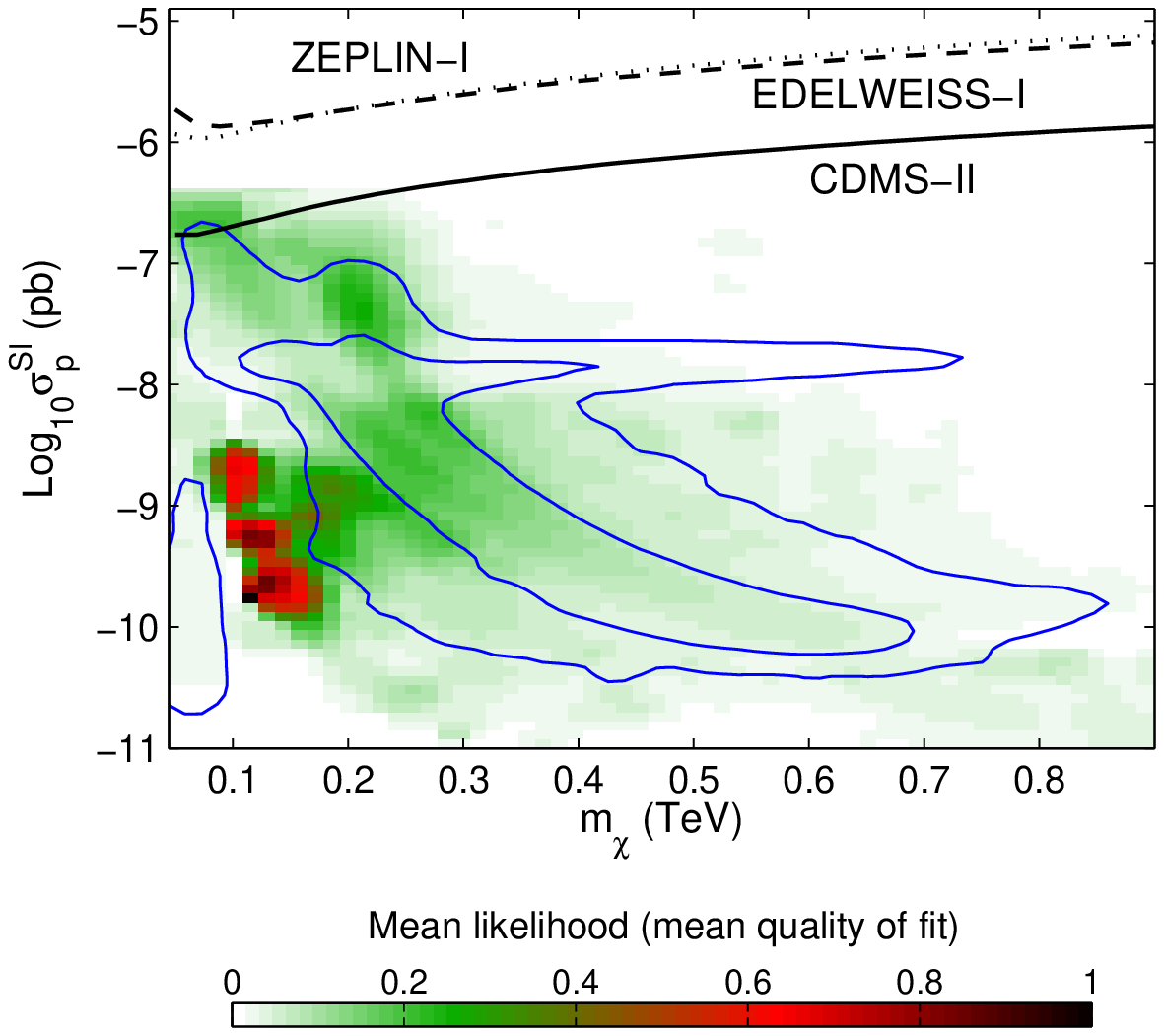,width=3.5in}
\end{minipage}
\caption{\label{fig:dd} Top panel: the 2--dimensional relative probability
  density $\relprobtwo{\mchi}{\sigsip}$, with the contours containing
  68\% and 95\% probability also marked. Bottom panel: the mean
  quality of fit (likelihood) with the same 68\% and 95\% probability
  contours as in the top panel to facilitate the comparison.  Current
  90\% experimental upper limits are also shown.  }
\end{center}
\end{figure}

%
\subsection{Direct detection of DM} \label{sec:dd}
Predictions for $\sigsip$ are usually determined
as a function of the CMSSM parameters by rigidly enforcing
relevant constraints, \eg, 1 or $2\sigma$ ranges of $\abundchi$
or $\BRg$, \etc. In our analysis, we present a posterior pdf which
simultaneously accounts for all the constraints and sources of
uncertainties. 

In Fig.~\ref{fig:5g} we plot the posterior pdf for the
spin--independent elastic cross section $\sigsip$ and the CMSSM
parameters $\mzero$, $\mhalf$ and $\tanb$. In the left panel one can
see three well--separated high probability regions. One is centered at
$0.8\tev\lsim \mzero \lsim 3\tev$ and $\sigsip\simeq10^{-10}\pb$,
although values almost as large as $10^{-8}\pb$ are also allowed. It
comes from the bulk, the stau coannihilation and the pseudoscalar
resonance regions. To the left, and almost adjacent to it, there is a
fairly narrow vertical band resulting from a light Higgs resonance
mentioned above.  The last main region is at large $\mzero\gsim 2\tev$
and almost constant $\sigsip\simeq1.6\times10^{-8}\pb$ which results
from the FP region. Comparing with the right panel in
Fig.~\ref{fig:5g} we can see that $\sigsip$ from the FP region and the
light Higgs resonance are fairly independent of $\tanb$ while the
bulk, coannihilation and resonance region requires large
$\tanb$. Finally, the middle panel is added for completeness and it
closely resembles the top panel of Fig.~\ref{fig:dd} (because
$\mchi\simeq0.4\mhalf$) where we present the 2--dimensional pdf for
$\sigsip$ and $\mchi$. For comparison, we also show current CDMS--II,
Edelweiss--I and UKDMC ZEPLIN--I 90\%~\cl\ limits, but we stress that
this constraint has not been used in the likelihood.

The features discussed above are visible in Fig.~\ref{fig:dd}
(top). Firstly, the biggest, banana--shaped region of high probability
($68\%$ regions delimited by the internal solid, blue curve) shows a
well--defined anti--correlation between $\sigsip$ and $\mchi$. It comes
from the bulk and stau coannihilation region (larger $\sigsip$) and from the
$\ha$--resonance (smaller $\sigsip$) at large $\tanb$. This region
covers roughly the range $10^{-10} \lsim \sigsip\lsim10^{-8}\pb$ and
$0.2\tev\lsim\mchi\lsim0.7\tev$. In both cases the dominant contribution
to $\sigsip$ typically comes from a heavy Higgs exchange.

Secondly, at small $\mchi\lsim0.1\tev$ we can again see a small vertical
band of fairly low relative probability density ($\lsim0.2$) at small
$\sigsip$. This region is allowed by the light Higgs resonance
contribution to reducing $\abundchi$ at small $\mhalf$. This
region essentially never features in usual fixed--grid scans,
which do not smear out experimental bounds. This region would also
disappear with a fair improvement in the lower bound on $\mhl$.

Thirdly, we can see a well pronounced region of high relative probability at
fairly constant $\sigsip\simeq 1.6\times10^{-8}\pb$ for
$\mchi\lsim0.42\tev$ which at low $\mchi$ partly overlaps with the
previous region. At $95\%$ this region extends up to
$\mchi\lsim 0.72\tev$ for fairly constant $\sigsip$. This ``high''
$\sigsip$ band is a result of the FP region, basically independently
of $\tanb$, as discussed above. This result has to be interpreted
carefully, since there are large uncertainties associated with FP
region, in particular with its location in the $(\mhalf,\mzero)$ plane
mentioned earlier. Hopefully, associated uncertainties in $\sigsip$
are going to be much smaller since it depends on low energy quantities
like Higgs masses and the $\mu$ parameter. Despite those outstanding
questions, we believe that it is probably safe to expect that the FP will be
the first to be probed by DM search experiments.

Finally, after marginalizing over all other parameters, we obtain
the following 1--dimensional regions encompassing 68\% and 95\% of
the total probability:
\be
 \begin{aligned}
  1.0\times10^{-10}\pb   < \sigsip < 1.0\times10^{-8}\pb \quad & (68\% \text{ region}), \\
  0.5\times10^{-10}\pb < \sigsip < 3.2\times10^{-8}\pb \quad & (95\% \text{ region}).
 \end{aligned}
\ee

Currently running experiments (most notably CDMS--II but also
Edelweiss--II and ZEPLIN--II) should be able to reach down to a
few $\times 10^{-8}\pb$, on the edge of exploring the FP region.
A future generation of ``one--tonne'' detectors are expected to reach
down to $\sigsip\gsim 10^{-10}\pb$, thus exploring almost the
whole $68\%$ region and much of the $95\%$ interval as well.

As discussed in section~\ref{sec:fitquality}, the posterior pdf
may be rather different from the mean quality of fit.  In the
bottom panel of Fig.~\ref{fig:dd} we plot the quality of fit
(defined in Eq.~\eqref{eq:effchisquare}) for $\sigsip$ and $\mchi$.
The best fit points are found in the region
$0.1\tev\lsim\mchi\lsim 0.2\tev$ and $1\times 10^{-10}\pb
\lsim\sigsip \lsim 3\times10^{-9}\pb$, but other good
fitting points (quality of fit about 0.4, dark green regions) lie
right near the top of the high probability region, at rather large
$\sigsip$.

%
\subsection{Correlations among observables} \label{sec:correlationsobs}
%
\EPSFIGURE[t]{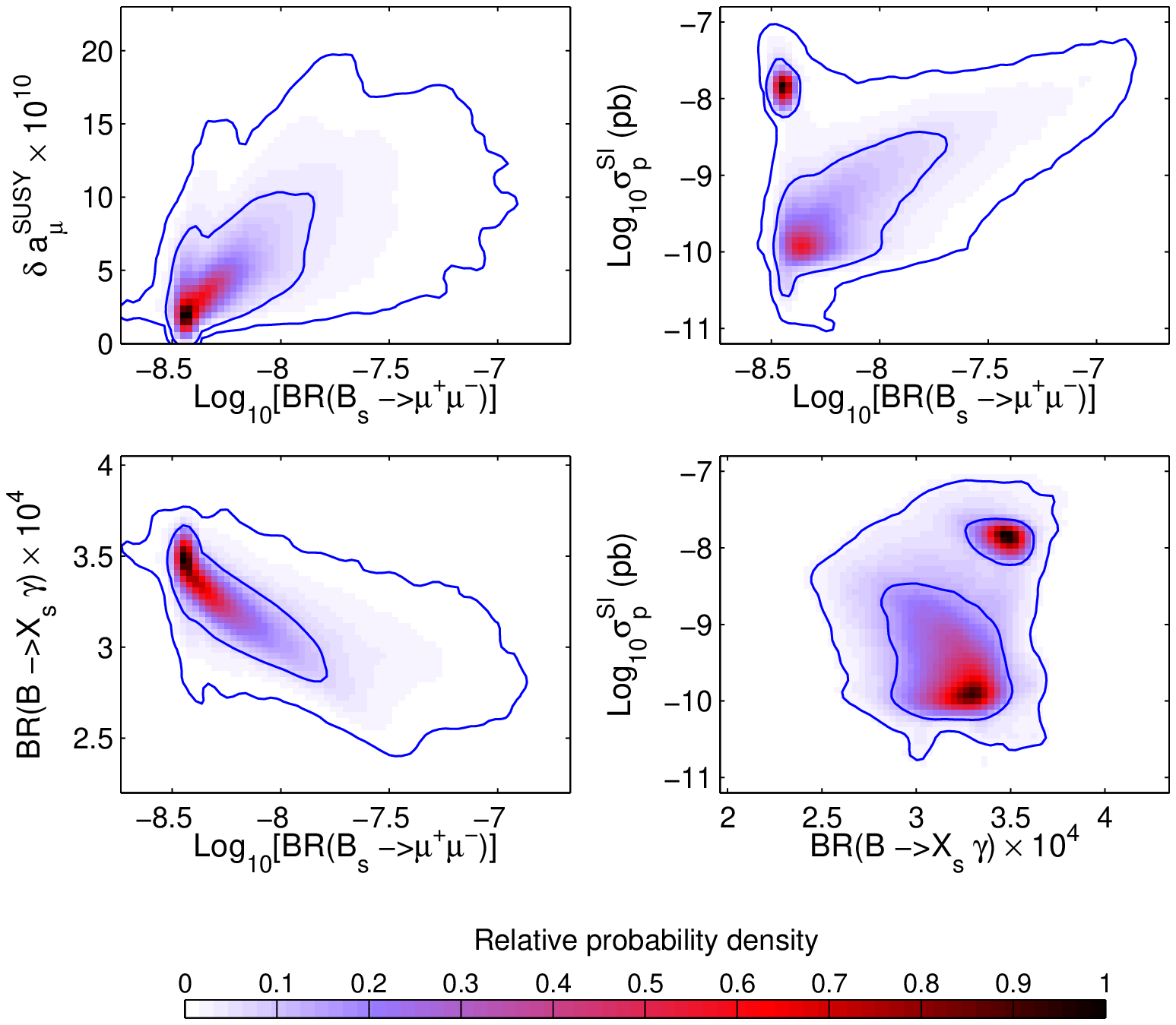,width=5in} {The 2--dimensional relative probability
density $\relprobtwo{\xi_i}{\xi_j}$ for the pairs of observables
marked along horizontal and vertical
axes.\label{fig:correlations1} }

We now proceed to examine various correlations among the
observables discussed above. In the upper left panel of
Fig.~\ref{fig:correlations1} we show the 2--dimensional pdf for
$BR(B_s \rightarrow \mu^+ \mu^-)$ and $\dasusy$.
We can clearly see a rather strong correlation between those two
variables~\cite{ddn01} but, as pointed out above, the most
probable ranges of both variables are on a low side. A positive
measurement of $BR(B_s \rightarrow \mu^+ \mu^-)$ at the Tevatron
(above some $2\times10^{-8}$) would lead to a very strong
tension with the current result for $\dasusy$.

In the upper right panel of Fig.~\ref{fig:correlations1} we show
the 2--dimensional pdf for $BR(B_s \rightarrow \mu^+ \mu^-)$ and
$\sigsip$.
We can see two high relative probability regions here showing an
interesting pattern. The one at smaller 
values of both variables comes from the coannihilation and Higgs
resonance regions and has been pointed out in~\cite{bkk05}.
In addition, the FP region allows for a second
``island'' where $\sigsip$ is not far below the current CDMS limit
while $BR(B_s \rightarrow \mu^+ \mu^-)$ is very small, far below the reach
of the Tevatron. Thus, because of the FP region, a signal in
measuring $\sigsip$ in current generation of DM search detectors
would not necessarily imply a high probability of measuring $BR(B_s
\rightarrow \mu^+ \mu^-)$ at the Tevatron, (or vice versa, contrary to the claim
of~\cite{bkk05}).

In the two bottom panels of Fig.~\ref{fig:correlations1} we show
high probability regions for $BR(B_s \rightarrow \mu^+ \mu^-)$ and
$\BRg$, and for $\BRg$ and $\sigsip$.
In the first case, the high relative probability region for $BR(B_s
\rightarrow \mu^+ \mu^-)$ lies at small values, just above the SM
prediction and corresponds to the $\BRg$ around the central value,
as noted above. It may eventually be possible to verify this
correlation experimentally. Finally the variables $\BRg$ and
$\sigsip$ show two separate peaks in the relative probability density.
Again, the peak at smaller $\sigsip$ comes from the coannihilation
and Higgs resonance regions while the other one from the FP
region. In principle, a measurement of $\sigsip$ above some
$10^{-8}\pb$ would point towards a range of $\BRg$ above the current
central value. Unfortunately, until a full NLO SUSY contribution
is available, error bars in the latter quantity will remain at the
level of some 10\%, which would make it difficult to confirm the
FP origin of the $\sigsip$ measurement.

\EPSFIGURE[t]{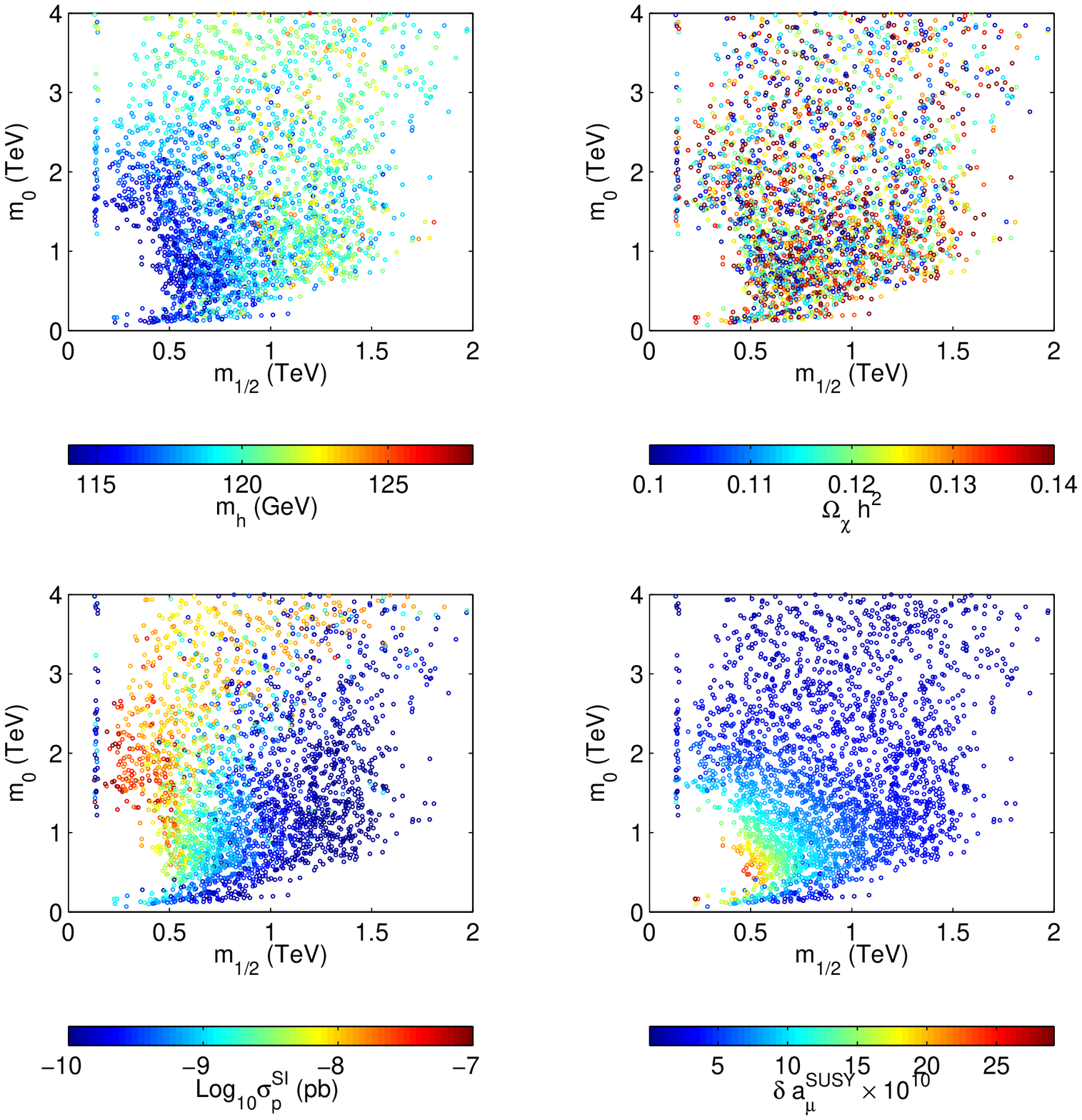,width=5in} {Values of $\mhl$ (upper
left), $\abundchi$ (upper right), $\sigsip$ (lower left) and
$\dasusy$ (lower right) in the $(\mhalf,\mzero)$ plane for
samples uniformly selected from our MC chains. While $\sigsip$ and
$\dasusy$ show a clear correlation with the values of $(\mhalf,
\mzero)$, the CDM abundance $\abundchi$ can take any value within
the $2\sigma$ WMAP range in the plane. Narrow ``WMAP strips'' in
the $(\mhalf,\mzero)$ plane for fixed values of $\tanb$ and
$\azero$ thus disappear when these parameters are allowed to vary
and all other parameters are correctly marginalized over.
\label{fig:3d} }

In the four panels of Fig.~\ref{fig:3d} we plot in the
$(\mhalf,\mzero)$ plane values of $\mhl$ (upper left), $\abundchi$
(upper right), $\dasusy$ (lower left) and $\sigsip$ (lower right).
The points have been drawn uniformly from our MC chains. To
highlight the values of interest for the observables, the range of
the color scales has been reduced, and points with values above
(below) the scale have been plotted in red (blue). One can see
that $\mhl$ increases with increasing $\mhalf$ or $\mzero$, as
expected. In the region where $\mhalf\lsim1\tev$ and/or
$\mzero\lsim2\tev$ (roughly the reach of the LHC) one
predominantly finds $\mhl\lsim117\gev$ (although larger values are
not excluded), which is encouraging for Higgs
searches~\cite{chww05}. In all the panels, at $\mhalf\simeq
0.2\tev$, there is a vertical favored region due to a narrow light
Higgs resonance contribution to $\abundchi$. It is interesting
that in the rest of the $(\mhalf,\mzero)$ plane one finds that the
WIMP relic density (upper right window) can take any value within
about the $2\sigma$ range of \eqref{eq:abundchitot}. In other
words, even though for each particular choice of the CMSSM and
nuisance SM parameters there are only a few narrow regions
consistent with the DM constraint~\eqref{eq:cdmrange}, by
performing the appropriate marginalization over all other
parameters it appears to be fairly easy to satisfy the WMAP value
of the DM abundance at almost any point in the $(\mhalf,\mzero)$
plane. This feature can also be seen in Fig.~\ref{fig:5b}.

The values of $\sigsip$ (bottom left panel of
Fig.~\ref{fig:3d}) cover a large range but are
generally larger for smaller $\mhalf$ and intermediate and large
$\mzero$ (compare with Figs.~\ref{fig:5g} and~\ref{fig:dd} (top)), with the
exception of the light Higgs boson resonance region at small
$\mhalf$. This means that DM
direct detection searches will in general explore the large
$\mzero$ region of FP first -- an interesting complementarity with collider
searches.

On the other hand, the claimed discrepancy between experiment and
the SM value~\eqref{dasusy:eq} of the anomalous magnetic moment,
when taken at face value, clearly points towards a different
region of $\mhalf\lsim 0.8\tev$ (at small $\mzero$) and
$\mzero\lsim 1.5\tev$ (at small $\mhalf$), as can be seen in the
bottom right panel of Fig.~\ref{fig:3d}. Thus, similarly to the
case of $BR(B_s \rightarrow \mu^+ \mu^-)$ described above, a positive
measurement of $\sigsip$ in currently running DM detectors (above
some $2\times10^{-8}\pb$) would lead to a strong tension with the
current result for $\dasusy$.

\subsection[Sensitivity to $\gmtwo$]{Sensitivity to \boldmath $\gmtwo$ } \label{sec:nogm2}
\EPSFIGURE[t]{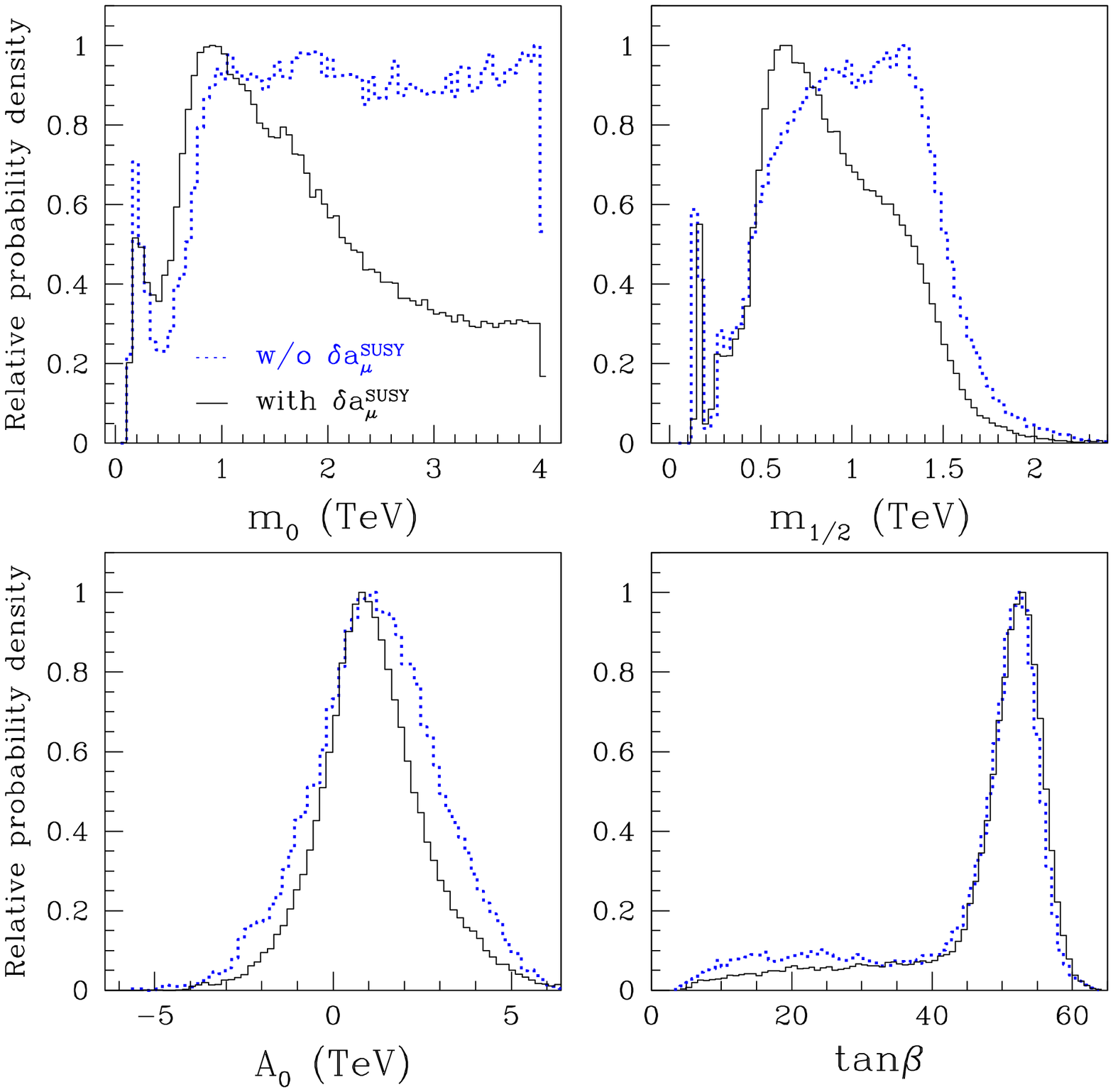,width=6in} {The 1--dimensional
relative posterior pdf's as in
Fig.~\protect\ref{fig:susyinputs2vs4tev}. The  black solid (blue
dotted) line is for the analysis including (excluding) the
measurement of the anomalous magnetic moment of the muon,
Eq.~\protect\eqref{dasusy:eq}. Both cases assume the ``4\tev\
range'' priors.  \label{fig:susyinput_ng2} }

\EPSFIGURE[t]{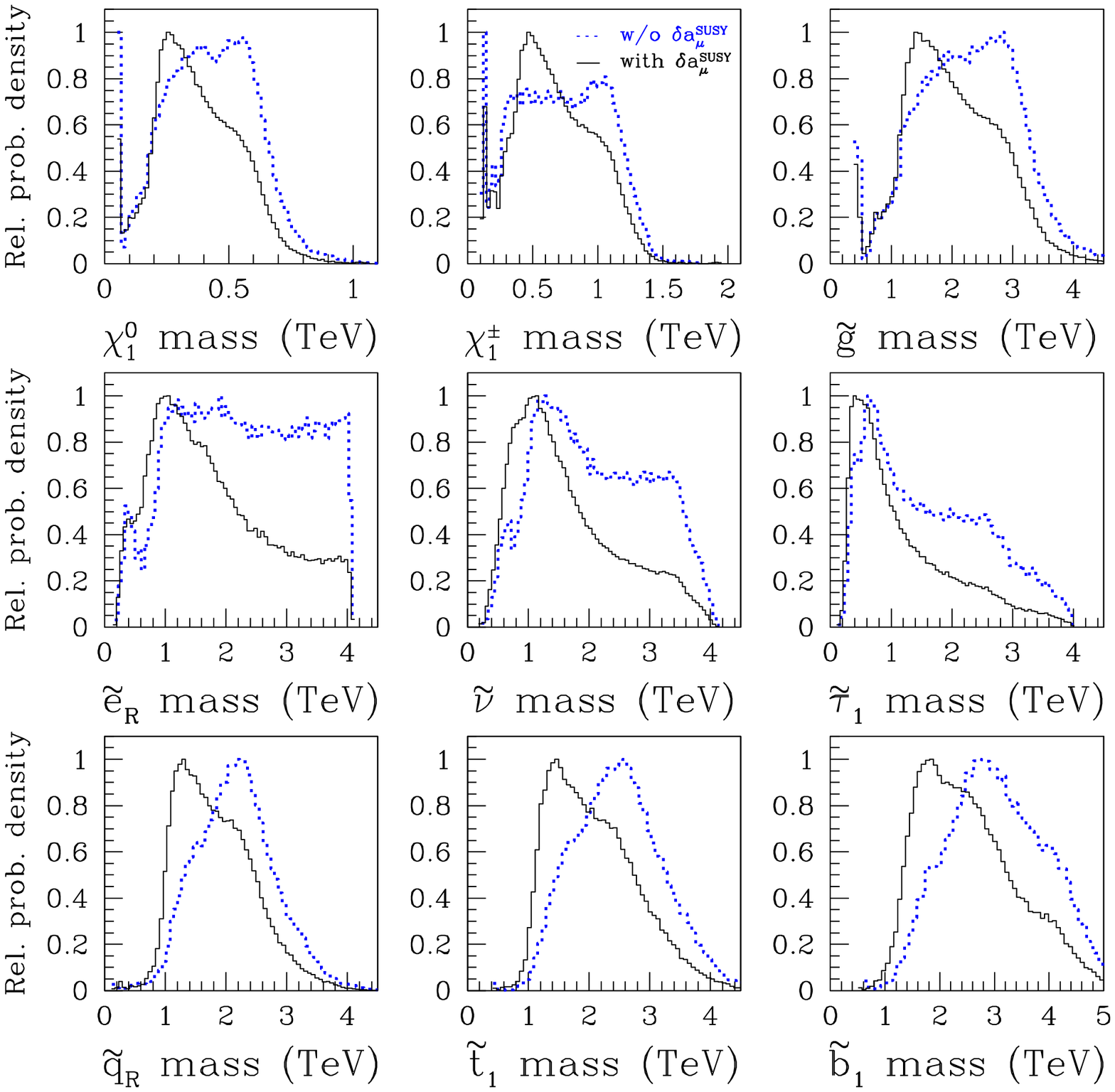,width=6in} {As in
Fig.~\protect\ref{fig:susymasses2vs4tev}. The  black solid (blue
dotted) line is for the analysis including (excluding) the
measurement of the anomalous magnetic moment of the muon,
Eq.~\protect\eqref{dasusy:eq}. Both cases assume the ``4\tev\
range'' priors. \label{fig:susymasses_ng2}}

\EPSFIGURE[t]{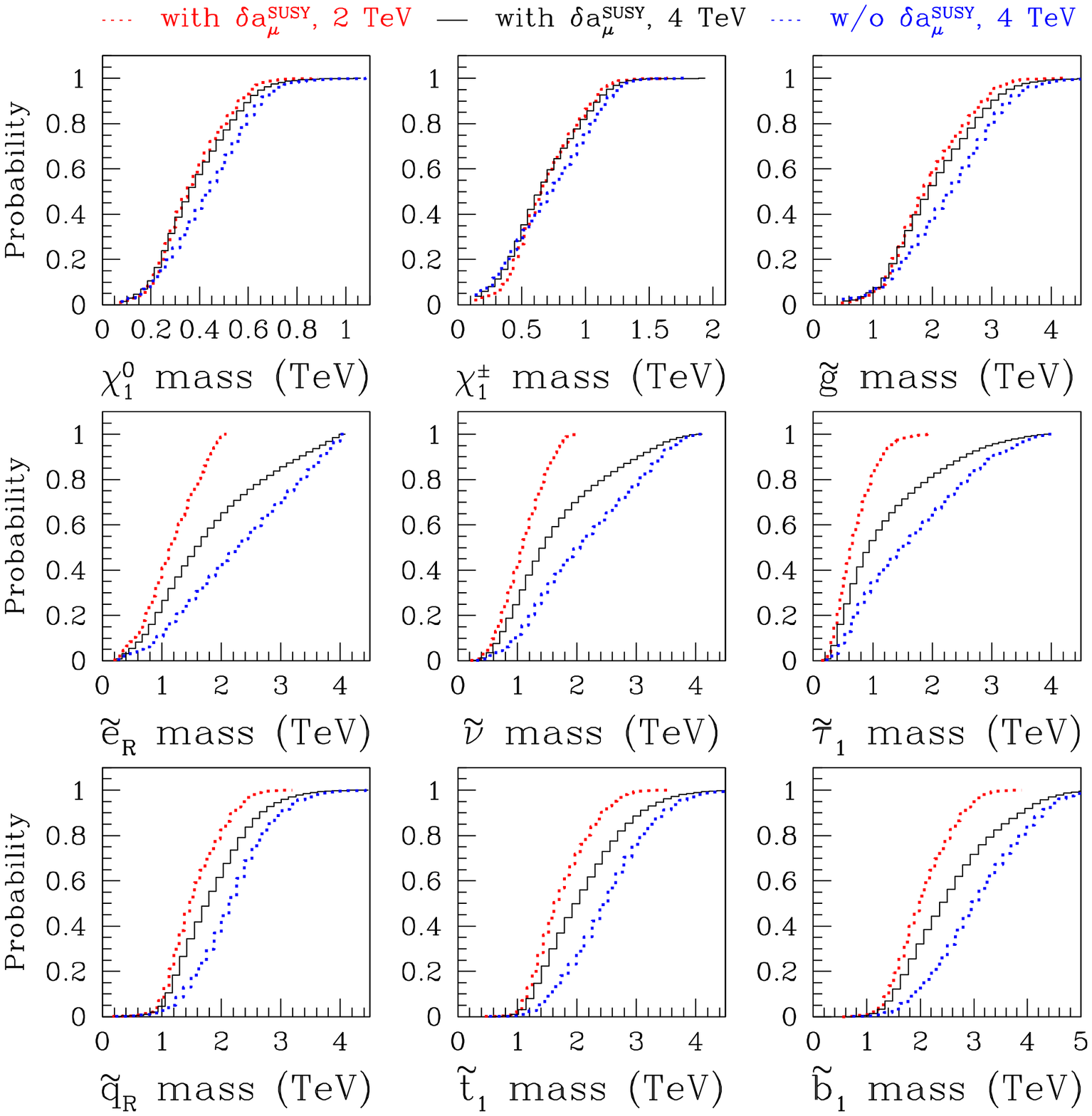,width=6in} {The probability for the
superpartner masses to lie below a given mass. All other
parameters have been marginalized over. \label{fig:intprob} }

As we have shown above, the $2.7\sigma$ deviation of the
anomalous magnetic moment of the muon from the SM prediction appears
to be in tension with some of the other observables. To investigate to
what extent our statistical inferences on superpartner masses rely on
the inclusion of $\dasusy$, in this section we present the posterior
pdf's obtained by dropping it from the likelihood. To be as general as
possible, we have adopted the ``4 \tev\ range'' of the priors.

Fig.~\ref{fig:susyinput_ng2} compares the 1--dimensional
marginalized posterior pdf $\relprobone{\theta_i}$, where
$\theta_i=\mzero$, $\mhalf$, $\azero$ and $\tanb$ with and without the
inclusion of $\dasusy$ in the likelihood. If we drop the anomalous
magnetic moment measurement we loose essentially all constraint on
$\mzero$, whose pdf becomes flat above $1\tev$. The impact on
$\mhalf$ is very mild, with a slight shift to larger values of the
bulk of the pdf. There is also almost no change in the pdf's for
$\azero$ and $\tanb$. Intervals encompassing 68\% and 95\% of
probability for the CMSSM parameters are given in
Table~\ref{table:CMSSMtableNogm2} (compare with
Table~\ref{table:CMSSMtable}).

The implications for several representative superpartner masses
are presented in Fig.~\ref{fig:susymasses_ng2}, which should be
compared with Fig.~\ref{fig:susymasses2vs4tev}. The corresponding
probability intervals are summarized in
Table~\ref{table:massesTableNogm2}. One should remember that the
dominant contribution to $\dasusy$ comes from the
sneutrino--chargino exchange~\cite{cn96}. Since $m_{\widetilde
\nu}$ depends much more strongly on $\mzero$ than on $\mhalf$,
while $m_{\charone}$ is more dependent on $\mhalf$, both soft
parameters are affected but $\mzero$ more strongly. 

An experimental implication for the LHC is rather obvious. If the
$\dasusy$ anomaly is not confirmed, the probability of finding
sleptons and squarks will be reduced. This can be see in
Fig.~\ref{fig:intprob} where we plot the total probability as a
function of mass for several superpartners.  We show three cases: the
``2\tev\ range'' (dotted red), as well as our default ``4\tev\ range''
case with (black solid) and without (dotted blue) the $\gmtwo$
constraint~\eqref{dasusy:eq}. It is clear that the total probability
that the mass of squarks or sleptons lies below a certain value
depends rather strongly on the choice priors and/or the $\gmtwo$
constraint. On the other hand, this is not the case for the gauginos.
By comparing Tables~\ref{table:massesTableNogm2}
and~\ref{table:massesTable} it would seem that the upper bound of the
95\% interval does not change much for the squark and gluino masses.
However, this is probably a manifestation of the upper cut induced
around 4 \tev\ by the prior, and therefore should not be taken as a
robust result of the inference, analogously to what has been discussed
above for $\mzero$.

\begin{table}
\centering
\begin{tabular}{|l |c c|}
 \hline
           & \multicolumn{2}{|c|}{``4\tev\ range'', no $\gmtwo$} \\
 Parameter & 68\% region             & 95\% region   \\
\hline
 $\mzero$~(TeV)  & $<2.87$       & $<3.85 $      \\
 $\mhalf$~(TeV)  & $(0.56, 1.38)$ & $(0.14, 1.73)$\\
 $\azero$~(TeV)  & $(-0.60, 2.89)$& $(-2.46, 4.59)$  \\
 $\tanb $        & $(28.2, 53.8)$     & $(10.2, 57.0)$     \\ \hline
 \end{tabular}
\caption{CMSSM parameter ranges corresponding to 68\% and 95\% of
posterior probability (with all other parameters marginalized over) for the
 ``4\tev\ range'' but without the constraint from $\gmtwo$. These ranges should be
compared with Table~\protect\ref{table:CMSSMtable}. }
\label{table:CMSSMtableNogm2}
\end{table}
\begin{table}
\centering
\begin{tabular}{|c |c c|}
 \hline
  Super--  & \multicolumn{2}{|c|}{``4\tev\ range'', no $\gmtwo$} \\
 partner  & 68\% & 95\%  \\
           \hline
 $\chi_1^0$     & $(0.23, 0.60)$  & $(0.57, 0.76)$ \\
 $\charone$     & $(0.33, 1.10)$  & $(0.11, 1.31)$ \\
 $\gluino$      & $(1.36, 3.05)$  & $(0.42, 3.79)$ \\
 $\tilde{e}_R$  & $(1.12, 3.48)$  & $(0.40, 3.95)$ \\
 $\tilde{\nu}$  & $(1.11, 3.18)$  & $(0.56, 3.75)$ \\
$\tilde{\tau}_1$& $(0.61, 2.71)$  & $(0.32, 3.58)$ \\
 $\tilde{q}_R$  & $(2.08, 4.00)$  & $(1.43, 4.75)$ \\
 $\tilde{t}_1$  & $(1.47, 2.76)$  & $(1.00, 3.48)$ \\
 $\tilde{b}_1$  & $(1.70, 3.19)$  & $(1.21, 3.95)$ \\
\hline
 \end{tabular}
\caption{Selected superpartner mass ranges (in TeV) containing
68\% and 95\% of posterior probability (with all other parameters
marginalized) for the ``4\tev\ range'' but without the constraint
from $\gmtwo$. These intervals should be compared with
Table~\protect\ref{table:massesTable}. }
\label{table:massesTableNogm2}
\end{table}
\begin{table}
\centering
\begin{tabular}{|l |c c | c c|}
 \hline
  Observable  & \multicolumn{2}{|c|}{``4\tev\ range'', with $\gmtwo$} & \multicolumn{2}{|c|}{``4\tev\ range'', w/o $\gmtwo$} \\
  & 68\% & 95\%  & 68\% & 95\% \\
           \hline
$\abundchi$  & $(0.107,0.138)$ & $(0.946,0.157)$ & $(0.107,0.138)$ & $(0.949,0.158)$ \\
$\dasusy \times 10^{10}$    & $(1.9,9.9)$     & $(0.8,17.1)$    & N/A             & N/A\\
$BR(B_s \rightarrow \mu^+ \mu^-) \times 10^9$ & $(3.5,16.8)$
&$(3.3,75.0)$                                                   &$(3.5, 8.5)$     & $(3.3,41.0)$\\
$\BRg \times 10^4$          & $(2.93,3.44)$   & $(2.62,3.61)$   &$(3.08,3.49)$    & $(2.77,3.62)$\\
$\sigsip (\pb \times 10^{10})$& $(1,100)$     & $(0.5,320)$     &$(0.8,117)$      & $(0.3,208)$\\
 \hline
 \end{tabular}
\caption{Intervals encompassing 68\% and 95\% of posterior
probability for selected observables (with all other parameters
marginalized). The two columns compare results for the ``4\tev\
range'' with (left column) and without (right column) inclusion of
the constraint from $\gmtwo$. } \label{table:ResultsObservables}
\end{table}

In Table~\ref{table:ResultsObservables} we summarize the effect of the
$\gmtwo$ constraint on some of our other observables. While the relic
abundance is unaffected, $BR(B_s \rightarrow \mu^+ \mu^-)$ and
$\sigsip$ get shifted to lower values as a consequence of a less
constrained $\mzero$ when the $\gmtwo$ anomaly is discarded.

\section{Summary and conclusions}\label{sec:summary}

In this work we performed a detailed investigation of the CMSSM
parameter space using a MCMC method and analyzed our results in terms
of Bayesian statistics. The power and flexibility of the approach
allowed us to probe many previously unexplored ranges of
parameters. Furthermore, we were able to improve on previous analyses
in several important aspects. We fully incorporated the effects of
remaining uncertainties in relevant SM parameters (including
$\alphaemmz$), which are usually fixed to their central values.  We
incorporated often neglected theoretical uncertainties in computing
mass spectra and observables. Finally, we improved upon the usual
practice of applying sharp experimental limits and (typically)
$1\sigma$ uncertainties by smearing them out.

By carefully constructing the likelihood function we constrained the
CMSSM parameters by comparing several collider and astrophysical
observables (except for $\sigsip$) with the available data. For all
these variables we computed the posterior probability density functions --
our main tool in analyzing our findings -- in terms of which we
delimited the favored regions of the CMSSM parameters, and further derived the
ranges of the observables themselves that are favored by a combination
of currently available data. We emphasized the difference between the
posterior probability density and the quality of fit statistics. The
latter is formulated in terms of an average effective $\chi^2$, and is
akin to what is used in fixed--grid scans. For example, we concluded
that the strong preference for a large $\tanb$ shown by the posterior
pdf does not imply that all the best fitting points lie in that region
of parameter space. This issue can only be resolved by acquiring
better data.

We explored in detail the robustness and sensitivity of our
results to the assumed initial ranges of CMSSM and SM parameters
(priors). To this end we compared our findings for the default
``4\tev\ range'' prior (which include the somewhat uncertain FP
region) extending above the LHC reach, with the more restrictive
``2\tev\ range'' prior. We emphasized that much care
must be exercised in interpreting our inferences whenever
boundaries of high probability regions lie close to the prior
ranges. This applies mainly to $\mzero$, and to the superpartner
masses that primarily depend on it, while all other variables
appear robust to a change in the range of the priors.

We furthermore examined various correlations among the relevant
observables. Some have been pointed out before, \eg, between
$BR(B_s \rightarrow \mu^+ \mu^-)$ and $\dasusy$ or $\sigsip$ (the
last one showing a new feature due to the presence of the FP
region). A more subtle correlation between $\BRg$ and $BR(B_s
\rightarrow \mu^+ \mu^-)$ emerged, which may eventually be tested
experimentally. We note that at present none of the observables
appear to be in conflict with observations or with each other,
with the possible exception of $\gmtwo$.  In particular, the cosmological
constraint on $\abundchi$ appears less severe than what has been
previously thought.

Our findings in the CMSSM strongly support the idea of low energy
SUSY, although of course they do not disfavor other possibilities,
like split SUSY~\cite{splitsusy}. Quantitatively, at 68\% probability,
we have found $0.52\tev<\mhalf< 1.26\tev$, $\mzero<2.10\tev$,
$-0.34\tev<\azero<2.41\tev$ and $38.5<\tanb<54.6$. The corresponding
ranges of superpartner masses are given in column two of
Table~\ref{table:massesTable}.  A significant fraction of the 68\%
probability ranges of superpartner masses falls within the LHC reach,
and typically outside the Tevatron reach.  The same applies to $BR(B_s
\rightarrow \mu^+ \mu^-)$ for which most favored CMSSM values are not
far above the SM prediction. On the other hand, a positive measurement
of $BR(B_s \rightarrow \mu^+ \mu^-)$ at the Tevatron would strongly
disfavor large $\mzero$, including the focus point region. The WIMP DM
direct detection elastic scattering cross section $\sigsip$ shows a
wide spread of values (below today's limits) at around
$10^{-9\pm1}\pb$ and a strong anti--correlation with $\mchi$. In
addition, a relatively large $\sigsip\simeq 1.6\times10^{-8}\pb$, 
fairly independent of $\mchi$, appears to be a feature of the FP
region (despite large theoretical uncertainties) and will probably be the first
to be tested in direct detection experiments.

The $\gmtwo$ anomaly still remaining the subject of some controversy,
we re--examined its impact on the CMSSM parameter space.  We showed
the inference to be substantial on $\mzero$, and any superpartner
masses that primarily depend on it, while much less so with the other
CMSSM parameters. The chance for the LHC to detect superpartners
reduces somewhat but still remains strong. 

\acknowledgments R.RdA thanks B.~Allanach, G.~B\'{e}langer,
A.~Dedes, J.~Foster, G.S.~~Heinemeyer and K.~Okumura. R.T.\ is
grateful to G.~Nicholls for useful discussions. R.RdA is supported
by the program ``Juan de la Cierva'' of the Ministerio de
Educaci\'{o}n y Ciencia of Spain. R.RdA and R.T. would like to thank the European
Network of Theoretical Astroparticle Physics ILIAS/N6 under contract
number RII3-CT-2004-506222 for financial support.
R.T.\ is supported by the Royal
Astronomical Society through the Sir Norman Lockyer Fellowship. LR
acknowledges partial support from the EC 6th Framework
Programme MRTN-CT-2004-503369 ``The Quest for Unification''.
The use of the Glamdring cluster of Oxford University and the HEP
cluster of the University of Sheffield is acknowledged. Parts of the code
used are based on the publicly available package
\texttt{cosmomc}.\footnote{Available from
\texttt{cosmologist.info}.}

\appendix

\section{Markov chain Monte Carlo algorithm \label{appx:mcmc}}

 \subsection{Sampling}

The purpose of the Markov Chain Monte Carlo (MCMC) algorithm is to
construct a sequence of points in parameter space (called ``a
chain'') whose density is proportional to the posterior pdf of
Eq.~\eqref{eq:bayes}. Once such a chain has been produced, the
posterior probability for a given region of parameter space (a
bin) is obtained by simply counting the number of samples within
that region. Marginalization over nuisance parameters (see
Eq.~\eqref{eq:marginalization}) is trivial: the coordinates of the
parameters that one is not interested in are simply ignored when
counting the samples.

Several algorithms are available to construct Markov chains, which
are more or less suited to the structure of the parameter space
under consideration, see \eg~\cite{Neal93} for an introductory
review and references therein. We make use of the Metropolis
algorithm: from a starting point in parameter space $\basis_0$
with associated posterior probability $p_0$, a candidate point
$\basis_c$ for the next sample is proposed by sampling it from a
transition probability $T(\basis_0,\basis_c)$. The candidate
sample is then accepted with probability
 \be
\alpha = \min\(\frac{p_c}{p_0}, 1\),
 \ee
 where $p_c = p(\basis_c |\data)$. Notice that all steps for which the
 candidate sample has a better probability than the previous one
are accepted. If the candidate point is accepted, it becomes the
new starting point, and another candidate is drawn. Otherwise
the chain stays at the previous point (which is thus counted twice) and
a new attempt is made from there. It can be shown that the
sequence of samples $\basis_0, \basis_1, \dots, \basis_t, \dots$
constructed this way converges to the target distribution
$p(\basis | \data)$ for $t\rightarrow \infty$.

The transition probability $T(\basis_t,\basis_{t+1})$ must be
symmetric in its arguments (so called ``detailed balance''), a
sufficient condition which ensures that the Markov chain 
constructed this way is sampling from the target probability distribution.
In our case, the transition probability is given by the following
prescription, which we found strikes a good balance between
efficiency and robustness of the exploration. At each step, we
alternatively update the value of the CMSSM parameters $\params$
or of the nuisance parameters $\nuis$. In general, all of the
components of either $\params$ or $\nuis$ are updated at each
step. The candidate point is proposed along the direction $w$,
where the vector $w$ is restricted to either one of the two
subspaces (CMSSM or nuisance parameters) and it is given by
 \be
 {w} = {A} \cdot {u}.
 \ee
Here, ${A}$ is a random rotation matrix which is restricted to a given
subspace and which is renewed every 4 steps. The components of
${u}$ are initially chosen as a guess of the typical spread of the
posterior distribution along each direction of parameter space.
The results of a preliminary MCMC run are then used to estimate
the covariance matrix for the posterior pdf, whose eigenvectors
give directions of approximate degeneracies in the parameter
space. In the final run, ${w}$ is built analogously as above, but
this time by a random rotation in the space spanned by the
projections of the eigenvectors of the covariance matrix. This
procedure aims at aligning the directions of the proposals to
degeneracy lines in parameter space, thus improving the efficiency
of the MCMC walk.

Along the direction defined by ${w}$, the width of the step is chosen
by multiplying $|{w}|$ by a scaling factor $s$ and a factor $r$ drawn
from the distribution 
\be \label{eq:prop} 
p(r) \propto \frac{2}{3}
r^{n-1} \exp(-n r^2/2) + \frac{1}{3}\exp(-r), 
\ee 
with $n = 4$ and $s =
2.4$. The first term on the rhs of the proposal distribution
\eqref{eq:prop} tends to make the chain move away from $r=0$ for
$n>1$, while the second term increases the probability near the
origin. Thus this distribution tends to be robust even in situations
where the target pdf is strongly non--Gaussian. The above choices for
the proposal distribution and for the parameters $n,s$ are mostly a
matter of trial--and--error.  Our updating procedure follows closely
the recommendations of~\cite{BLcosmo}. Notice that since our choice of
the step direction and size do not depend on $\basis_t,\basis_{t+1}$
at any moment, the condition of detailed balance for the transition
probability holds true.

\subsection{Convergence}

We start $N=12$ or $N=16$ chains in randomly chosen points of
parameters space (within the boundary specified by our prior
range), making sure that they lie well apart from each other in
order to maximise the initial variance. A certain number of
samples have to be discarded at the beginning of the chain, since
the chain requires some time to equilibrate around the target
distribution. This ``burn--in period'' is assessed by inspecting
the evolution of the probability as a function of the number of
samples. We find that discarding $10^3$ initial samples is more
than sufficient.

The acceptance rate is defined as the percentage of accepted
proposals. For the runs using the covariance matrix the typical
acceptance rate is in the range $2-3\%$. This is rather low,
compared to optimal cases where the acceptance rate is typically
an order of magnitude larger. The reason for this is the
configuration of the posterior pdf in multi--dimensional parameter
space, which is strongly non--Gaussian and presents a challenging
combination of wide regions and narrow wedges of high probability.
There is no optimal strategy in this case, but we are confident
that our chains have appropriately sampled the whole parameter
space. We performed extensive checks by comparing runs obtained
with and without the covariance matrix in order to make sure that
the efficiency gain did not come at the expenses of a reduced
sampling ability.

Mixing and convergence of the chains is assessed with the Gelman
\& Rubin R--statistics~\cite{GRR}. This represents the variance of
the means divided by the mean of the variances between different
chains. The usual criterion is that $R-1 \lsim 0.2$, but to be
conservative we require that for our chains $R-1< 0.05$ for all
parameters (this means that our convergence criteria are more
stringent).

Typically we run in parallel two sets of $N=12$ or $N=16$ chains,
until each chain within the set has reached $3\times 10^4$ or
$4\times 10^4$ samples\footnote{Since each sample is obtained with
a typical acceptance rate of 3\%, this means that each chain
requires $\mathcal{O}(10^6)$ likelihood evaluations, each of which
takes about 1--2~\second\ on our machines.} (the exact numbers depending on
the computing power available and on the convergence status). We
check that each run has converged using the criterion outlined
above, and we then perform consistency checks between the two
runs. The final inferences are obtained after merging the two sets
of chains together. At this final stage, the Gelman and Rubin
criterion is again satisfied by the merged set consisting of $24
\leq N \leq 32$ chains, containing a total number of samples in
the range of $0.7 \times 10^6$ to $1.3 \times 10^6$.


\end{document}